\documentclass[11pt]{article}
\pdfoutput=1 
\usepackage{shorthand}

\usepackage{mathtools}
\usepackage{booktabs}
\usepackage[english]{babel}
\usepackage{amsmath,amssymb,amsbsy,amstext, amsthm, simplewick, amsfonts}
\usepackage{graphicx}
\usepackage[small]{caption}
\usepackage{siunitx}
\usepackage{upgreek}
\usepackage{framed}
\usepackage{wrapfig}
\usepackage{multirow}
\usepackage{bbm}
\usepackage[numbers,sort&compress]{natbib}
\usepackage[svgnames,dvipsnames,x11names]{xcolor}
\usepackage[utf8x]{inputenc}
\usepackage{selinput}
\usepackage{bm}
\usepackage{braket}
\usepackage{float}
\usepackage{dsfont}
\usepackage{yfonts}
\usepackage{caption}
\usepackage{subcaption}
\usepackage{sidecap}
\usepackage{longtable}
\usepackage{anyfontsize}

\usepackage{epstopdf}
\usepackage{cancel}
\usepackage{tcolorbox}

\usepackage{jcapmod}

\def\xyma{\xymatrix@M.7em}
\def\xymas{\xymatrix@M.1em}

\newcommand{\Comment}[1]{{}}

\definecolor{blue3}{RGB}{31, 119, 180}
\definecolor{red3}{RGB}{	214, 39, 40}
\definecolor{darkblue}{rgb}{0.15,0.35,0.55}
\definecolor{reddish}{rgb}{0.65, 0.2, 0.2}
\definecolor{darkgreen}{RGB}{50,150,0}
\definecolor{greyish2}{rgb}{.96,.96,.96}
\definecolor{powderblue(web)}{rgb}{0.69, 0.88, 0.9}
\definecolor{palerobineggblue}{rgb}{0.59, 0.87, 0.82}

\usepackage[linktocpage=true]{hyperref}
\hypersetup{
colorlinks=true,
citecolor=darkblue,
linkcolor=reddish,
urlcolor=darkblue,
pdfauthor={},
pdftitle={},
pdfsubject={}
}

\DeclareFontFamily{OT1}{rsfs10}{}
\DeclareFontShape{OT1}{rsfs10}{m}{n}{ <-> rsfs10 }{}
\DeclareMathAlphabet{\mathscript}{OT1}{rsfs10}{m}{n}

\def\gsim{ \lower .75ex \hbox{$\sim$} \llap{\raise .27ex \hbox{$>$}} }
\def\lsim{ \lower .75ex \hbox{$\sim$} \llap{\raise .27ex \hbox{$<$}} }
\def\be{\begin{equation}}
\def\ee{\end{equation}}
\def\bea{\begin{eqnarray}}
\def\eea{\end{eqnarray}}

\newcommand{\rd}{{\rm d}}
\newcommand{\vp}{\varphi}

\usepackage{latexsym,amsmath,amssymb,epsfig}

\usepackage[letterpaper,margin=1in]{geometry}
\usepackage{tikz}
\usetikzlibrary{decorations}
\pgfdeclaredecoration{complete sines}{initial}
{
    \state{initial}[
        width=+0pt,
        next state=upsine,
        persistent precomputation={\pgfmathsetmacro\matchinglength{
            \pgfdecoratedinputsegmentlength / int(\pgfdecoratedinputsegmentlength/\pgfdecorationsegmentlength)}
            \setlength{\pgfdecorationsegmentlength}{\matchinglength pt}
        }] {}
    \state{upsine}[width=\pgfdecorationsegmentlength,next state=downsine]{
        \pgfpathsine{\pgfpoint{0.25\pgfdecorationsegmentlength}{0.5\pgfdecorationsegmentamplitude}}
        \pgfpathcosine{\pgfpoint{0.25\pgfdecorationsegmentlength}{-0.5\pgfdecorationsegmentamplitude}}
    }
    \state{downsine}[width=\pgfdecorationsegmentlength,next state=upsine]{
        \pgfpathsine{\pgfpoint{0.25\pgfdecorationsegmentlength}{-0.5\pgfdecorationsegmentamplitude}}
        \pgfpathcosine{\pgfpoint{0.25\pgfdecorationsegmentlength}{0.5\pgfdecorationsegmentamplitude}}
}
    \state{final}{}
}

\definecolor{greyish}{rgb}{.90,.90,.90}
\definecolor{greyish2}{rgb}{.96,.96,.96}
\usepackage{xcolor,colortbl}
\usepackage{tcolorbox}

\usepackage[all]{xy}

\usepackage{ytableau}

\definecolor{labelkey}{rgb}{0,0,1}

\numberwithin{equation}{section}

\begin{document}
%

\begin{titlepage}

\renewcommand{\thefootnote}{\fnsymbol{footnote}}
\vspace{0truecm}
\begin{center}
{\fontsize{18}{18} \bf{Soft limits of the wavefunction}}\\[5pt]
{\fontsize{18}{18} \bf{in exceptional scalar theories}}
\end{center} 

\vspace{.3truecm}

\begin{center}
{\fontsize{12.5}{18}\selectfont
Noah Bittermann${}^{\rm a}$ and Austin Joyce${}^{\rm b}$
}
\end{center}
\vspace{.2truecm}

 \centerline{{\it ${}^{\rm a}$Center for Theoretical Physics, Department of Physics,}}
 \centerline{{\it Columbia University, New York, NY 10027}} 
 
  \vspace{.3cm}
 
 \centerline{{\it ${}^{\rm b}$Kavli Institute for Cosmological Physics, Department of Astronomy and Astrophysics,}}
 \centerline{{\it University of Chicago, Chicago, IL 60637}}
 
  \vspace{.3cm}

\vspace{1cm}

\begin{center}{\bf Abstract}
\end{center}
\noindent
We study the structure of the flat space wavefunctional in scalar field theories with nonlinearly realized symmetries. 
These symmetries imply soft theorems that are satisfied by wavefunction coefficients in the limit where one of the external momenta is scaled to zero.
After elucidating the structure of these soft theorems in the nonlinear sigma model, Dirac--Born--Infeld, and galileon scalar theories, we combine them with information about the singularity structure of the wavefunction to bootstrap the wavefunction coefficients of these theories.
We further systematize this construction through two types of recursion relations: one that utilizes the flat space scattering amplitude plus minimal information about soft limits, and an alternative that does not require amplitude input, but does require subleading soft information.

\end{titlepage}

\newpage
\setcounter{page}{2}

\setcounter{tocdepth}{2}

\tableofcontents

\newpage
\renewcommand*{\thefootnote}{\arabic{footnote}}
\setcounter{footnote}{0}

\section{Introduction}

Some of the deepest insights arising from the study of scattering amplitudes have been the discovery of structural motifs that appear in seemingly unrelated theories. In many cases this is a reflection of the underlying rigidity of consistent quantum field theories.  These recurring patterns in field theories take on various forms. A famous manifestation of this underlying structure is Weinberg's soft theorem~\cite{Weinberg:1965nx}, which shows that gauge theory amplitudes in the soft limit  have a universal form. Another important example is provided by BCFW recursion relations, which make it possible to systematically construct amplitudes for complicated processes from simpler building blocks~\cite{Britto:2005fq}. Similarly, the double copy makes precise the notion that some theories can be thought of as squares of others~\cite{Bern:2019prr}. Taken together, these examples are hints of deeper structures that relate different field theories. In some cases these relations can be made  more explicit via the double copy and other transmutation operations that transform theories into each other~\cite{Bern:2010ue,Cachazo:2013iea,Cachazo:2014xea,Cheung:2017ems}. Not only are  these relations conceptually illuminating, but they are practically useful in enabling computations that would otherwise be out of reach.

\vskip4pt
These rich structures also appear in certain scalar field theories. Consistent theories of massless spinning particles are highly constrained---with Yang--Mills and Einstein gravity being essentially unique at low energies~\cite{Boulanger:2000rq,Benincasa:2007xk,Schuster:2008nh,Porrati:2008rm,McGady:2013sga}---and this rigid structure partially explains the recurrence of various features. Massless scalar theories are somewhat less constrained, in the absence of any additional assumptions. However, requiring that the scalar field behave as a Nambu--Goldstone boson---nonlinearly realizing certain symmetries---is a sufficiently strong demand to make interesting structures reemerge. For example, scalar field theories can then be classified by their amplitudes' behavior in the soft limit~\cite{Cachazo:2014xea,Cheung:2014dqa,Cheung:2016drk}, with the nonlinear sigma model (NLSM), Dirac--Born--Infeld (DBI), and special galileon theories appearing as distinguished points in theory space. These exceptional scalar theories exhibit interesting relations to each other, and also to Yang--Mills and Gravity~\cite{Cachazo:2014xea,Cheung:2017yef,Cheung:2017ems}, and their amplitudes can be constructed by means of recursion relations, similar to BCFW~\cite{Cheung:2015ota,Padilla:2016mno,Elvang:2018dco,Bonifacio:2019rpv}. The interesting features and relative simplicity of these exceptional scalar field theories, along with their connections to gauge theory amplitudes, make them ideal places to explore the hidden structures in scattering amplitudes.

\vskip 4pt
In this paper, we explore the analogues of these on-shell structures in the wavefunctional of Nambu--Goldstone scalar theories. This is motivated by recent progress in the study of cosmological correlation functions.  Compared to our understanding of scattering amplitudes, our knowledge of correlators (or the wavefunction) is considerably less sophisticated, even at tree level. Nevertheless, much is now known about the singularity structure of cosmological correlators~\cite{Maldacena:2011nz,Raju:2012zr,Arkani-Hamed:2017fdk,Arkani-Hamed:2018kmz,Baumann:2020dch}, including how they encode locality. Other properties of bulk time evolution are captured by the way that correlators behave as we vary the kinematic parameters to move away from singular configurations~\cite{Arkani-Hamed:2018kmz,Baumann:2021fxj}. For example, bulk unitarity has been studied both perturbatively~\cite{Goodhew:2020hob,Cespedes:2020xqq,Benincasa:2020aoj,Meltzer:2020qbr} and non-perturbatively~\cite{Sleight:2020obc,Hogervorst:2021uvp,DiPietro:2021sjt,Sleight:2021plv} in the cosmological context. Aside from providing insights into the structure of quantum field theory, these formal developments have also enabled the calculation of inflationary signatures that would be otherwise intractable (see, e.g.,~\cite{Arkani-Hamed:2015bza,Arkani-Hamed:2018kmz,Baumann:2019oyu,Sleight:2019mgd,Sleight:2019hfp,Sleight:2021iix,Albayrak:2018tam,Albayrak:2019yve,Benincasa:2019vqr,Pajer:2020wxk,Bonifacio:2021azc,Cabass:2021fnw,Hillman:2021bnk}).\footnote{An important phenomenological motivation for these studies is the possibility of using the inflationary background as a sort of cosmological collider~\cite{Chen:2009zp,Noumi:2012vr,Arkani-Hamed:2015bza,Arkani-Hamed:2018kmz,Baumann:2019oyu,Assassi:2012zq,Lee:2016vti,An:2017hlx,Kumar:2017ecc,Alexander:2019vtb,Wang:2019gbi,Wang:2020ioa,Wang:2021qez,Lu:2021wxu,Tong:2021wai}.}

\vskip4pt
Despite recent progress, many deep mysteries remain in the study of cosmological correlators, and our investigation is aimed at shedding light on some of these issues. Much in the same way that exceptional scalar field theories have been useful in the study of flat space scattering, it is natural to expect that there will be hidden structures in the wavefunction of these exceptional scalar theories. Though our ultimate interest is in cosmology, in this paper we specialize to the study of the wavefunction in flat space. The study of the flat space wavefunction has already proven to be useful for the understanding of de Sitter correlators, leading to many insights that can be imported into the cosmological setting~\cite{Arkani-Hamed:2017fdk,Arkani-Hamed:2018bjr,Benincasa:2018ssx,Hillman:2019wgh}. Even more concretely, in many cases of interest the cosmological wavefunction can be obtained from these flat space expressions by acting with appropriate transmutation operations~\cite{Arkani-Hamed:2017fdk,Benincasa:2019vqr,Baumann:2020dch,Hillman:2021bnk}. We therefore anticipate that the lessons learned for exceptional scalar theories in the flat space context can similarly be abstracted into the cosmological setting.

\paragraph{Summary:}
\noindent For convenience, we summarize our main results:
\begin{itemize}
	\item We derive the soft theorems that wavefunction coefficients satisfy as a consequence of nonlinearly realized shift symmetries.  The general soft theorem is given by equation~\eqref{eq:generalsoftthm}, which we apply to a number of theories exhibiting these symmetries.  Evidently, the Ward identity is far more complicated than the  (enhanced) Adler zero condition enjoyed by the analogous scattering amplitudes.  Ultimately this is because, compared to scattering amplitudes, wavefunction coefficients depend on a single additional variable---the total energy involved in the relevant process. Though this difference might seem somewhat innocuous, it leads to important structural differences between these two objects.  In particular wavefunction coefficients obey soft theorems relating $n$-point wavefunction coefficients to lower point wavefunction coefficients when one of their external momenta is taken to be soft.\footnote{Scattering amplitudes can also satisfy soft theorems when there are cubic vertices present that are compatible with the nonlinearly realized symmetries, for example in the conformal dilaton~\cite{Callan:1970yg}, or in non-relativistic cases~\cite{Grall:2020ibl} (see also~\cite{Kampf:2019mcd} for another relativistic example).}

	\item We organize soft information into a bootstrap-like construction and use this to fix the wavefunction coefficients of theories with shift symmetries.  A conceptually interesting question is whether the wavefunction contains more, less, or the same information as the flat space $S$-matrix. In some cases it is known that these two objects can be constructed from each other~\cite{Benincasa:2018ssx}.  In these soft scalar theories, in addition to the scattering information, one requires information about the soft theorems that the wavefunction satisfies in order to reconstruct wavefunction coefficients uniquely.\footnote{This is perhaps unsurprising because there are ambiguities related to (position space) contact terms and field redefinitions that have to be fixed in order to uniquely specify a wavefunction coefficient. See~\cite{Green:2020ebl} for a similar discussion in the context of inflation.}

\end{itemize}

\begin{itemize}
	\item We systematize the  bootstrap  by deriving recursion relations for wavefunction coefficients.  The general recursion formula is given in equation \eqref{eq:correlatorrecurse}, and is obtained by deforming the kinematics that the wavefunction depends on into the complex plane.  The recursion relations are conceptually different from those in~\cite{Arkani-Hamed:2017fdk,Jazayeri:2021fvk,Baumann:2021fxj}, which deform the energy variables. In order to input information about the soft structure of the wavefunction, it is important to 
deform the momentum variables directly.  An interesting feature of the complexified wavefunction in this case is that it has branch cuts.
In a sense, these branch cuts are avatars of particle exchange, and it turns out that the wavefunction factorizes into a product of lower point shifted wavefunctions along these branch cuts.  This is analogous to how scattering amplitudes factorize into lower point amplitudes on their poles. We recursively construct the wavefunction with two different sets of inputs.
First, we recurse wavefunction coefficients from their respective scattering amplitudes and soft theorems.  Then, we demonstrate that for theories exhibiting higher order soft theorems (NLSM, DBI, special galileon), it is possible to construct the wavefunction without knowledge of any scattering information at all, though at the expense of complicating the procedure.  An interesting feature of these constructions it that we are able to effectively give a definition of these exceptional scalar theories directly at the level of the wavefunction without referring to the underlying action. 

	\item There are a number of technical intermediate results that may be of independent interest. Many of these details are given in the appendices. In particular, it is interesting to note that the classical canonical momentum is a generating functional for tree level wavefunction coefficients, which is expressed in~\eqref{canonicalgenerator}. Perhaps more surprisingly, we also show that the classical field profile at early times is a generating functional for tree-level shifted wavefunction coefficients.  This is expressed in~\eqref{earlytimegenerator}, and indicates that shifted wavefunction coefficients possess information about the system at early times. These statements are analogues of the fact that the classical field profile in the presence of a source with a Feynman pole prescription is a generating functional of in/out correlators.  We expect that both of these formulae hold at loop order as well, and also have de Sitter analogues. 

\end{itemize}

\vskip4pt
\paragraph{Outline:}
In Section~\ref{sec:wavefunctionandsoft} we first review the definition and perturbative calculation of the quantum field theory wavefunctional, which is the object of interest. We then describe how wavefunction coefficients in scalar theories with nonlinearly realized symmetries obey soft theorems. In Section~\ref{sec:WFfromSmatrix} we first derive the relevant soft theorems for the NLSM, DBI, and the special galileon. We then show how information about the wavefunction's singularities (including the fact that the residue of one of its singularities is the corresponding scattering amplitude) along with partial information about the soft limit is sufficient to uniquely reconstruct the wavefunction.  In Section~\ref{sec:softrecurs}, we systematize the construction of the wavefunction in these theories by deriving recursion relations that input information about singularities and soft limits in two ways. The first is a systematic implementation of the arguments of Section~\ref{sec:WFfromSmatrix}, which relies on scattering information as one of the inputs. It is reasonable to ask if it is possible to replace the amplitude information with knowledge of the full soft theorems that a given theory obeys, and indeed we show that this is the case by explicitly constructing recursion relations relying only on soft information, but restricting our discussion to the NLSM and DBI cases for simplicity. We conclude in Section~\ref{sec:conclusions}. A number of appendices collect technical information that is somewhat outside the main line of development. In Appendix~\ref{app:scalartheories} we provide a brief review of the exceptional scalar theories that we study in this paper. In Appendix~\ref{app:technical}, we discuss many of the important technical subtleties that must be addressed in order to give a boundary definition of the wavefunction of a higher-derivative bulk theory. In Appendix~\ref{app:fieldasgenfunct} we show how the early time classical field profile can be viewed as the generating functional of shifted wavefunction coefficients. In Appendix~\ref{app:cuts} we discuss some details of the analytic structure of the wavefunction. Finally in Appendix~\ref{app:Dcopy} we briefly discuss some aspects of the wavefunction double copy for the theories of interest.

\paragraph{Conventions \& Notation: } We work with the mostly plus metric signature in four spacetime dimensions, use Greek letters, e.g., $\mu,\nu,\rho,\cdots$ to indicate spacetime indices, and use Roman letters from the middle of the alphabet, e.g., $i,j,k,\cdots$ to indicate spatial indices. We index various particles/lines/operators by Roman letters from the beginning of the alphabet, e.g., $a,b,c,\cdots$. We label spatial momenta by $\vec k_a$ (or $\vec{p}_a$) with magnitude $k_a \equiv \sqrt{\vec k_a^2}$ (or $p_a \equiv \sqrt{\vec p_a^{\,2}}$), which we will refer to as ``energies". We Fourier transform with the convention
\be
f(x) = \int\frac{\rd^3k}{(2\pi)^3}e^{-i\vec k\cdot\vec x} f_{\vec k}\,.
\ee
We denote sums of energies by $k_{12\cdots n} = k_1+k_2+\cdots k_n$. For a given process, we denote the total energy involved in the process by $E$ (irrespective of the number of external lines). In many cases we use the following partial energies: $E_{a_1a_2\cdots a_n} \equiv k_{a_1}+k_{a_2}+\cdots+k_{a_n}+\lvert \vec k_{a_1}+\vec k_{a_2}+\cdots +\vec k_{a_n}\rvert$, which denote the energy flowing into a vertex of an exchange diagram. We also define the exchanged momenta $\vec s_{a_1\cdots a_n}\equiv  \vec k_{a_1}+\vec k_{a_2}+\cdots +\vec k_{a_n}$ and its corresponding energy $ s_{a_1\cdots a_n}\equiv \lvert \vec k_{a_1}+\vec k_{a_2}+\cdots +\vec k_{a_n}\rvert$. Combined, we see that for example, $E_{12} = k_{12}+s_{12} \equiv E_L^{(s)}$ flows into the left vertex of a four-point tree level exchange diagram in the $s$-channel.  When describing scattering amplitudes, we will also often make use of generalized Mandelstam variables, which we define to be $S_{i_1...i_n} = -(P_{i_1} + ... + P_{i_n})^2$.  The variable $\phi$ denotes a field propagating in the four-dimensional bulk spacetime, while we use $\varphi$ to denote its profile on the time slice where we compute the wavefunction. Other notational conventions are introduced as they arise.

\newpage
\section{Wavefunction soft theorems}
\label{sec:wavefunctionandsoft}

Our goal is to explore the features of the soft limit, where one of the external momenta is taken to zero, in the wavefunction of scalar theories. We are particularly interested in understanding the extent to which the wavefunction is fixed by its singularity structure and soft limit. In the context of scattering amplitudes, enhanced Adler zeroes---where amplitudes vanish faster than expected in the soft limit---correspond to nonlinearly realized shift symmetries from the field theory perspective~\cite{Cheung:2014dqa,Cheung:2016drk}. We therefore begin by exploring the consequences of these symmetries for the wavefunction.  As we will show, an important difference between the wavefunction and the $S$-matrix is that wavefunction coefficients arising from exceptional scalar theories typically obey soft theorems rather than having Adler zeroes.  A simple way to understand the difference between the amplitude and wavefunction case is to note that the highest-order enhanced Adler zeroes of amplitudes are a result of a cancellation between exchange and contact contributions. In the wavefunction context, these two contributions have a different analytic structure with respect to internal energies, $s_I$, so they cannot cancel.
 As a result, one must understand how to translate the symmetries enjoyed by these theories into the relevant soft theorems. We will later utilize these soft theorems as a bootstrap input to generate wavefunction coefficients.

\subsection{Review of the wavefunction}

We begin by briefly reviewing the definition of the quantum field theory wavefunctional and its associated wavefunction coefficients, which will be the objects of central interest. (For more details see, e.g.,~\cite{Anninos:2014lwa,Arkani-Hamed:2017fdk,Goon:2018fyu,Baumann:2020dch}.)

\vskip4pt
The wavefunctional of interest is a representation of the ground state of an interacting field theory given by projecting onto the Heisenberg-picture eigenstates of the fields, $\phi$, as $\Psi[\varphi(\vec x), t_*] \equiv \langle \varphi\rvert 0\rangle$.\footnote{The field eigenstates $\lvert\varphi\rangle$ satisfy $\phi(\vec x, t)\lvert\varphi\rangle= \varphi(\vec x)\lvert\varphi\rangle$.} The wavefunctional is therefore naturally a function of the field profile $\varphi(\vec x)  \equiv \phi(\vec x, t_*)$ at the time $t_*$. Given the wavefunctional, we can recover correlation functions of the field $\phi$ at time $t_*$ by employing the usual quantum mechanics formula:
\be
\langle\vp(\vec x_1)\cdots \vp(\vec x_n)\rangle =\frac{ \displaystyle\int{\cal D} \vp\,\vp(\vec x_1)\cdots \vp(\vec x_n) \left\lvert\Psi[\vp]\right\rvert^2}{\displaystyle\int{\cal D} \vp \left\lvert\Psi[\vp]\right\rvert^2} \, .
\label{eq:corrfromwf}
\ee
It is convenient to organize the late-time wavefunctional in a series of connected \textit{wavefunction coefficients},  $\psi_n(\vec k_N)$, in Fourier space as:\footnote{We will often refer to these wavefunction coefficients---in a slight abuse of terminology---as wavefunctions.}
\be
\log\Psi[\vp, t_f] = \sum_{n = 2}\frac{1}{n!}\int \frac{\rd^{3}k_1\cdot\cdot\cdot \rd^{3}k_n}{(2\pi)^{3n}}\,\vp_{\vec{k}_1}\cdot\cdot\cdot \vp_{\vec{k}_n}\,(2\pi)^3  \delta(\vec k_1+\cdots+\vec k_n)\, \psi_n(\vec k_N)\,,
\label{coefficientexpansion}
\ee
where the wavefunction coefficients are functions of the set of momenta $\vec k_N \equiv \{\vec k_1, \ldots, \vec k_n \}$. We will also often make use of the canonical commutation relation
\be
[\phi(\vec{x}, t), \Pi^{(\phi)}(\vec{y}, t)] = i\delta^{(3)}(\vec{x} - \vec{y})\,,
\ee
 which given our Fourier convention reads as follows in momentum space
\be
[\phi_{\vec{k}_1}(t), \Pi^{(\phi)}_{\vec{k}_2}(t)] = i(2\pi)^3\delta^{(3)}(\vec{k}_1 + \vec{k}_2)\,.
\ee
In this representation, the canonical momentum is realized as a functional derivative $\Pi^{(\phi)}_{\vec{k}} = (2\pi)^3\frac{\delta}{i\delta \phi_{-\vec{k}}}$.

\subsubsection{Perturbation theory}

In many cases it is convenient to express the wavefunction at some late time $t_f$ as a path integral that evolves a wavefunction from an initial time $t_i$, which we denote by $\Psi_i$, in the following way
\be
 \Psi[\vp_f, t_f] =\braket{\vp_f|0}= \int\mathcal{D}\vp_{i}\braket{\vp_f|\vp_{i}}\braket{\vp_{i}|0} =\int\mathcal{D}\vp_{i}
\int\limits_{\substack{\hspace{-0.4cm}\phi(t_f) \,=\,\vp_f\\ \hspace{-0.4cm}\phi(t_i)\,=\,\vp_i}} 
\hspace{-0.5cm} \raisebox{-.05cm}{ ${\cal D} \phi\, e^{iS[\phi]}$ }
\Psi_i[\vp_i, t_i]\,.
\label{eq:latetimepathintegral}
\ee
The path integration is done over all field configurations connecting the profile $\vp_i$ at the initial time to $\vp_f$ at the final time.
In the cases of interest, we will take the initial time to be in the infinite past $(t_i\to-\infty)$ with an initial wavefunctional that is a gaussian
\be
\Psi_i[\vp_i, t_i]=  \braket{\vp_i|0} \propto \exp\left(-\frac{1}{2}\int \frac{\rd^3k}{(2\pi)^3}\,\mathcal{E}(E_k)\,\vp_i{}_{\vec k}\,\vp_i{}_{-\vec{k}}\right)\,,
\label{eq:GaussianInitial}
\ee
where $\mathcal{E}(E_{k})$ is a kernel capturing the statistics of the initial fluctuations, which depends on the energy of a mode with momentum $\vec k$, denoted as $E_k = \sqrt{\vec k^2+m^2}$. In what follows, we will always set $m^2 = 0$ and consider massless fields, so that $E_k = k$.\footnote{In the massless case, the vacuum wavefunction is actually highly degenerate due to the free action having an infinite tower of nonlinearly realized symmetries which are all spontaneously broken.  We will always take \eqref{eq:GaussianInitial} to be the initial state in our computations, and routinely abuse language by referring to this state as \textit{the} vacuum state.}
We will primarily be interested in wavefunction coefficients in  Minkowski spacetime, where ${\cal E}(E_k) = E_k$.
Sometimes when performing the path integral~\eqref{eq:latetimepathintegral} it is convenient to trade the dependence on the initial state for an unconstrained path integral, at the cost of introducing $i\epsilon$ terms into the action. However, since the initial state is typically not invariant under the symmetries of interest---and thus contributes to the Ward identities we will derive---we will often keep it explicit.

\vskip4pt
Similar to scattering amplitudes, the computation of the wavefunction coefficients appearing in~\eqref{coefficientexpansion} can be organized into a diagrammatic perturbative expansion. There are two essential differences from the computation of $S$-matrix elements. The first is that energy is not necessarily conserved, because we have broken time translation invariance by choosing a surface on which to compute the wavefunction. Relatedly, the second difference is that there are now two different kinds of propagators that appear in diagrams.  First, there are those that connect bulk vertices to the boundary $t= t_f$ surface:
\be
{\cal K}(k,t) = e^{iE_kt}\,, \label{eq:bboundary}
\ee
which we refer to as the {\it bulk-to-boundary propagator} in analogy to AdS/CFT. In the expression~\eqref{eq:bboundary} we have taken $t_f = 0$, without loss of generality. In contrast, lines that connect bulk vertices to each other represent the {\it bulk-to-bulk propagator}\footnote{This Green's function satisfies $(\partial_t^2+E_k^2){\cal G}(k;t,t') = -i\delta(t-t')$.}
\be
{\cal G}(k;t,t') = \frac{1}{2E_k}\left(e^{i E_k(t'-t)}\theta(t-t')+e^{i E_k(t-t')}\theta(t'-t)-e^{iE_k(t+t')}\right)\,,
\label{eq:bbulkprop}
\ee
which differs from the usual Feynman propagator by an un-time-ordered piece that enforces the boundary condition ${\cal G}(k;0,t')=0$.

\vskip4pt
In order to compute wavefunction coefficients in perturbation theory, we follow a recipe that is quite similar to the computation of scattering amplitudes. We derive Feynman rules from the vertices in the action in the same way that we would do for the $S$-matrix (only Fourier transforming in the spatial directions), and call the corresponding vertex factors $iV$. Then, we draw all possible Feynman--Witten diagrams with the desired number of lines ending on the $t=0$ surface (for concreteness, $n$), associate to the bulk vertices factors of $iV$, use ${\cal G}$ to connect bulk vertices to each other, and use ${\cal K}$ to connect bulk vertices to the boundary at $t=0$. We then integrate over all the bulk vertex insertion times to produce the wavefunction coefficient $\psi_n$ on the $t=0$ surface.\footnote{In cases that involve loops of internal lines, one should also integrate over the undetermined loop momenta. However, we will restrict ourselves to tree level computations in the following.}

\vskip4pt
Effectively this Feynman diagram expansion is computing the saddle-point approximation of the path integral~\eqref{eq:latetimepathintegral} by first constructing the classical solution to the nonlinear equations of motion with vacuum initial conditions and a given field profile $\varphi_{\vec k}  = \phi_{\vec k}(0)$ at time $t=0$, and then evaluating the action on-shell. The wavefunctional is then a functional of the boundary field profile $\varphi_{\vec k}$. From the bulk perspective the computation of wavefunction coefficients is completely algorithmic, but becomes quite complicated even in flat space as the multiplicity of external lines increases, which is part of the motivation to search for more efficient computational methods.

\subsubsection{Singularities and cuts}
\label{sec:sing}

An important lesson about the structure of wavefunction coefficients is that their singularities largely control their behavior~\cite{Arkani-Hamed:2017fdk,Arkani-Hamed:2018kmz,Baumann:2020dch,Baumann:2021fxj}. Indeed, in some cases the singularity structure completely specifies the wavefunction~\cite{Benincasa:2018ssx,Baumann:2020dch}. Information about the singularities of the wavefunction and their residues therefore serves as useful input from the boundary perspective. The characteristics of the singularities can be thought of as a boundary manifestation of bulk locality. There are also boundary manifestations of bulk unitarity that provide important constraints on the wavefunction~\cite{Goodhew:2020hob,Cespedes:2020xqq,Benincasa:2020aoj,Meltzer:2020qbr}. Specifically, bulk unitarity implies that the wavefunction satisfies an analogue of Cutkosky rules~\cite{Meltzer:2020qbr,Melville:2021lst,Goodhew:2021oqg,Baumann:2021fxj}, that also serve as a useful input to reconstruct the wavefunction~\cite{Jazayeri:2021fvk,Baumann:2021fxj,Cabass:2021fnw}. We will utilize both of these pieces of information in the following, so we briefly review both the singularities and cuts of the wavefunction.

\vskip4pt
\paragraph{Singularities:} We first review the possible singularities of the wavefunction and their residues. Interestingly, this information can be specified in a general way, without specializing to a specific model. Essentially, from the bulk perspective wavefunction coefficients can become singular when the energy flowing into a subgraph happens to add up to zero~\cite{Arkani-Hamed:2017fdk,Baumann:2020dch}. An important special case is when the total energy involved in a process adds up to zero. At this kinematic location, the wavefunction has a singularity whose residue is the corresponding flat space scattering amplitude~\cite{Maldacena:2011nz,Raju:2012zr}:
\be
\lim_{E\to 0}\psi_n = \frac{A_n}{E}\,,
\ee
where we have denoted the total energy as $E\equiv k_1+k_2+\cdots k_n$.\footnote{The intuition for this singularity is that the diagrammatics involved in computing the wavefunction and the $S$-matrix are very similar. However, the different integration region for time---in the amplitude case we go to frequency space which involves integrating over all times---leads to a pole in the total energy rather than an energy-conserving delta function $\delta(E)$.}
This {\it total energy singularity} provides a beautiful connection between the wavefunction and the $S$-matrix: wavefunction coefficients are in a precise sense deformations of scattering amplitudes. Physically, the $E\to 0$ divergence arises from integrating the bulk vertices all the way into the infinite past, which typically is suppressed by an oscillatory factor $\sim e^{iEt}$. When the total energy vanishes, this integration is unsuppressed and diverges.

\vskip4pt
The coefficients of singularities where the energy flowing into a subgraph vanishes---so-called {\it partial energy singularities}---can also be understood in generality. At these locations, the wavefunction factorizes into a product of a lower-point amplitude and a shifted wavefunction. As a concrete example, consider an $(n+m)$-point wavefunction coefficient in the limit that the energies flowing into some particular $n$-point subgraph, $E_{1\cdots n}$, add up to zero. We will assume for simplicity that there is a single internal line connected to the subgraph. Pictorially, we can write this as
\be
\raisebox{-20pt}{
\begin{tikzpicture}[line width=1. pt, scale=2]
\draw[fill=black] (0,0) -- (1,0);
\draw[lightgray, line width=1.pt] (0,0) -- (-0.3,0.75);
\draw[lightgray, line width=1.pt] (0,0) -- (0.3,0.75);
\draw[lightgray, line width=1.pt] (0,0) -- (-0.1,0.75);
\draw[lightgray, line width=1.pt] (0,0) -- (0.1,0.75);

\draw[lightgray, line width=1.pt] (1,0) -- (0.7,0.75);
\draw[lightgray, line width=1.pt] (1,0) -- (1.3,0.75);
\draw[lightgray, line width=1.pt] (1,0) -- (0.9,0.75);
\draw[lightgray, line width=1.pt] (1,0) -- (1.1,0.75);

\draw[lightgray, line width=2.5pt] (-0.5,0.75) -- (1.5,0.75);
\draw[fill=lightgray] (0,0.16) circle (.15cm);
\draw[fill=black] (0,0) circle (.03cm);
\draw[fill=lightgray] (1,0.16) circle (.15cm);
\draw[fill=black] (1,0) circle (.03cm);
\node[scale=1] at (.01,-.155) {$E_{1\cdots n}$};
\node[scale=1] at (1.15,-.155) {$E_{n+1\cdots m+n}$};
\node[scale=1] at (0.52,.115) {$s_{I}$};
\draw[scale=.7,dotted,darkblue] (0,.06) circle (.565cm);
\end{tikzpicture}
}
=\lim_{E_{1\cdots n}\to 0}\psi_{n+m} = \frac{A_n \times \tl\psi_m}{\color{darkblue}{E_{1\cdots n}}}\,.
\label{eq:partialEsing}
\ee
The residue of this singularity is a product of the scattering amplitude, $A_n$, corresponding to the subgraph whose energy is conserved, multiplied by a shifted version of the wavefunction coefficient corresponding to the rest of the graph. Here the shifting is with respect to the internal energy, $s_I$, and is defined by
\be
\tl\psi_m(k_{n+1},\cdots, k_{n+m}, s_I) \equiv \frac{1}{2s_I}\Big(\psi_m(k_{n+1},\cdots, k_{n+m}, -s_I)-\psi_m(k_{n+1},\cdots, k_{n+m}, s_I)\Big)\,.
\ee
The expression~\eqref{eq:partialEsing} can straightforwardly be understood from the form of the bulk-to-bulk propagator with the terms corresponding to the left subgraph taken to the infinite past, where the divergence is localized. These partial energy subgraph singularities are signatures of exchange---wavefunctions arising from contact interactions have only a total energy singularity.\footnote{Though we have focused for simplicity on a graph with a single internal line, the fact that the wavefunction has a pole when the energy flowing into any subgraph vanishes is true for an arbitrary graph and at arbitrary order in perturbation theory, and the corresponding residues can be characterized.}

\vskip4pt
Importantly, the total energy and partial energy singularities are the {\it only} tree level singularities of the wavefunction, which places strong constraints on its analytic structure. In flat space, all of these singularities are simple poles, so that wavefunction coefficients are rational functions in the total and partial energies.\footnote{In de Sitter space, or other cosmological backgrounds, the nature of the singularities can change (for example there are sometimes branch cuts), but the presence of singularities at these---and only these---locations is robust (at least at tree level).}

\paragraph{Cuts:} We can get some further insight into the structure of the wavefunction from the form of the bulk-to-bulk propagator~\eqref{eq:bbulkprop}. If we add to it its complex conjugate, all the time-ordering disappears:
\be
\widetilde {\cal G}(k;t,t') \equiv
 {\cal G}(k;t,t')+{\cal G}^*(k;t,t') = -\frac{1}{2E_k}\big(e^{-iE_kt}-e^{iE_kt}\big)\big(e^{-iE_kt'}-e^{iE_k t'}\big)\,.
 \label{eq:cutprop}
\ee
This suggests that certain combinations of wavefunction coefficients should simplify and be writeable in terms of shifted lower-point wavefunctions. This is indeed the case, and such relations can be systematized as a set of cutting rules satisfied by the wavefunction~\cite{Meltzer:2020qbr,Melville:2021lst,Goodhew:2021oqg,Baumann:2021fxj}, which are consequences of unitarity~\cite{Goodhew:2020hob,Cespedes:2020xqq,Benincasa:2020aoj,Meltzer:2020qbr}. Schematically, the statement is that
\be
\psi_n(X)+ \psi_n^*(-X) = -\sum_{\rm cuts} \psi_n\,,
\label{cutcoskycuts}
\ee
where $X$ is a multi-index standing for all the external energies of a given wavefunction.\footnote{In general in a cosmological spacetime one has to be careful about the precise analytic continuation to negative energies, but this subtlety is unimportant in flat space.} On the right hand side, the sum runs over all partitions of the graph in two, flipping the signs of the external energies of all the vertices to the right of the cut and replacing any internal lines that the cut crosses with the cut propagator~\eqref{eq:cutprop}.  This transforms the original graph into a pair of graphs, each of which computes a shifted wavefunction coefficient. 

\vskip4pt
In practice we will only require the simplest of these cutting rules.  For contact diagrams, the right hand side of \eqref{cutcoskycuts} is zero, indicating that the wavefunction added to itself with its external energies flipped will vanish.  We will also make use of the case of single exchange, which takes the pictorial form (for an $(n+m)$-point function)
\vspace{1pt}
\be
\raisebox{-33pt}
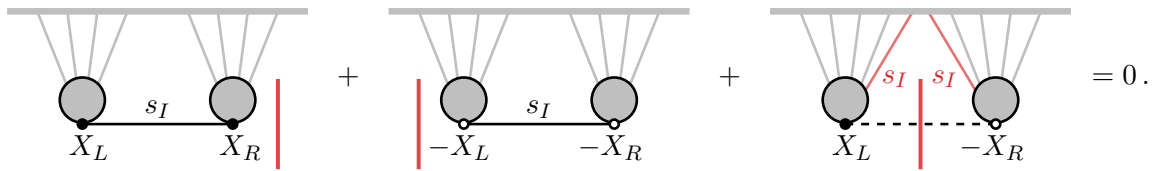
{
\begin{tikzpicture}[line width=1. pt, scale=2]

%
%

\draw[fill=black] (0,0) -- (1,0);
\draw[lightgray, line width=1.pt] (0,0) -- (-0.3,0.75);
\draw[lightgray, line width=1.pt] (0,0) -- (0.3,0.75);
\draw[lightgray, line width=1.pt] (0,0) -- (-0.1,0.75);
\draw[lightgray, line width=1.pt] (0,0) -- (0.1,0.75);

\draw[lightgray, line width=1.pt] (1,0) -- (0.7,0.75);
\draw[lightgray, line width=1.pt] (1,0) -- (1.3,0.75);
\draw[lightgray, line width=1.pt] (1,0) -- (0.9,0.75);
\draw[lightgray, line width=1.pt] (1,0) -- (1.1,0.75);

\draw[lightgray, line width=2.5pt] (-0.5,0.75) -- (1.5,0.75);
\draw[fill=lightgray] (0,0.16) circle (.15cm);
\draw[fill=black] (0,0) circle (.03cm);
\draw[fill=lightgray] (1,0.16) circle (.15cm);
\draw[fill=black] (1,0) circle (.03cm);
\node[scale=1] at (.045,-.155) {$X_L$};
\node[scale=1] at (1.05,-.155) {$X_R$};
\node[scale=1] at (0.5,.1) {$s_{I}$};

\draw[red3, line width=1.5pt,opacity=.85] (1.3,-.3) -- (1.3,0.3);
\end{tikzpicture}
}~+~
\raisebox{-33pt}{
\begin{tikzpicture}[line width=1. pt, scale=2]
%
%

\draw[fill=black] (0,0) -- (1,0);
\draw[lightgray, line width=1.pt] (0,0) -- (-0.3,0.75);
\draw[lightgray, line width=1.pt] (0,0) -- (0.3,0.75);
\draw[lightgray, line width=1.pt] (0,0) -- (-0.1,0.75);
\draw[lightgray, line width=1.pt] (0,0) -- (0.1,0.75);

\draw[lightgray, line width=1.pt] (1,0) -- (0.7,0.75);
\draw[lightgray, line width=1.pt] (1,0) -- (1.3,0.75);
\draw[lightgray, line width=1.pt] (1,0) -- (0.9,0.75);
\draw[lightgray, line width=1.pt] (1,0) -- (1.1,0.75);

\draw[lightgray, line width=2.5pt] (-0.5,0.75) -- (1.5,0.75);
\draw[fill=lightgray] (0,0.16) circle (.15cm);
\draw[fill=white] (0,0) circle (.03cm);
\draw[fill=lightgray] (1,0.16) circle (.15cm);
\draw[fill=white] (1,0) circle (.03cm);
\node[scale=1] at (-.03,-.155) {$-X_L$};
\node[scale=1] at (.975,-.155) {$-X_R$};
\node[scale=1] at (0.5,.1) {$s_{I}$};

\draw[red3, line width=1.5pt,opacity=.85] (-.3,-.3) -- (-0.3,0.3);
\end{tikzpicture}
}~+~
\raisebox{-33pt}{
\begin{tikzpicture}[line width=1. pt, scale=2]
%

%
%
%

\draw[red3, line width=1.pt,opacity=.65] (0,0) -- (0.45,0.75);
\draw[red3, line width=1.pt,opacity=.65] (1,0) -- (0.55,0.75);

\draw[fill=black,dashed] (0,0) -- (1,0);
\draw[lightgray, line width=1.pt] (0,0) -- (-0.3,0.75);
\draw[lightgray, line width=1.pt] (0,0) -- (0.3,0.75);
\draw[lightgray, line width=1.pt] (0,0) -- (-0.1,0.75);
\draw[lightgray, line width=1.pt] (0,0) -- (0.1,0.75);

\draw[lightgray, line width=1.pt] (1,0) -- (0.7,0.75);
\draw[lightgray, line width=1.pt] (1,0) -- (1.3,0.75);
\draw[lightgray, line width=1.pt] (1,0) -- (0.9,0.75);
\draw[lightgray, line width=1.pt] (1,0) -- (1.1,0.75);

\draw[lightgray, line width=2.5pt] (-0.5,0.75) -- (1.5,0.75);
\draw[fill=lightgray] (0,0.16) circle (.15cm);
\draw[fill=black] (0,0) circle (.03cm);
\draw[fill=lightgray] (1,0.16) circle (.15cm);
\draw[fill=white] (1,0) circle (.03cm);
\node[scale=1] at (.045,-.155) {$X_L$};
\node[scale=1] at (.975,-.155) {$-X_R$};

\draw[red3, line width=1.5pt,opacity=.85] (.5,-.3) -- (.5,0.3);
\node[scale=1] at (0.66,.3) {${\color{red3} s_I}$};
\node[scale=1] at (0.33,.3) {${\color{red3} s_I}$};

\end{tikzpicture}
} = 0\,.
\nonumber
\ee
Here $X_{L,R}$ stand schematically for all the energies in the left (right) subgraph, while $s_I$ is the energy of the internal line, and we denote the vertices with their energies flipped by white dots. Translating this into an equation we obtain
\be
\psi_{n+m}(X_L,X_R)+\psi_{n+m}^*(-X_L,-X_R) = -2s_I\,\tl\psi_{n+1}(X_L,\mp s)\,\tl{\psi}_{m+1}(-X_R, \mp s)\,.
\label{eq:cuteq}
\ee
Analogues of this formula can be found for more complicated graph topologies, but we will not need them.

\subsubsection{(Non)uniqueness of wavefunction coefficients}
An important issue that we have to face is that wavefunction coefficients are not completely uniquely defined. In particular, field redefinitions and boundary terms can change the wavefunction, and we must deal with these ambiguities.

\vskip 4pt
The field redefinition ambiguity is relatively straightforward to resolve. Essentially it is fixed by demanding that the wavefunction coefficients satisfy soft theorems in a particular form. Intuitively, the soft theorems follow from certain symmetry transformations. If we were to perform a field redefinition, this would change the form of the symmetry transformation, and the wavefunction would correspondingly satisfy a different soft theorem.\footnote{A similar point was made in the context of the inflationary curvature perturbation $\zeta$ in~\cite{Green:2020ebl}.} There is, of course, the residual ambiguity that there could exist field redefinitions that preserve the form of the symmetry transformation but nevertheless change the wavefunction coefficients.  However, we will see that this possibility does not arise because the wavefunctions of interest are fixed uniquely by the soft information.

\vskip4pt
The question of boundary terms is more subtle. Given a wavefunction coefficient bootstrapped via some set of criteria, it must correspond to some action, with some particular choice of boundary terms. How are we to know which one? It turns out that the relevant action is the one that has a well-posed variational principle. That is, the boundary terms are such that the interactions have only a single time derivative per field. This is a necessary condition for the on-shell action to actually be computing the transition amplitude of interest. Any other choice of boundary terms will change the states involved or, equivalently, correspond to matrix elements with operator insertions. This is a somewhat technical point that we elaborate on in Appendix~\ref{app:technical}, but the takeaway is fairly simple to state: there is a distinguished choice of boundary terms---those that make the variational principle well-posed---and interestingly it is the wavefunction coefficients in this presentation that are most naturally generated by the soft bootstrap.

\subsection{Derivation of soft theorems}
\label{sec:generalsoftthm}
We now derive the various soft theorems that control the wavefunction coefficients of derivatively coupled scalar field theories. Here we present the general formalism before specializing to the relevant theories of interest in the following sections.

\vskip4pt
The philosophy is to start from the algebra of symmetries and their representation on fields in the theory, $\delta\phi$. We avoid as much as possible directly using the action that is invariant under the relevant symmetries (though we do assume that one exists). Instead, we want to extract the consequences of these symmetries for wavefunction coefficients directly, without passing through some intermediate Lagrangian (see Appendix \ref{app:generators}). The most essential fact that we use is that there is a conserved charge that generates the symmetry $\delta\phi$, which we denote $Q(t)$. Conservation of this charge implies the following equality of matrix elements
\be
\bra{\vp_{f}}Q(t_f)\ket{0} = \bra{\vp_{f}}Q(t_{i})\ket{0} \,.
\label{conservationlaw}
\ee
Our goal in this Section is to express this as a relation between wavefunction coefficients.

\vskip4pt
In order to simplify the left hand side of this equation, we take advantage of the fact that the charge at time $t_f$ can be expressed in terms of the field $\varphi_f$ and its conjugate momentum, which in the field basis takes the form $\Pi^{(\phi)}(t_f) = -i\delta/\delta\vp_f$. We can then write the charge acting on the late-time wavefunctional abstractly as
\be
\bra{\varphi_{f}}Q(t_f)\ket{0} = Q\Big[\vp, \frac{\delta}{i\delta\vp}, t_f\Big]\Psi[\vp, t_{f}]\,.
\label{LHSgeneralWard}
\ee
In explicit examples it will be useful to perform further manipulations to simplify this expression, but for now we leave it abstract.

\vskip4pt
To simplify the right hand side of the expression~\eqref{conservationlaw} it is convenient to split the charge into a piece that generates the nonlinear part of the relevant symmetry---denoted by $Q^{\rm NL}$---and the rest, denoted by $Q^{\rm L}$:
\be
Q = Q^{\text{NL}} + Q^{\text{L}}.
\ee
The nonlinear part of the charge can always be written in terms of the canonical momentum as
\begin{equation}
    Q^{\text{NL}} = \int \rd^{3}x\,\delta^{\text{NL}}\phi\,\Pi^{(\phi)}\, = \int \frac{\rd^3k}{(2\pi)^3}\,\delta^{\text{NL}}\phi_{-\vec{k}}\Pi^{(\phi)}_{\vec{k}},
\end{equation}
where $\delta^{\text{NL}}\phi$ denotes the nonlinear part of the symmetry transformation on the field variable.\footnote{More properly, this should be called the {\it sub-linear} part of the symmetry---i.e., the part independent of the fields themselves.}  In momentum space, we may write this as
\begin{equation}
\delta^{\text{NL}}\phi_{\vec k}(t) =(2\pi)^3 D^{n}_{\vec{k}}(t)\delta^{3}(\vec{k})\,,
\end{equation}
where $D^{n}_{\vec{k}}(t)$ is some (possibly time dependent) order $n$ differential operator. It is then convenient to evaluate $ \bra{\vp_{f}}Q^{\rm NL}(t_{i})\ket{0} $ by introducing a complete set of early-time field eigenstates
\be
\bra{\vp_{f}}Q^{\rm NL}(t_i)\ket{0} = \int \mathcal{D}\varphi_{i}\braket{\varphi_f|\varphi_i}\bra{\varphi_i}Q^{\rm NL}(t_i)\ket{0} = \int\mathcal{D}\varphi_{i}\int\limits_{\substack{\hspace{-0.4cm}\phi(t_f) \,=\,\vp_f\\ \hspace{-0.4cm}\phi(t_i)\,=\,\vp_i}} 
\hspace{-0.5cm} \raisebox{-.05cm}{ ${\cal D} \phi\, e^{iS[\phi]}$ }Q^{\rm NL}\Big[\varphi_i, \frac{\delta}{i\delta\varphi_i}\Big]\Psi[\varphi_i, t_i]\,.
\label{conservationcompletelbasis}
\ee
We have written the charge $Q^{\rm NL}(t_i)$ in the field basis as we did for the charge at $t_f$, now acting on the initial wavefunctional. A benefit is that we know the initial wavefunction---we work in the infinite past so that it is just a gaussian (as in~\eqref{eq:GaussianInitial})
\be
Q^{\text{NL}}\Big[\varphi_i, \frac{\delta}{i\delta\varphi_i}, t_i\Big]\Psi[\varphi_i, t_i] =i(-1)^{n}\Psi[\varphi_i, t_i] \int \rd^3k\,\delta^{(3)}(\vec{k})D^{n}_{-\vec{k}}(t_i)\Big[\mathcal{E}(k)\varphi_i{}_{\vec{k}}\Big]\,,
\ee
where $\mathcal{E}(k)$ is a kernel parameterizing the two-point function of $\varphi_i$, which is given by $\langle\varphi_i{}_{\vec k}\varphi_i{}_{-\vec k}\rangle = 1/2\Re {\cal E}(k)$. We can then remove the $\lvert\varphi_i\rangle$ eigenstates to obtain
\be
\bra{\varphi_{f}}Q^{\rm NL}(t_i)\ket{0} =  i(-1)^{n}\int\rd^3k\,\delta^{(3)}(\vec{k})D_{-\vec{k}}^n(t_i)\Big[\mathcal{E}(k)\bra{\varphi_f}\phi_{\vec k}( t_i)\ket{0}\Big]\,.
\label{eq:QNLmatrix}
\ee
We still have to simplify the matrix element $ \bra{\vp_{f}}Q^{\rm L}(t_{i})\ket{0} $. In order to do this, it is convenient to further split the linear part of the charge as
\begin{equation}
    Q^{\text{L}} = -\int \rd^3x\,\Delta_t\phi(\vec{x}, t) + Q_{(2)}^{\text{L}}\,.
\end{equation}
This first piece, proportional to $\Delta_t$, is present only when the free Lagrangian shifts by a temporal boundary term under the symmetry of interest, and $\Delta_t\phi$ is precisely this boundary term.
It is worth noting that this type of contribution is special to nonlinearly realized symmetries, and this is the only part of the charge linear in the field $\phi$.  The remaining part of the charge $Q_{(2)}^{\text{L}}$ captures all of the pieces that start at quadratic order and higher in $\phi$ and $\Pi^{(\phi)}$.  Then, we can write
\be
 \bra{\vp_{f}}Q^{\rm L}(t_{i})\ket{0} =  -\Delta_t\int \rd^3k\,\delta^{(3)}(\vec{k})\bra{\varphi_f}\phi_{\vec k}( t_i)\ket{0}
    + \bra{\varphi_f}Q_{(2)}^{\text{L}}(t_i)\ket{0}\,.
\ee
Putting this together with~\eqref{eq:QNLmatrix} we obtain
\be
\begin{aligned}
\bra{\vp_{f}}Q(t_{i})\ket{0} =  \int \rd^3k\,\delta^{(3)}(\vec{k}) \bigg(i(-1)^{n}D_{-\vec{k}}^n(t_i)\Big[\mathcal{E}(k)\bra{\varphi_f}\phi_{\vec k}( t_i)\ket{0}\!\Big]&-\Delta_t\bra{\varphi_f}\phi_{\vec{k}}(t_i)\ket{0}\bigg)\\
&\,+ \bra{\varphi_f}Q_{(2)}^{\text{L}}(t_i)\ket{0}.
\label{eq:RHShalfway}
\end{aligned}
\ee
Now we need to evaluate each of these pieces separately.

\vskip4pt
First, we will study $\bra{\varphi_f}\phi(\vec{k}, t_i)\ket{0}$, which will allow us to simplify the first and second terms in~\eqref{eq:RHShalfway}.  In general, the full quantum object is difficult to compute. However, at tree level this object (properly normalized) solves the classical equations of motion with boundary sources $\varphi_f$:\footnote{The factor in the denominator is simply the late time wavefunction itself, $\braket{\varphi_f|0}$, and serves to cancel the disconnected contributions from computing the path integral in the numerator.  }
\begin{equation}
  \phi^{\text{cl}}_{\vec{k}} (t)
\simeq  \frac{\displaystyle\int^{ \varphi_f}\mathcal{D}\phi\,e^{i S_\epsilon[\phi]}\phi_{\vec{k}}( t)}{\displaystyle\int^{\varphi_f}\mathcal{D}\phi\,e^{i S_\epsilon[\phi]}} = \frac{\bra{\varphi_f}\phi(\vec{k}, t_i)\ket{0}}{\braket{\varphi_f|0}} \,,   \label{onepoint}
\end{equation}
where we have deformed the action $S_\epsilon[\phi]$ by $i\epsilon$ terms that project onto the vacuum in the far past, and the path integral is done subject to the boundary condition that $\phi$ approaches $\varphi_f$ at $t = t_f$. In~\eqref{eq:RHShalfway} we are interested more specifically in the classical field profile at early times $t\to -\infty$. It turns out that this object is a generating functional for shifted wavefunction coefficients (see Appendix~\ref{app:fieldasgenfunct}): 
\be
\begin{aligned}
&\phi^{\text{cl}}_{\vec{k}}(-\infty) = \mathcal{K}(k, -\infty)\sum_{n=2}\frac{1}{n!}\int \frac{\rd^{3}p_1\cdots\rd^3p_{n-1}}{(2\pi)^{3(n-2)}}\,\varphi_{\vec{p}_1}\cdots\varphi_{\vec{p}_{n-1}}\\
&\hspace{7 cm}\times\delta^{(3)}(\vec{p}_1 + ... + \vec{p}_{n-1} - \vec{k})\tl{\psi}_n(\vec{p}_1,\cdots,\vec{p}_{n-1},-\vec{k})\,,
\end{aligned}
\label{generator}
\ee
where we have defined the shifted wavefunction coefficient
\be
\tl{\psi}_n(\vec{k}) \equiv\frac{1}{2k} \Big( \psi_{n}(\vec{p}_1,\cdots,\vec{p}_{n-1},\vec{k};-k)-\psi_{n}(\vec{p}_1,\cdots,\vec{p}_{n-1},\vec{k};k)\Big),
\label{eq:shiftedcoeff}
\ee
which is the difference of two wavefunction coefficients with the sign of the energy corresponding to $\vec{k}$ flipped.  We will only need the classical field profile in the soft limit, which reduces to 
\begin{align}
&\lim_{\vec{k}\rightarrow 0} \phi_{\vec{k}}^{\text{cl}}(-\infty) = -\sum_{n=2}\frac{1}{(n-1)!}\int \frac{\rd^{3}p_1\cdots\rd^3p_{n-1}}{(2\pi)^{3(n-2)}}\,\varphi_{\vec{p}_1}\cdots\varphi_{\vec{p}_{n-1}}\nonumber\\
&\hspace{7 cm}\times\delta^{(3)}(\vec{p}_1 + ... + \vec{p}_{n-1})\partial_{k}\psi_n(\vec{p}_1,\cdots,\vec{p}_{n-1},0)\,.
\label{eq:earlytimegen}
\end{align}

Using this expression, we can simplify the first two terms on the right hand side of~\eqref{eq:RHShalfway}. Further simplification will require specifying the charge in a particular theory.

\vskip4pt
Finally we have to consider the third term in~\eqref{eq:RHShalfway}. Schematically its contribution will take the form
\begin{equation}
\bra{\varphi_f}Q_{(2)}^{\text{L}}(t_i)\ket{0}\sim \int \rd^3p_1\cdots \rd^3p_m\,\delta^{(3)}(\vec{p}_1 + \cdots + \vec{p}_m)\bra{\varphi_f}\phi_{\vec{p}_1}( t_i)\cdots\phi_{\vec{p}_m}( t_i)\ket{0}\,,
\end{equation}
for some integer $m$.  
At tree level, we can substitute in the classical field profile, sourced by the boundary field $\varphi_f$.  From~\eqref{generator}, we see that at early times all of the time dependence is contained in $\mathcal{K}(p, t_i)\sim e^{ipt_i}$.  Thus, we will arrive at an integral of the form
\begin{equation}
    \bra{\varphi_f}Q_{(2)}^{\text{L}}(t_i)\ket{0}\sim \int \rd^3p_1\cdots \rd^3p_m\,e^{i(p_1 + \cdots+p_m)t_i}F(\vec{p}_1,\cdots,\vec{p}_m)\delta^{3}(\vec{p}_1 + \cdots + \vec{p}_m)\,.
\end{equation}
For generic kinematics, the integrand is highly oscillatory as
 $t_i\rightarrow \infty$.  Thus the integral will vanish so long as the function $F$ is sufficiently smooth in $p_1 + \cdots + p_m$, which we will assume to be the case.  Thus, the matrix element $\bra{\varphi_f}Q_{(2)}^{\text{L}}\ket{0}$ simply vanishes.\footnote{Importantly, the same is not true of the contribution from $Q^{\text{L}}$ with a single $\phi$ because it is evaluated in the soft limit, which removes the oscillatory factor.}

\vskip4pt
Putting all of this together, we obtain the soft theorem
\begin{tcolorbox}[colframe=white,arc=0pt,colback=greyish2]
\vspace{-8pt}
\be
 Q\Big[\vp, \frac{\delta}{i\delta\vp}, t_f\Big]\Psi[\vp_f, t_{f}] =  \lim_{\vec k\to 0} \Big(i(-1)^{n}D_{-\vec{k}}^n(t_i)\Big[\mathcal{E}(k)\phi^{\rm cl}_{\vec k}(t_i)\Big]-\Delta_t\phi^{\rm cl}_{\vec k}(t_i)\bigg)\Psi[\vp_f, t_{f}]\,,
 \label{eq:generalsoftthm}
\ee
\end{tcolorbox}
\noindent
where we have evaluated the matrix elements involving a single field in~\eqref{eq:RHShalfway} using~\eqref{generator}. Note that this implies that the soft theorem~\eqref{eq:generalsoftthm} is only valid at tree level, but this is sufficient for our purposes.\footnote{Of course, this expression could be corrected at higher orders in perturbation theory by evaluating $\bra{\varphi_f}\phi(\vec{k}, t_i)\ket{0}$ at higher order.} This expression should be utilized by substituting in the expansion of the wavefunction into wavefunction coefficients~\eqref{coefficientexpansion} and replacing the classical field profile with~\eqref{generator}. Then, to isolate the soft theorem for a particular wavefunction coefficient one simply acts repeatedly with $\delta/\delta\varphi$, setting $\varphi$ to zero in the end.  We now turn to applying this general formalism to a few specific theories of interest.

\newpage
\section{The wavefunction from the $S$-matrix}
\label{sec:WFfromSmatrix}
As a first step toward constructing the wavefunction in theories with enhanced soft limits, we take as an input the $S$-matrix of the relevant theory and ask: {\it how much more information is contained in the wavefunction?} Interestingly, we find that if one is willing to input information about the singularities of the wavefunction coefficients, the remaining part of the wavefunction is fixed by a soft theorem at one lower order in the soft momentum than the final wavefunction actually satisfies. Further, this lower-order soft theorem is actually an Adler zero-like vanishing condition in general, which is simpler to implement.\footnote{The two-point and four-point wavefunctions for the (special) galileon are exceptions to this, see~\eqref{2pointgalsoft} and~\eqref{4pointgalsoft}.}

\subsection{The nonlinear sigma model}
\label{nlsmbootstrapsection}
We begin by considering the wavefunction of the nonlinear sigma model (NLSM). This theory is the effective description of the Nambu--Goldstone modes arising from the spontaneous breaking of ${\rm SU}(N)_L\times {\rm SU}(N)_R$ global symmetries to its diagonal subgroup. (For more details, see Appendix~\ref{app:NLSM}.)
 
\vskip4pt
The nonlinear sigma model is a somewhat exceptional case because one does not need a proper soft theorem to bootstrap its wavefunction. Its amplitudes exhibit an Adler zero and vanish like ${\cal O}(p)$ in the soft limit, and the wavefunction coefficients correspondingly satisfy a soft theorem that constrains their ${\cal O}(p^0)$ behavior.\footnote{To be clear, we say that a theory has an $\mathcal{O}(p^n)$ soft theorem if there is a Ward identity that controls the behavior of the wavefunction or amplitude at order $\mathcal{O}(p^n)$ in the soft limit.  Note that the this slightly different than the language used for Adler zeros---if an object possesses an order $\mathcal{O}(p^{n + 1})$ Adler zero, then it vanishes like $p^{n+1}$ in the soft limit.}
However for a general wavefunction coefficient this soft theorem is difficult to write down without knowing the precise form of the symmetry itself to all orders. Fortunately, the wavefunction soft theorem is not necessary---the wavefunction can be reconstructed from its singularities (including the total energy scattering pole) along with the requirement that the $\rm{U}(1)$ mode decouples (see Appendix \ref{app:flavor} for details about flavor ordering and $\rm{U}(1)$ decoupling).

\subsubsection{Wavefunction coefficients}
\label{nlsmbootstrap}
In order to construct NLSM wavefunction coefficients, our strategy is to parameterize the most general wavefunction, subject to the constraints that it has the correct singularity structure and obeys the $\rm{U}(1)$ decoupling identity (see Appendix~\ref{app:flavor}).

\paragraph{Four points:} 
To see explicitly how this works, we begin by bootstrapping the flavor ordered four-point wavefunction, which is generated by a contact interaction in the bulk. 
The most general ansatz for the four-point wavefunction coefficient, having only a total energy singularity and a vanishing cut, is of the form
\begin{equation}
\psi_4^{({\rm nlsm})}(\vec{p}_1, \vec{p}_2, \vec{p}_3, \vec{p}_4) = \frac{A_4}{E} + R,
\end{equation}
where $A_4$ is some representation of the corresponding scattering amplitude, and $R$ is the most general polynomial with mass dimension one in the variables $p_{a}, s_{ab}$ for $a,b = 1, 2, 3, 4$.  For brevity, we will often suppress the argument of the wavefunction.  When this is the case, the momenta are ordered as indicated above.  Without loss of generality, we may choose a representation for $A_4$ which is manifestly Lorentz invariant. The difference between this choice and any other choice can be absorbed into $R$.
The symmetries of the flavor decomposition imply that $\psi_4$ must be even under cyclic permutations.
This is most easily achieved by defining $A_4$ and $R$ to separately be invariant under cyclic permutations.
Since the amplitude has two powers of momenta in each term, it will be constructible out of the building blocks
\be
\begin{aligned}
    C_1 &= P_1\cdot P_2 + P_2\cdot P_3 + P_3\cdot P_4 + P_4\cdot P_1\,,\\
    C_2 &= P_1\cdot P_3 + P_2\cdot P_4\,,
\end{aligned}
\ee
where we have defined the four-momentum-like object $P_a^\mu \equiv (p_a,\vec p_a)$ and the dot products are the contraction $P_a\cdot P_b = \eta_{\mu\nu}P_a^\mu P_b^\nu$ where $\eta_{\mu\nu}$ is the ordinary Minkowski space metric (mostly plus signature). The two building blocks $C_1$ and $C_2$ are not actually independent, but satisfy
\begin{equation}
    C_1 + C_2 = \frac{1}{2}\Big(P_1 + P_2 + P_3 + P_4\Big)^2 = -\frac{1}{2}E^2\,.
\end{equation}
Among many, one viable representation of $A_4$ is (temporarily setting the coupling $1/f^2$ to 1)
\begin{equation}
    A_4 = \frac{1}{2}C_1\,.
\end{equation}
Next we construct $R$. Since we are studying a contact interaction, the cut of $\psi_4$ must vanish.  The scattering part of our ansatz already has this property, so it must be obeyed separately by $R$.  Therefore, each term in $R$ must have an odd number of $p_a$'s, which by dimensional analysis means that $s_{ab}$ cannot appear.
Therefore, the only cyclic permutation-invariant combination of mass dimension one is the total energy: $R = a_1 E$ where $a_1$ is some constant that remains to be determined.  
Our ansatz for the wavefunction is then
\begin{equation}
\psi_4 = \frac{C_1}{2E} + a_1E\,.
\label{NLSMonefree}
\end{equation}
This ansatz manifestly has the correct total energy singularity and has a vanishing cut. We must finally impose the constraint that the ${\rm U}(1)$ mode decouples, which is enforced by the ${\rm U}(1)$ decoupling identity (see Appendix~\ref{app:flavor}).  At four points, this is 
\begin{equation}
 \psi_4(\vec{p}_1, \vec{p}_2, \vec{p}_3, \vec{p}_4) + \psi_4(\vec{p}_1, \vec{p}_3, \vec{p}_4, \vec{p}_2) + \psi_4(\vec{p}_1, \vec{p}_4, \vec{p}_2, \vec{p}_3) = \Big(3a_1 - \frac{1}{2}\Big)E = 0\,.
\end{equation}
Requiring that this vanishes imposes $a_1 = -1/6$.  Thus after restoring the coupling, we have completely determined the flavor ordered four-point wavefunction to be
\begin{tcolorbox}[colframe=white,arc=0pt,colback=greyish2]
\begin{equation}
\psi_4^{({\rm nlsm})} = \frac{1}{2f^2E}\Big(P_1\cdot P_2 + P_2\cdot P_3 + P_3\cdot P_4 + P_4\cdot P_1\Big) +\frac{1}{6f^2}E\,.
\label{eq:NLSM4ptwf}
\end{equation}
\end{tcolorbox}
\noindent It is straightforward to check that this matches a direct perturbative computation.   
Using 3-momentum conservation, we may also bring the wavefunction to the form
\be
	\psi_4^{({\rm nlsm})} = \frac{1}{3f^2E}\Big(P_1\cdot P_2 + P_2\cdot P_3 - 2P_1\cdot P_3\Big) - \frac{1}{6f^2}\Big(p_1 - 2p_2 + p_3\Big)\,.
	\label{eq:NLSM4pt}
\ee
By writing the wavefunction in this form, we have sacrificed cyclic invariance in order to eliminate all instances of $\vec{p}_4$ and $p_4$ (except in the total energy singularity).  As we will see momentarily, writing the wavefunction in this way is useful for bootstrapping wavefunction coefficients that involve exchange interactions.\footnote{There is also a more economical way of writing the wavefunction, which takes the form of an NLSM scattering amplitude divided by the total energy:
\be
	\psi_4^{(\text{nlsm})} = \frac{1}{6f^2E}\Big(P_1\cdot P_2 + P_2\cdot P_3 + P_3\cdot P_4 + P_4\cdot P_1 - 2P_1\cdot P_3 - 2P_2\cdot P_4\Big)\,.
\ee
}

\paragraph{Six points:} As a more nontrivial example, one can construct the six-point wavefunction coefficient from knowledge of its singularities along with the ${\rm U}(1)$ decoupling identity. Since we are constructing the flavor ordered wavefunction, the only possible singularities occur when adjacent sums of three energies add up to zero. There are three such factorization channels: one is when $E_{123} = p_1+p_2+p_3 + s_{123}$ or $E_{456} = p_4+p_5+p_6 + s_{456}$ vanish, or the analogous partial energy singularities in the other factorization channels. 
A natural ansatz is then of the form (again setting the coupling $1/f^2$ to 1)
\be
\psi_6^{({\rm nlsm})}(\vec{p}_1,\vec{p}_2,\vec{p}_3,\vec{p}_4,\vec{p}_5,\vec{p}_6) = \frac{1}{E}\left(\frac{N_{123}N_{456}}{E_{123}\,E_{456}}+\frac{N_{561}N_{234}}{E_{561}\,E_{234}}+\frac{N_{612}N_{345}}{E_{612}\,E_{345}}
\right)+\frac{C}{E}+{\rm regular}\,,
\label{eq:6ptansatz}
\ee
where the regular terms do not have any singularities.  There are various ways to fix the form of the kinematic numerators.  The simplest is to note that the expression~\eqref{eq:6ptansatz} must factorize appropriately into products of shifted four-point functions when we take cuts according to the prescription~\eqref{eq:cuteq}.\footnote{Equivalently, we can require that the residues of the partial energy singularities are the appropriate combination of shifted four-point wavefunction coefficients and four-point amplitudes.} These shifted four-point functions can be computed using~\eqref{eq:NLSM4ptwf}:
\be
\tl\psi_4^{({\rm nlsm})}(\vec{p}_1,\vec{p}_2, \vec{p}_3, \vec{p}_I) = \frac{P_1\cdot P_2 + P_2\cdot P_3 - 2P_1\cdot P_3}{3(p_{123}^2-s_{123}^2)}\,,
\ee
where we have shifted the internal line, $I$.
This implies that we should take
\be
N_{123}\equiv \frac{1}{3}\Big(P_1\cdot P_2 + P_2\cdot P_3 - 2P_1\cdot P_3\Big)\,,
\ee
and similarly for the other permutations.
With this choice~\eqref{eq:6ptansatz} will have the correct cuts (and by extension the correct partial energy singularities). We must then fix the $C$ and regular terms in~\eqref{eq:6ptansatz}. The $C$ terms have mass dimension two and can be built from the cyclic permutation-invariant building blocks\footnote{There is another possible cyclic-invariant building block with the right mass dimension:  $C_3 = P_1\cdot P_4 + P_2\cdot P_5 + P_3\cdot P_6$. As in the four-point function case, these quantities are not independent, but are related by $C_1 + C_2 + C_3 = -\frac{E^2}{2}$
so we can eliminate $C_3$ in terms of $C_1,C_2$ at the cost of shifting around the regular terms with no singularities.}
\be
\begin{aligned}
    C_1 &= P_1\cdot P_2 + P_2\cdot P_3 +P_3\cdot P_4 + P_4\cdot P_5 + P_5\cdot P_6  +P_6\cdot P_1\,,\\
    C_2 &= P_1\cdot P_3 + P_2\cdot P_4 +P_3\cdot P_5 + P_4\cdot P_6 + P_5\cdot P_1 + P_6\cdot P_2\,\,.
\label{cyclic6}
\end{aligned}
\ee
Finally, the regular terms in~\eqref{eq:6ptansatz} must also be cyclic permutation invariant, have mass dimension one, and have a vanishing cut, which implies they must be proportional to $E$. We then have
\be
\psi_6^{({\rm nlsm})} = \frac{1}{E}\left(\frac{N_{123}N_{456}}{E_{123}E_{456}}+\frac{N_{561}N_{234}}{E_{561}E_{234}}+\frac{N_{612}N_{345}}{E_{612}E_{345}}
\right)+\frac{a_1 C_1 + a_2 C_2}{E} + a_3 E\,.
\ee
The coefficients $a_1$ and $a_2$ can be fixed from the $E\to 0$ limit, whose residue must be the scattering amplitude
\be
A_6= -\left(\frac{N_{123}N_{456}}{S_{123}}+\frac{N_{561}N_{234}}{S_{561}}+\frac{N_{612}N_{345}}{S_{612}}
\right)- \frac{1}{18}\Big(C_1 + 2C_2\Big).
\ee
We can then  immediately read off the coefficients
\begin{equation}
    a_1 = -\frac{1}{18}, \hspace{1.5 cm}a_2 = -\frac{1}{9}\,.
\end{equation}
To fix the last coefficient, we use the ${\rm U}(1)$ decoupling identity, which at six points reads
\be
\begin{aligned}
0 = &~\psi_6(\vec{p}_1, \vec{p}_2, \vec{p}_3, \vec{p}_4, \vec{p}_5, \vec{p}_6) + \psi_6( \vec{p}_2,\vec{p}_1, \vec{p}_3, \vec{p}_4, \vec{p}_5, \vec{p}_6) + \psi_6(\vec{p}_2, \vec{p}_3,\vec{p}_1, \vec{p}_4, \vec{p}_5, \vec{p}_6)\nonumber\\
&\, + \psi_6(\vec{p}_2, \vec{p}_3, \vec{p}_4,\vec{p}_1, \vec{p}_5, \vec{p}_6) + \psi_6( \vec{p}_2, \vec{p}_3, \vec{p}_4,\vec{p}_5,\vec{p}_1, \vec{p}_6)=\Big(5a_3 + \frac{1}{6}\Big)E\,.
\end{aligned}
\ee
This must vanish, which implies $a_3 = -1/30$, so that the six-point wavefunction is given by
\begin{tcolorbox}[colframe=white,arc=0pt,colback=greyish2]
\be
\psi_6^{({\rm nlsm})} = \frac{1}{E}\left(\frac{N_{123}N_{456}}{E_{123}E_{456}}+\frac{N_{561}N_{234}}{E_{561}E_{234}}+\frac{N_{612}N_{345}}{E_{612}E_{345}}
\right)-\frac{ C_1 + 2 C_2}{18E} -\frac{1}{30} E\,,
\label{eq:NLSM6pwfans}
\ee
\end{tcolorbox}
\noindent which can again be matched to a bulk perturbative calculation.  At higher points, the bootstrap  procedure generalizes in a straightforward fashion.\footnote{A natural question is to understand how this discussion changes when studying double trace theories---in particular the ${\rm SO}(N+1)/{\rm SO}(N)$ NLSM---where the Goldstones transform in the fundamental representation of ${\rm SO}(N)$.  In particular, there is no analogue of the ${\rm U}(1)$ decoupling identity for such theories, and it would interesting to understand what (if any) piece of data replaces it.}

\subsubsection{Soft theorem} 

Though the NLSM wavefunction is fixed by the residues of its singularities along with the ${\rm U}(1)$ decoupling identity, it is nevertheless interesting to explore the soft theorem that the wavefunction satisfies at ${\cal O}(p^0)$. 
The NLSM wavefunction soft theorem is somewhat less powerful than its amplitude counterpart. Essentially this is because it is necessary to know the precise form of the symmetry~\eqref{nlsmsymmetry} (or, equivalently, the corresponding Noether charge) to a given order in fields in order to derive the relevant soft theorem. Further, because the symmetry transformation has  infinitely many terms, there is no universal form of the soft theorem that holds for a general $n$-point wavefunction.  Instead, one must work out the soft theorem order by order.

\vskip4pt
We begin by considering the NLSM symmetry transformation: 
\be
\delta\phi^{c} = B^{c} - \frac{1}{3f^{2}}B^{a_1}\phi^{b_1}\phi^{b_2}f^{a_1 b_1 b_3}f^{c\, b_2 b_3} - \frac{1}{45f^{4}}B^{a_1}\phi^{b_1}\phi^{b_2}\phi^{b_3}\phi^{b_4}f^{a_1 b_1 \,c}f^{a_2 b_2 a_3}f^{b_3 c a_4}f^{b_4 a_3 a_4} + \mathcal{O}(\phi^6)\,,
\ee
where $f$ is the symmetry breaking scale and $f^{abc}$ are the Lie algebra structure constants (hopefully the difference is clear from context).
The corresponding symmetry charge is given by
\begin{equation}
    Q = \int \rd^3x\,\delta\phi^{b}\Pi^{(\phi)}_{b}\,.
\end{equation}
We want to apply the formalism of Section~\ref{sec:generalsoftthm} to this particular charge. First, note that $Q$ only has a piece which is linear in $\Pi$, so computing the left hand side of~\eqref{eq:generalsoftthm} is straightforward.  Moreover, the differential operator coming from the nonlinear part of the symmetry is just a constant $D_{\vec{k}}^{n}(t) = 1$. 
Finally, note that $Q$ does not contain a term with an isolated $\phi$, so $\Delta_t = 0$.\footnote{This is because the symmetry transformation $\delta\phi^{a}$ does not induce a temporal boundary term on the kinetic part of the action.}  

\vskip4pt
With these considerations, we can write the soft theorem~\eqref{sec:generalsoftthm} as
\be
    -i\lim_{\vec{k}\rightarrow 0}\delta\varphi_{\vec{k}}^{a}\,\frac{\delta \log\Psi[\varphi_f, t_f]}{\delta\varphi^{a}_{\vec k}} =i\lim_{\vec{k}\rightarrow 0} B^{a}\,\mathcal{E}(k)\phi_{\vec k}^{\text{cl}}{}^{\,a}( t_i)\,,
\ee
where we have divided through by $\Psi[\varphi_f, t_f]$ in order to write the LHS in terms of $\log \Psi$.
Next we expand both sides in terms of wavefunction coefficients, which gives
\begin{align}
   &- \lim_{\vec{k}\rightarrow 0}\sum_{n=2}\frac{1}{(n-1)!}\int \frac{\rd^{3}p_1\cdots \rd^3p_{n-1}}{(2\pi)^{3(n - 1)}}\,\varphi^{b_1}_{\vec{p}_1}\cdots\varphi^{b_{n-1}}_{\vec{p}_{n-1}}\delta\varphi^{a}_{\vec{k}}\delta^{(3)}(\vec{p}_1 + ... + \vec{p}_{n-1})\nonumber\\
&\hspace{9.2 cm}\times\psi_n^{b_1\cdots b_{n-1}a}(\vec{p}_1,\cdots,\vec{p}_{n-1},\vec{k})\nonumber\\
    &~~~= \lim_{\vec{k}\rightarrow 0}B^a\mathcal{E}(k) \sum_{n=2}\frac{1}{(n-1)!}\int \frac{\rd^{3}p_1\cdots \rd^3p_{n-1}}{(2\pi)^{3(n-2)}}\,\varphi^{b_1}_{\vec{p}_1}\cdots \varphi^{b_{n-1}}_{\vec{p}_{n-1}}\delta^{(3)}(\vec{p}_1 + ... + \vec{p}_{n-1})\nonumber\\
&\hspace{9.2 cm}\times\partial_{k}\psi^{b_1\cdots b_{n-1}a}_n(\vec{p}_1,\cdots,\vec{p}_{n-1},\vec{k})\,.
    \label{nlsmcoeffon}
\end{align}
Note that the RHS vanishes because $\mathcal{E}(k)$ vanishes in the soft limit, and $\partial_{k}\psi$ is finite.  At this point, we may take functional derivatives with respect to $\varphi$ to extract particular wavefunction coefficients from these sums.

\vskip4pt
Since $\delta\phi$ has infinitely many terms, there is no uniform way to write the resulting soft theorem for a general $n$-point wavefunction coefficient.  Therefore, one must derive results coefficient by coefficient.  This is straightforward to do; after stripping the flavor indices, one finds that the four-point soft theorem is
\begin{tcolorbox}[colframe=white,arc=0pt,colback=greyish2]
\vspace{-6pt}
\begin{align}
	\lim_{\vec{k}\rightarrow 0}\psi_4(\vec{k}, \vec{p}_2, \vec{p}_3, \vec{p}_4) = -\frac{1}{6f^2}\Big(\psi_{2}(p_4) - 2\psi_2(p_3) + \psi_2(p_2)\Big)\,.
\end{align}
\end{tcolorbox}
\noindent It is easy to verify that~\eqref{eq:NLSM4ptwf} satisfies this identity. We can similarly work out  the six-point soft theorem
\begin{tcolorbox}[colframe=white,arc=0pt,colback=greyish2]
\vspace{-6pt}
\be
\begin{aligned}
	\lim_{\vec{k}\rightarrow 0}&\,\psi_6(\vec{k}, \vec{p}_2, \vec{p}_3,\vec{p}_4,\vec{p}_5,\vec{p}_6) =  \\
& -\frac{1}{6f^2}\Big(\psi_{4}(\vec{p}_2 + \vec{p}_3, \vec{p}_4, \vec{p}_5, \vec{p}_6)- 2\psi_{4}(\vec{p}_6 + \vec{p}_2, \vec{p}_3, \vec{p}_4, \vec{p}_5) + \psi_{4}(\vec{p}_5 + \vec{p}_6, \vec{p}_2, \vec{p}_3, \vec{p}_4)\Big)\\
& + \frac{1}{180f^4}\Big(\psi_2(p_6) - 4\psi_2(p_5) + 6\psi_2(p_4) - 4\psi_2(p_3) + \psi_2(p_2)\Big)\,,
\end{aligned}
\label{nslm6pointsofttheorem}
\ee
\end{tcolorbox}
\noindent 
and check that it is satisfied by the true answer~\eqref{eq:NLSM6pwfans}. It is worth emphasizing that in contrast to the Adler zero that amplitudes satisfy, these soft theorems are comparatively less useful, because we do not know how to write them down without knowing precisely form of the full symmetry transformation. Nevertheless, given knowledge of these soft theorems, it is possible to construct the NLSM wavefunction recursively, without using any scattering information, as we show in Section~\ref{nlsmrecursefromsoft}.

\subsection{$P(X)$ and Dirac--Born--Infeld}
The next example we consider is that of a Dirac--Born--Infeld (DBI) scalar. In addition to the ordinary shift symmetry $\delta_C\phi = 1$, the theory is invariant under the symmetry
\be
\delta_{B_\mu}\phi = x_\mu+\frac{1}{f^4}\phi\partial_\mu\phi\,.
\label{eq:DBIsymm}
\ee
In the context of scattering amplitudes, this symmetry causes the $S$-matrix elements to vanish as ${\cal O}(p^2)$ in the soft limit.

\vskip4pt
We will require input from these symmetries in the form of soft theorems in order to fix the wavefunction. Since we are utilizing scattering information, we only need the soft theorem corresponding to the shift symmetry, which in this case implies that  the wavefunction also has an Adler zero.\footnote{Later in Section~\ref{sec:softrecurs} we will construct two types of recursion relations, one that uses the same input as this section, but also one that uses the soft theorem  for the DBI symmetry~\eqref{eq:DBIsymm} instead of the scattering amplitude.}

\subsubsection{Soft theorems}
In the $P(X)$ and DBI cases, bootstrapping wavefunction coefficients requires input from the lowest-order soft theorem. This soft theorem is particularly simple---it is just a vanishing statement like the Adler zero. We therefore first derive this identity.  We will also derive the corresponding statement for the DBI symmetry. This latter statement is not directly needed to bootstrap the wavefunction from the $S$-matrix, but can be used to recursively construct the wavefunction via a different method (as we will do in Section~\ref{sec:softrecurs}).

\paragraph{Wavefunction Adler zero:} We first derive the lowest-order soft theorem satisfied by $P(X)$ wavefunctions (and in particular DBI). This is the soft theorem corresponding to the shift symmetry  $\delta_C\phi = 1$. This symmetry is generated by the charge
\be
Q_C = \int \rd^3x\,\Pi^{(\phi)}\,,
\ee
and the differential operator appearing in~\eqref{eq:generalsoftthm} is $D^{0}_{\vec{k}}(t_i) = 1$.  From here we proceed exactly as in the case of the NLSM.  However in this case the shift symmetry does not have quadratic and higher terms, so there is a simple expression holding for all wavefunction coefficients: 
\begin{tcolorbox}[colframe=white,arc=0pt,colback=greyish2]
\be
\lim_{\vec{k}\rightarrow 0 }\psi_n(\vec{k},\vec{p}_2,\cdots,\vec{p}_{n}) = 0\,.
\label{simpleshiftward}
\ee
\end{tcolorbox}
\noindent 
That is, the wavefunction coefficients possess an Adler zero.

\paragraph{DBI symmetry soft theorem:} We now sketch the derivation of the higher-order soft theorem associated to the DBI symmetry~\eqref{eq:DBIsymm}. Notice that this symmetry has a full Lorentz vector of charges. Since we are breaking Lorentz symmetry by choosing a time slice on which to define the wavefunction, the spatial and temporal components of this charge (and the corresponding soft theorems) behave slightly differently. Explicitly, these charges are given by
\begin{align}
Q_B^{i} &= \int \rd^3x\,\Big(x^i + \frac{1}{f^4}\phi\partial^{i}\phi\Big)\Pi^{(\phi)}\,,\label{eq:spatialDBIcharge}\\
Q_B^{t} &= \int \rd^3x\,t\Pi^{(\phi)} -  \phi\sqrt{\Big(1 + \frac{1}{f^4}(\nabla\phi)^2\Big)\Big(1 + \frac{1}{f^4}(\Pi^{(\phi)})^2\Big)}\,.
\label{eq:temporalDBIcharge}
\end{align}
Knowing only the symmetry transformation $\delta \phi$, we may immediately write down the spatial charge by appealing to the logic outlined in Appendix \ref{app:generators}.  On the other hand, the temporal charge must be computed by explicitly working out the boundary term that the Lagrangian shifts by under this symmetry.  We now consider the soft theorems arising from each of these charges in turn.

\vskip4pt
\noindent
{\it Spatial soft theorem:} First consider the spatial charge~\eqref{eq:spatialDBIcharge}. The spatial DBI symmetry does not have a boundary term, so $\Delta_t = 0$ in~\eqref{eq:generalsoftthm}. The differential operator corresponding to the $x^i$ part of the symmetry is $D_{\vec{k}}^{1} = -i\rd/\rd \vec k$.\footnote{A helpful fact to remember is $\vec{x} = i\int \rd^3k\,e^{i\vec{k}\cdot\vec{x}}\vec{\partial}_k\delta^{(3)}(\vec{k})\,.$} The general soft theorem~\eqref{eq:generalsoftthm} then takes the form 
\be
\begin{aligned}
    &(2\pi)^3\lim_{\vec k\to0}\frac{\rd}{\rd\vec{k}}\Big[\frac{\delta\log\Psi[\varphi, t_f]}{\delta\varphi_{-\vec k}}\Big] +\frac{1}{f^4} \int \frac{\rd p_1^3 \rd p_2^3\rd p_3^3}{(2\pi)^{3}}\,\delta^{(3)}(\vec{p}_1 + \vec{p}_2 + \vec{p}_3)\,\Big(\frac{\vec{p}_1 + \vec{p}_2}{2}\Big)\,\varphi_{\vec{p}_1}\varphi_{\vec{p}_2}\frac{\delta\log\Psi[\varphi, t_f]}{\delta\varphi_{-\vec{p}_3}}\\
	&~~~~~~~~~~~~~~~~
	= -\lim_{\vec k\to 0}\frac{\rd}{\rd\vec{p}}\Big[\mathcal{E}(k)\phi^{\text{cl}}_{\vec{k}}( t_i)\Big]\,,
    \label{eq:DBIsymmintermediate}
\end{aligned}
\ee
where the two terms on the LHS come from the two terms in the charge~\eqref{eq:spatialDBIcharge}.
Now we may expand both sides in terms of wavefunction coefficients and take functional derivatives to obtain (after relabelling)
\be
\begin{aligned}
\lim_{\vec{k}\rightarrow 0}\frac{\rd}{\rd\vec{k}}\psi_{n}(-\vec{k}, \vec{p}_2,\cdots,\vec{p}_{n}) &+ \sum_{\text{pairs }a,b} \frac{1}{f^4}(\vec{p}_a + \vec{p}_b)\psi_{n-2}(\vec{p}_a + \vec{p}_b, \vec{p}_2,\cdots,\overline{p}_a,\cdots,\overline{p}_b,\cdots,\vec{p}_{n})\\
    & = \lim_{\vec{k}\rightarrow 0}\frac{\rd}{\rd\vec{k}}\Big[\mathcal{E}(q)\psi_{n}( -\vec{k}, \vec{p}_2,\cdots,\vec{p}_{n})\Big]\,,
    \label{DBIRHS}
\end{aligned}
\ee
where the bar over a particular momentum indicates that it should be removed.  Using the fact that $\lim_{\vec{k}\rightarrow 0}\psi_n = -\partial_k\psi_n|_{0}$ along with the chain rule, we can simplify~\eqref{DBIRHS} to the schematic form\footnote{Specifically we use the relation $
\frac{\rd\psi_n}{\rd\vec{k}} = \vec{\partial}_{k}\psi_n + \hat{k}\partial_k\psi_n$.}
\begin{equation}
    \lim_{\vec{k}\rightarrow 0}\left(\vec{\partial}_k\psi_n(-\vec{k}) + \hat{k}\partial_k\psi_n(-\vec{k})\right) + \psi^{ab}_{n-2} = \lim_{\vec{k}\rightarrow 0}\left(\hat{k}\partial_k\psi_n(-\vec{k}) - \mathcal{E}(k)\frac{\rd}{\rd\vec{k}}\left[\psi_n(-\vec{k})\right]\right)
\end{equation}
where $\psi^{ab}_{n-2}$ is the sum in the first line of~\eqref{DBIRHS}.  Note that the terms involving $\lim_{
\vec{k}\rightarrow 0}\hat{k}$---which strictly speaking are ill-defined  because they are direction dependent---cancel between the two sides.  The second term on the RHS vanishes in the soft limit.  Thus we are left with
\begin{tcolorbox}[colframe=white,arc=0pt,colback=greyish2]
\vspace{-2pt}
\begin{equation}
    \lim_{\vec{k}\rightarrow 0}\vec{\partial}_k\psi_{n}(\vec{k}, \vec{p}_2,\cdots,\vec{p}_{n})= \sum_{\text{pairs }a,b} \frac{1}{f^4}(\vec{p}_a + \vec{p}_b)\psi_{n-2}(\vec{p}_a + \vec{p}_b, \vec{p}_2,\cdots,\overline{p}_a,\cdots,\overline{p}_b,\cdots,\vec{p}_{n})\,.
\end{equation}
\end{tcolorbox}
\noindent This implies that the DBI wavefunction coefficients at order ${\cal O}(p)$ satisfy a soft theorem rather than an Adler zero.

\vskip4pt
\noindent
{\it Temporal soft theorem:} Finally, we want to derive the soft theorem associated to the temporal DBI charge~\eqref{eq:temporalDBIcharge}. Notice that this charge does have a term linear in $\phi$, which means that $\Delta_t = 1$.  Moreover, the differential operator for the temporal symmetry is $D^{0}_{\vec{k}}(t_i) = t_i$. The soft theorem~\eqref{eq:generalsoftthm} then can be written as
\be
\bar Q_B^t\Big[\varphi, \frac{\delta}{i\delta\varphi}, t_f\Big]\Psi[\varphi, t_f] = \lim_{\vec{k}\rightarrow 0}\Big(it_i\mathcal{E}(k)-1\Big)\phi^{\text{cl}}(\vec{k}, t_i)\Psi[\varphi, t_f]\,,
\ee
where $\bar Q_B^t$ is only the square root term in~\eqref{eq:temporalDBIcharge}.\footnote{At late times, we may ignore the nonlinear piece if we set $t_f = 0$.} Now we expand this expression in wavefunction coefficients and take functional derivatives to extract their soft theorems. Unfortunately, there is no simple way to write the LHS in general, but for particular examples everything can be straightforwardly worked out.
For example, we have at four-points
\begin{tcolorbox}[colframe=white,arc=0pt,colback=greyish2]
\vspace{-6pt}
\begin{align}
	\lim_{\vec{k}\rightarrow 0}\partial_k\psi_4(\vec{k}, \vec{p}_2, \vec{p}_3, \vec{p}_4) = -\frac{1}{f^4}\Big(\vec{p}_2\cdot\vec{p}_3 + \psi_2(p_2)\psi_2(p_3)\Big) + \text{perms.}\,,
\end{align}
\end{tcolorbox}
\noindent where the overall $\delta$-function on each side is $\delta^{(3)}(\vec{p}_2 + \vec{p}_3 + \vec{p}_4)$, and ``perms." indicates a symmetrization over these momenta. At six points, the soft theorem reads

\begin{tcolorbox}[colframe=white,arc=0pt,colback=greyish2]
\be
\begin{aligned}
	\lim_{\vec{k} \to  0}\partial_k\psi_6&(\vec{k}, \vec{p}_2, \vec{p}_3, \vec{p}_4, \vec{p}_5, \vec{p}_6) = \frac{1}{f^4}\psi_4(-\vec{p}_{234}, \vec{p}_2, \vec{p}_3, \vec{p}_4)\psi_2(p_5)\\
& + \frac{1}{f^8}\Big(-\vec{p}_2\cdot\vec{p}_3\psi_2(p_4)\psi_2(p_5) + 3\psi_2(p_2)\psi_2(p_3)\psi_2(p_4)\psi_2(p_5)+ \vec{p}_2\cdot\vec{p}_3\vec{p}_4\cdot\vec{p}_5\Big) + \text{perms.}\,
\end{aligned}
\ee
\end{tcolorbox}
\noindent As in the case of the nonlinear sigma model, these soft theorems may be used to recursively construct the six-point wavefunction coefficient if the scattering amplitude is not known, see Section \ref{DBIrecursionfromsoft}. However, it is often simpler to use the scattering amplitude plus the Adler zero that wavefunction coefficients satisfy as a consequence of the ordinary shift symmetry, as we now demonstrate.

\subsubsection{Wavefunction coefficients}
Much like the NLSM, we can construct wavefunction coefficients of DBI from their singularities, but now supplemented with the Adler zero condition. We will demonstrate this procedure for a number of simple examples.

\paragraph{Four-point wavefunction:} To bootstrap the four-point function, we will take an ansatz of the form
\begin{equation}
    \psi_4^{({\rm dbi})} = \frac{A_4}{E} + R\,,
\end{equation}
where $A_4$ is an arbitrary representation of the scattering amplitude and $R$ is analytic in the total energy $E$.  For convenience, we will choose a manifestly Lorentz invariant representation of the scattering amplitude which is also manifestly Bose-symmetric (temporarily setting the coupling $1/f^4$ to $1$):
\begin{equation}
    A_4 = \frac{1}{2}\Big((P_1\cdot P_2)^2 + (P_1\cdot P_3)^2 + (P_1\cdot P_4)^2 + (P_2\cdot P_3)^2 + (P_2\cdot P_4)^2 + (P_3\cdot P_4)^2\Big)\,.
    \label{DBIamplitude}
\end{equation}
To parametrize $R$, we write down the most general Bose-symmetric polynomial which is cubic in $p_a, s_{ab}$. As in the NLSM case, we are assuming that the wavefunction coefficient is purely generated by a contact interaction in the bulk, which means that the cut of the wavefunction vanishes.  This implies that each term in $R$ must have an odd number of external energies, so by dimensional analysis $s_{ab}$ can only appear in even powers.  Modulo dimension-dependent Gram identities, the most general polynomial of this form is 
\be
    R = a_1\Big(p_1 p_2 p_3 + p_1 p_2 p_4 + p_1 p_3 p_4 + p_2 p_3 p_4 \Big) + a_2\Big(p_1^2 p_2 + \text{perms.}\Big)+a_3\Big(p_1^3 + p_2^3 + p_3^3 + p_4^3\Big)\,.
\ee
All of the unfixed coefficients are completely determined by the simple Adler zero.  Sending $\vec{p}_1\rightarrow 0$ gives:
\be
\lim_{\vec{p}_1\rightarrow 0}\psi_4=\Big(a_1 - \frac{3}{4}\Big)p_2p_3p_4 + \Big(a_2 + \frac{3}{8}\Big)\Big(p_2^3 + p_3^3 + p_4^3\Big) + \Big(a_3 + \frac{1}{8}\Big)\Big(p_2^2 p_3 + p_2^2 p_4 + p_3^2 p_4\Big)\,.
\ee
Since all the terms on the right hand side are independent, their coefficients must each vanish, so that
\begin{equation}
    a_1 = \frac{3}{4}\,, \hspace{1.5 cm}a_2 =- \frac{3}{8}\,, \hspace{1.5 cm}a_3 = -\frac{1}{8}\,.
\end{equation}
Restoring the coupling, this fixes the wavefunction to be
\begin{tcolorbox}[colframe=white,arc=0pt,colback=greyish2]
\be
\begin{aligned}
    \psi_4^{({\rm dbi})} = ~&\frac{1}{2f^4E}\Big((P_1\cdot P_2)^2 + (P_1\cdot P_3)^2 + (P_1\cdot P_4)^2 + (P_2\cdot P_3)^2 + (P_2\cdot P_4)^2 + (P_3\cdot P_4)^2\Big)\\
    & + \frac{3}{4f^4}\Big(p_1 p_2 p_3 + p_1 p_2 p_4 + p_1 p_3 p_4 + p_2 p_3 p_4 \Big) -\frac{3}{8f^4}\Big(p_1^2 p_2 + \text{perms.}\Big) \\
&\hspace{7.3 cm}\,- \frac{1}{8f^4}\Big(p_1^3 + p_2^3 + p_3^3 + p_4^3\Big)\,.
\end{aligned}
\ee
\end{tcolorbox}
\noindent Later on, when it comes to performing recursion, the following form of the wavefunction will also be useful:\footnote{Even this is not the most economical representation of the wavefunction.  Using 3-momentum conservation, it may be reduced to
\begin{equation}
    \psi_4^{({\rm dbi})}  = \frac{1}{f^4E}\Big(P_1\cdot P_2\, P_3\cdot P_4 + P_1\cdot P_3\, P_2\cdot P_4 + P_1\cdot P_4\, P_2\cdot P_3\Big)\,.
    \label{DBI4pointniceform}
\end{equation}
This is, of course, a particular representation of the scattering amplitude divided by the total energy. 
Notice that this form of the amplitude in the numerator of~\eqref{DBI4pointniceform} manifestly has the ${\cal O}(p)$ Adler zero, while this is not manifest in~\eqref{DBIamplitude}.
Noting that the different forms of the amplitude correspond to different forms of the Lagrangian related by an integration by parts, we can see that this simple way of writing the answer corresponds to the Lagrangian with one time derivative per field (see Appendix \ref{app:boundaryterms}).}
\be
\begin{aligned}
	\psi_4^{({\rm dbi})} =& -\frac{2}{f^4 E}\Big(P_1\cdot P_2 ~P_2\cdot P_3 + P_2\cdot P_3~P_3\cdot P_1 + P_3\cdot P_1 ~P_1\cdot P_2\Big) \\
&\hspace{4 cm}- \frac{1}{f^4}\Big(P_1\cdot P_2p_3 + P_1\cdot P_3p_2 +  P_2\cdot P_3p_1\Big)\,.
\end{aligned}
\label{eq:simplepx4pt}
\ee
It is possible to build the six-point wavefunction in the same systematic way, but in Section~\ref{sec:softrecurs} we give a more elegant recursive construction using the same input.

\subsection{Galileon theories}
As a final example, we consider galileon field theories. These are theories that have a shift symmetry similar to the DBI symmetry~\eqref{eq:DBIsymm}, but which is field independent
\be
\delta_{B^{\mu}}\phi = x^{\mu},
\ee
along with the ordinary shift symmetry $\delta_{C}\phi =1$.  In this section we will focus solely on the quartic galileon vertex. However, our results should generalize to any galileon theory.  In addition, among the class of galileon theories, there is a distinguished subset~\cite{Cachazo:2014xea,Cheung:2014dqa,Hinterbichler:2015pqa}---often called the special galileon---that has an additional symmetry of the form
\be
\delta_{S_{\mu\nu}}\phi = s_{\mu\nu}\left(x^\mu x^\nu + \frac{1}{f^6}\partial^\mu\phi\partial^\nu\phi\right)\,.
\label{eq:sgalsymm}
\ee
As before, we first derive the soft theorems associated to these symmetries and then use them to bootstrap the wavefunction.

\subsubsection{Soft theorems}
As was mentioned before, the special galileon has three different symmetries: an ordinary shift symmetry, a symmetry linear in $x^\mu$ (galileon), and a symmetry quadratic in $x^\mu$ (special galileon). We will treat each of these in turn.

\paragraph{Shift symmetry:} Like the DBI case, the shift symmetry, $\delta_C\phi = 1$ leads to an Adler zero---the wavefunction coefficients vanish in the soft limit as in~\eqref{simpleshiftward}.

\paragraph{Galileon symmetry:} We next consider the $\delta_{B^\mu}\phi = x^\mu$ symmetry.  We first need the generators of the spatial and temporal parts of this symmetry:\footnote{The spatial charge may be written down immediately if $\delta\phi$ is known.  The temporal charge is more subtle, but can be obtained by following the procedure outlined in Appendix~\ref{app:generators}. For the charge to act correctly, it must be of the form
\be
Q_{B}^{0} = \int \rd^3x\,t\Pi^{(\phi)}(\vec{x}, t) - K^{0}[\phi(\vec{x}, t)]\,.
\label{eq:temporalchargegal}
\ee
Using the explicit expression for the charges, we can compute the commutator $[Q_B^i, Q_B^0]= \int \rd^3x\,\delta_{B^i}K^{0}[\phi(\vec{x}, t)]$. We know from the algebra of symmetries that this commutator should vanish, which requires that $K^0$ is invariant under a spatial galileon symmetry transformation. The terms with this symmetry and with the right derivative counting and number of fields are themselves (three-dimensional euclidean) galileon terms: 
\be
K^{0}[\phi] = \phi - \frac{N}{f^6}\phi\Big[(\nabla^2\phi)^2 - (\partial_i\partial_j\phi)^2\Big]\,.
\ee
Fourier transforming~\eqref{eq:temporalchargegal} then produces~\eqref{eq:galB0symfourier}.
}
\begin{align}
\label{eq:galBisymfourier}
Q_B^i &= -i\lim_{\vec k\to 0} \frac{\rd}{\rd k^i}\Pi^{(\phi)}(\vec k)\,,\\
Q_B^0 &= \lim_{\vec k\to 0}\,\Big(t\Pi^{(\phi)}(\vec{k}) - \phi_{\vec{k}}( t)\Big)\nonumber\\
    &\hspace{1 cm}+\frac{N}{f^6}\int \frac{\rd^3p_1\rd^3p_2\rd^3p_3}{(2\pi)^{3\cdot2}}\,\delta^{(3)}(\vec{p}_1 + \vec{p}_2 + \vec{p}_3)\phi_{\vec{p}_1}( t)\phi_{\vec{p}_2}(t)\phi_{\vec{p}_3}(t)\Big(p_2^2p_3^2 -(\vec{p}_2\cdot\vec{p}_3)^2 \Big)\,.
    \label{eq:galB0symfourier}
\end{align}
Here, $N$ is a constant which fixes the normalization of the interaction term in the action.  It may be absorbed into the coupling parameter $1/f^6$. However we will keep it explicit for convenience later on when we bootstrap wavefunction coefficients.  Interestingly, the  $K^0$ term in $Q_B^0$ only has a term with one $\phi$ and one with three $\phi$s. Eventually we will see that these terms only contribute to the Ward identities satisfied by the two-point and four-point wavefunction coefficients. All other wavefunction coefficients will actually have higher-order Adler zeroes, vanishing like ${\cal O}(p^2)$.

\vskip4pt
\noindent
{\it Spatial soft theorem:} We first consider the soft theorem associated to the symmetry~\eqref{eq:galBisymfourier}. Everything proceeds essentially identically to the DBI case, except that there is no $\phi\partial\phi$ part of the symmetry. The resulting soft theorem is the same as~\eqref{eq:DBIsymmintermediate} with the middle term removed, because it comes from the $\phi\partial\phi$ part of the symmetry. In terms of wavefunction coefficients, we find

\begin{tcolorbox}[colframe=white,arc=0pt,colback=greyish2]
\be
\lim_{\vec{k}\rightarrow 0} \vec{\partial}_{k}\psi_n(\vec{k}, \vec{p}_2,\cdots,\vec{p}_{n}) = 0\,,
\ee
\end{tcolorbox}
\noindent which implies that the galileon wavefunction exhibits a spatial enhanced Adler zero.

\vskip4pt
\noindent
{\it Temporal soft theorem:} We next turn to the soft theorem that is a consequence of the temporal symmetry. In this case the general Ward identity~\eqref{eq:generalsoftthm} can be written as
\be
 Q_B^0\Psi[\phi, t_f]  =  \lim_{\vec q\to 0} ~\Big[it_i\mathcal{E}(k)-1\Big]\phi^{\rm cl}_{\vec k}(t_i)\Psi[\phi, t_f]\,,
\ee
where $Q_B^0$ is given by~\eqref{eq:galB0symfourier}. Now we may expand both sides in terms of wavefunction coefficients. After relabelling we have
\begin{align}
	&-\int \rd^3k\,\delta^{(3)}(\vec{k})\varphi_{\vec{k}}(t_f) + \frac{N}{f^6}\int\frac{\rd p_1^3\rd p_2^3 \rd p_3^3}{(2\pi)^{3\cdot 2}}\delta^{(3)}(\vec{p}_1 + \vec{p}_2 + \vec{p}_3)\varphi_{\vec{p}_1}(t)\varphi_{\vec{p}_2}(t)\varphi_{\vec{p}_3}(t)\Big(p_2^2p_3^2 - (\vec{p}_2\cdot\vec{p}_3)^2\Big)\nonumber\\
	& = -\lim_{\vec k\to 0} ~\Big[\frac{it_i}{(2\pi)^3}\mathcal{E}(k)-1\Big]\sum_{n = 2}\frac{1}{(n-1)!}\int \frac{\rd^{3}p_1\cdots \rd^3p_{n-1}}{(2\pi)^{3(n-2)}}\,\delta^{(3)}(\vec{p}_1 + ... + \vec{p}_{n-1})\nonumber\\
&\hspace{8 cm}\times\varphi_{\vec{p}_1}\cdots \varphi_{\vec{p}_{n-1}}\partial_{k}\psi_n(\vec{p}_1,\cdots,\vec{p}_{n-1},\vec{k})\,.
\end{align}
In the soft limit, the $\mathcal{E}(k)$ term vanishes. We may take functional derivatives to extract the soft theorems obeyed by the wavefunction coefficients:

\begin{tcolorbox}[colframe=white,arc=0pt,colback=greyish2]
\vspace{-8pt}
\begin{align}
    & \lim_{\vec{k}\rightarrow 0}\partial_k\psi_2(\vec{k}) =-1 \,,\label{2pointgalsoft}\\
    & \lim_{\vec{k}\rightarrow 0}\partial_k\psi_4(\vec{k}, \vec{p}_2, \vec{p}_3, \vec{p}_4) = \frac{2N}{f^6}\Big(p_2^2p_3^2 -(\vec{p}_2\cdot\vec{p}_3)^2\Big) + \text{perms.}\,, \label{4pointgalsoft}\\
    &\lim_{\vec{k}\rightarrow 0}\partial_k\psi_n(\vec{k}, \vec{p}_2,...,\vec{p}_n) =0, \hspace{1 cm}n>4\,.
\end{align}
\end{tcolorbox}
\noindent Notice that $n=2,4$ are exceptional cases where the wavefunction's energy derivative in the soft limit does not vanish, but rather obeys a soft theorem.

\vskip4pt
\paragraph{Special galileon symmetry:} Finally we consider the soft theorems associated to the special galileon symmetry~\eqref{eq:sgalsymm}.  This symmetry has a full symmetric traceless tensor of charges.  As in the DBI case, the spatial and temporal components of the charges behave differently.  The spatial (traceless) charge is given by
\begin{align}
	s_{ij}Q_{S}^{ij} = \int \rd^3x\,\Big(x^ix^j + \frac{1}{f^6}\partial^i\phi\partial^j\phi\Big)\Pi^{(\phi)}
\end{align} 
where $s_{ij}$ is a symmetric and traceless tensor.  This charge may be immediately written down from the symmetry transformation following the logic in Appendix~\ref{app:generators}, and does not require knowledge of the action. To derive the temporal charges, one needs to compute the boundary term the action develops by brute force. Since we do not require the temporal charges for our purposes, we leave these details for future work.

\vskip4pt
\noindent \textit{Spatial Soft Theorem}: To derive the special galileon spatial soft theorem, we essentially follow the same steps as in the derivation of the DBI spatial soft theorem.  The end result is

\begin{tcolorbox}[colframe=white,arc=0pt,colback=greyish2]
\be
	\lim_{\vec{k}\rightarrow 0}s_{ij}\partial_{k}^i\partial_{k}^j \psi_n(\vec{k}, \vec{p}_2,\cdots,\vec{p}_{n}) = -\sum_{\text{pairs }a,b}\frac{1}{f^6}s_{ij}p_a^ip_b^j\,\psi_{n-2}(\vec{p}_a + \vec{p}_b, \vec{p}_2,\cdots\bar{p}_a,\cdots, \bar{p}_b,\cdots,\vec{p}_{n})\,.
\ee
\end{tcolorbox}

\subsubsection{Wavefunction coefficients}
We now want to use the soft theorems discussed in this section to bootstrap the wavefunction of the galileon.  We begin with an ansatz for the wavefunction of the form
\be
    \psi_4^{({\rm gal})} = \frac{A_4}{E} + R\,,
\ee
where $A_4$ is some representation of the scattering amplitude and $R$ is a remainder term without any singularities. Concretely, we can write the amplitude as (temporarily setting the coupling $1/f^6$ to 1)
\be
 A_4 
 = -\frac{1}{2}\Big(P_1\cdot P_2 P_1\cdot P_3 P_1\cdot P_4 + {\rm perms.}\Big)\,.
\label{galamp}
\ee
This way of writing the amplitude manifestly vanishes as ${\cal O}(p)$ in the soft limit.  However when energy is conserved,  this object actually vanishes as 
 ${\cal O}(p^2)$, though this cannot be made manifest.\footnote{Actually it vanishes like $\mathcal{O}(p^3)$ due to the special galileon symmetry, but we will not utilize this property.}   As in the case of NLSM and DBI, the cut of the wavefunction coefficient must vanish, which means that each term must have an odd power of external energies.  Then by dimensional analysis, $s_{ab}$ can only appear in even powers.  Up to dimension-dependent Gram identities, the most efficient ansatz takes the form
\be
\begin{aligned}
    R =~ a_1\,p_1^2 p_2 p_3 p_4 + a_2\,p_1^3 p_2 p_3 + a_3\,p_1^4 p_2 + a_4\,p_1^5 &+ a_5\,p_1^2 p_2^2 p_3 + a_6p_1^3p_2^2\\
&+ a_7\,p_1 s_{12}^4 + a_8\,p_1^2p_2s_{12}^2 + {\rm perms.}
\end{aligned}
\ee
In order to restrict this ansatz, we first
check the simple Adler zero coming from the ordinary shift symmetry. The amplitude part manifestly has this property, but in the soft limit, $R$ becomes
\be
\lim_{\vec{p}_1\rightarrow 0}R= a_2\,p_2^3 p_3 p_4 + \Big(a_3 + a_7\Big)\,p_2^4 p_3 + \Big(a_4 + a_7\Big)\,p_2^5 + \Big(a_5 + 2a_8\Big)\,p_2^2 p_3^2 p_4 + a_6p_2^3p_3^2 + {\rm perms.}
\ee
Since this is an independent basis of polynomials, in order for this quantity to vanish we have to set
\begin{equation}
    a_2 = a_6 = 0,\hspace{1.5 cm}a_3 = a_4 = -a_7, \hspace{1.5 cm}a_5 + 2a_8 = 0\,.
\end{equation}
Next, we impose the galileon symmetry soft theorem. First, we impose the ${\cal O}(p)$ spatial soft theorem, which is an Adler zero condition: 
\be
\begin{aligned}
\lim_{\vec{p}_1\rightarrow 0}\vec{\partial}_{p_1}\psi_4^{({\rm gal})} = &~\vec{p}_2\frac{E}{4}(p_2 - p_4)\Big[(1 + 16 a_7)(p_2+p_4) - (1 + 8 a_8)p_3\Big] \\
	&+ \vec{p}_3\frac{E}{4}(p_3 - p_4)\Big[(1 + 16 a_7)(p_3+p_4) - (1 + 8 a_8)p_2\Big]\,.
\end{aligned}
\ee
In order for this quantity to vanish we must set
\be
	a_7 = -\frac{1}{16}, \hspace{1 cm}a_8 = -\frac{1}{8}\,.
\ee
Now we proceed to impose the ${\cal{O}}(p)$ temporal soft theorem, which reads:\footnote{Recall that $N$ corresponds to the normalization of the interaction term in the action.  However, we have already fixed this normalization by our choice of the overall constant in the scattering amplitude in~\eqref{galamp}.  Thus in this way of framing our input data, $N$ is an additional parameter we must solve for. }
\be
\begin{aligned}
	\lim_{\vec{p}_1\rightarrow 0}\partial_{p_1}\psi_4^{({\rm gal})} &= \frac{1}{4}p_2^4 - \frac{1}{2}p_2^2p_3^2 + \Big(a_1 + \frac{1}{2}\Big)p_2^2p_3p_4 + \text{perms.}\\
	& =-\frac{3N}{2}E\Big(p_2 - p_3 - p_4\Big)\Big(p_4 - p_2 - p_3\Big)\Big(p_3 - p_2 - p_4\Big)\,.
\end{aligned}
\ee
Note that there is no choice of $a_1$ which will cause the top line to vanish.  This implies that there is no theory that matches the galileon scattering amplitudes when $E=0$, has an ${\cal O}(p^2)$ spatial Adler zero, and also exhibits an ${\cal O}(p^2)$ temporal Adler zero. 
The best one can do is fix $a_1$ so that the wavefunction coefficient satisfies the soft theorem in  \eqref{4pointgalsoft}.  The equation above is uniquely solved by
\be
	a_1 = \frac{1}{2}, \hspace{2 cm} N = \frac{1}{6}\,.
\ee
Thus, we have successfully bootstrapped the wavefunction. The final result is given by
\begin{tcolorbox}[colframe=white,arc=0pt,colback=greyish2]
\vspace{-4 pt}
\be
\begin{aligned}
	 \psi_4^{({\rm gal})} = &\frac{1}{2f^6E}\Big(P_1\cdot P_2P_1\cdot P_3P_1\cdot P_4 + \text{perms.}\Big)\\
    & + \frac{1}{f^6}\Big(\frac{1}{2}\,p_1^2 p_2 p_3 p_4 + \frac{1}{16}\,p_1^4 p_2 + \frac{1}{16}\,p_1^5 + \frac{1}{4}\,p_1^2 p_2^2 p_3 -\frac{1}{16}\,p_1 s_{12}^4 -\frac{1}{8}\,p_1^2p_2s_{12}^2\Big) +\text{perms.}
\end{aligned}
\label{eq:gal4pt}
\ee
\end{tcolorbox}
\noindent Contrary to the NLSM and DBI cases, there does not exist a way of writing the wavefunction which takes the form of a manifestly Lorentz invariant scattering amplitude divided by the total energy. This is related to the fact that after accounting for the boundary term, the interaction term in the action is not manifestly Lorentz invariant (see Appendix \ref{galreview}).  This form of the wavefunction is somewhat cumbersome to work with when it comes to performing recursion. A nicer form of the four-point function, which does not have manifest Bose symmetry, is given by 
\begin{equation}
    \psi_4^{({\rm gal})} = \frac{2}{f^6E}P_1\cdot P_2 \,P_2\cdot P_3 \,P_3\cdot P_1 +\frac{1}{f^6}\Big( p_1\big(p_2^2p_3^2 -(\vec{p}_2\cdot\vec{p}_3)^2\big) + p_2\big(p_1^2p_3^2 - (\vec{p}_1\cdot\vec{p}_3)^2\big) + p_3\big(p_1^2p_2^2 - (\vec{p}_1\cdot\vec{p}_2)^2\big)\Big)\,.
    \label{eq:simplegal4pt}
\end{equation}
This form of the wavefunction coefficient may be straightforwardly derived by using three-momentum conservation to eliminate all instances of $\vec{p}_4$ and $p_4$ in~\eqref{eq:gal4pt}, except in the total energy singularity.

\newpage
\section{Recursion relations}
\label{sec:softrecurs}
So far, we have explored how the wavefunction in exceptional scalar theories can be fixed in terms of the corresponding scattering amplitudes, supplemented by some information about soft limits. However, the brute-force approach we have followed quickly becomes cumbersome, motivating us to search for a more efficient algorithm. In the $S$-matrix context, similar problems have been overcome via the construction of powerful recursion relations---the most famous of these being the BCFW relations~\cite{Britto:2005fq}---which have been applied to scalar theories~\cite{Cheung:2015ota,Cheung:2016drk,Padilla:2016mno,Bonifacio:2019rpv}. It is therefore natural to look for a similar recursive construction of the wavefunction.

\vskip4pt
Recursion relations for the  wavefunction both in flat space and in de Sitter space have been studied already by~\cite{Arkani-Hamed:2017fdk,Jazayeri:2021fvk,Baumann:2021fxj}. These relations are obtained by deforming the energy variables that the wavefunction depends on into the complex plane and writing the true wavefunction as a sum over residues of the poles of the complex function. This is natural because, as we reviewed in Section~\ref{sec:sing}, all the singularities of the wavefunction occur at loci in energy space. However, the wavefunction is fundamentally a function of {\it momenta}, rather than only energies. We are therefore motivated to look for a construction where the 3-momenta themselves are deformed. Another motivation for this approach is that we will need additional input from soft theorems in our recursion relations. This is information about the behavior of the wavefunction and its derivatives in the limit $\vec{p}\rightarrow 0$, and shifts that only deform the energies do not allow us to access this regime.  

\subsection{Recursion generalities}
Here we describe the philosophy underlying the recursion relations that we are going to construct. 
As orientation---and to contrast with the wavefunction case---it is useful to quickly review the construction for scattering amplitudes.
At the most basic level, the idea is to deform the amplitude into the complex plane by shifting some of its kinematic variables.  This deformed amplitude, $A(z)$, now is a complex function of the parameter of this shift, $z$, while the undeformed (true) amplitude is $A(0)$. We can then use Cauchy's formula to write
\be
A(0) = \frac{1}{2\pi i}\oint_{z = 0}\rd z \frac{A(z)}{z}\,,
\ee
where the subscript on the integral indicates that we are integrating around a small contour encircling the point $z = 0$. Then, we can deform the contour out to infinity. If all the singularities of $A$ are poles (which they are for amplitudes at tree level), we can write $A(0)$ as a sum of residues, plus a possible contribution from a pole at infinity:
\be
A(0) = -\sum_j \Res_{z=z_j}\left(\frac{A(z)}{z}\right) - B_\infty\,.
\ee
In many cases of interest, the pole at infinity vanishes and we can therefore reconstruct the amplitude from its residues, which are given by products of lower point amplitudes.

\vskip4pt
The situation for the wavefunction is conceptually similar.  We imagine analytically continuing the wavefunction coefficients into the complex plane by deforming the 3-momenta as
\be
    \vec{p}_a(z) = \vec{p}_a + z \vec{q}_a, \hspace{2 cm}\vec{p}_a\cdot\vec{q}_a = p_a q_a, \hspace{2 cm}\sum_{a}\vec{q}_a(z) = 0\,,
    \label{generalvectorshift}
\ee
where we have ensured that the $\vec q_a$ sum up to zero, so that momentum conservation continues to hold for the deformed variables. In addition, as a result of the middle constraint the shifted energies satisfy
\be
\sqrt{\vec{p}_a{}^2(z)} = p_a(z) = p_a + z q_a\,.
\ee
That is, the energies get deformed by the lengths of $\vec q_a$.

\vskip4pt
There is a very important difference between scattering amplitudes and the wavefunction: while analytically continued scattering amplitudes have {\it only} pole-like singularities, the wavefunction has branch cuts in the complex $z$-plane due to the presence of $s_{a_1\cdots a_m}$ factors, (where there are $m$ momenta adding up to the exchange momentum).\footnote{An exception is if we choose a special deformation such that $\vec{q}_1 +\cdots+\vec q_m = 0$.  In this case, the $z$ dependence under the square root falls out, and the branch cut issue is avoided, but only for the exchange channel where the internal momentum is $
\vec{p}_1 + \vec{p}_2+...+\vec p_m$. It is typically impossible to choose such set of constraints that simultaneously simplifies all of the possible channels in this way.}  
We must therefore  learn to deal with the branch cuts. This is equivalent to understanding the analytic structure of the deformed partial energies  $E_{1\cdots m}(z)$. 
Under the shifts in \eqref{generalvectorshift}, in the complex $z$ plane $E_{1\cdots m}(z)$ has a cut which extends between the branch points
\begin{equation}
   b_\pm = \frac{-\vec{p}_{1\cdots m}\cdot\vec{q}_{1\cdots m}\pm\sqrt{(\vec{p}_{1\cdots m}\cdot\vec{q}_{1\cdots m})^2 - \vec{p}_{1\cdots m}^{\,\,2}\vec{q}_{1\cdots m}^{\,\,2}}}{\vec{q}_{1\cdots m}^{\,\,2}}\,.
\end{equation}
In addition, because the partial energies generically appear in the denominator, one might also be concerned about the partial energy poles $E_{1\cdots m}(z) = 0$.  These lie at 
\begin{equation}
    z_{\pm} = \frac{-P_{1\cdots m}\cdot Q_{1\cdots m}\pm\sqrt{(P_{1\cdots m}\cdot Q_{1\cdots m})^2 - P_{1\cdots m}^2Q_{1\cdots m}^2}}{Q_{1\cdots m}^2}\,,
\end{equation}
where $P_{1\cdots m} = (p_{1\cdots m}, \vec{p}_{1\cdots m})$ and $Q_{1...m} = (q_{1\cdots m}, \vec{q}_{1\cdots m})$.   However the partial energies are multi-valued functions, and it turns out that we can always define $\psi(z)$ by choosing a branch which does not have zeros (see Appendix~\ref{app:cuts} for more details). Because of this, we only have to worry about partial energy branch cuts.

\vskip4pt
In addition to the partial energy branch cuts, a generic wavefunction will also have a singularity when $E(z) =0$, whose residue is the corresponding scattering amplitude:
\be
    \psi(z)\xrightarrow[z\rightarrow z_E]
    {}\frac{A(z_{E})}{E(z_{E})}\,, \hspace{.5 cm}{\rm where}\hspace{.5 cm}\sum_{i}p_{i}(z_E) = 0\,.
\ee

Now that we have catalogued all the relevant structures in the complex plane, we may arrange them into a recursion formula. The procedure is similar to the scattering amplitude case. We write the true wavefunction as
\be
\psi(0) = \frac{1}{2\pi i}\oint_{z = 0}\frac{\rd z}{z}\psi(z)\,,
\ee
and then deform the contour out to infinity, which picks up contributions from the total energy singularity, from the integration contour running along the partial energy branch cuts, and a possible contribution from infinity. All together, this means that we can write the wavefunction as
\be
    \psi(0)  =-\sum_I\frac{1}{2\pi i}\oint_{\text{cut $I$}}\frac{\rd z}{z}\psi(z) - \frac{1}{2\pi i}\oint_{E(z)=0}\rd z\frac{A(z)}{E(z)}  - B_{\infty}\,,
    \label{correlatorrecursestart}
\ee
where ``cut $I$" indicates that the contour encircles the branch cut associated to the partial energy singularities of factorization channel $I$.  We have also included a possible boundary contribution, $B_\infty$.\footnote{The integrand must vanish strictly faster than $1/z$ for large $z$ in order for this contribution to vanish.} Note that the integral around $E(z)=0$ just extracts the residue of the integrand at $z_E$---the point where $E(z_E)  =0$---which is  $- \frac{A(z_E)}{z_E} $. However, we will later see that it is often possible to do further contour deformations to simplify the evaluation of this expression.

\vskip4pt
Equation~\eqref{correlatorrecursestart} is true, but only useful if we know how to deal with the integrations along the branch cuts of $\psi(z)$, which appear because of the square roots in the partial energies. At tree level, we can gain some insight into this problem by examining where the branch cuts come from in the first place. In terms of bulk perturbation theory, partial energy singularities and their associated branch cuts  arise from exchanges of particles.  We can understand the analytic structure of these exchanges by inspecting the bulk-to-bulk propagator~\eqref{eq:bbulkprop} in frequency space:
\be
    {\cal G}( s_I ; t_1, t_2) 
    = \frac{1}{2}\int_{-\infty}^{\infty}\frac{\rd\omega}{2\pi i}\frac{\Big(e^{i\omega t_1} - e^{-i\omega t_1}\Big)\Big(e^{i\omega t_2} - e^{-i\omega t_2}\Big)}{\omega^2 - s^2_{I} + i\epsilon}\,,
\label{B2Bfrequency}
\ee
where $s_I$ is the magnitude of the exchanged momentum $\vec s_I$.  Consider a wavefunction with an arbitrary number of exchanges, and for now focus on just a single one: 
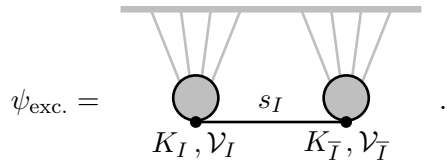
\begin{equation*}
\psi_{\rm exc.} = 
\raisebox{-23pt}{
\begin{tikzpicture}[line width=1. pt, scale=2]
\draw[fill=black] (0,0) -- (1,0);
\draw[lightgray, line width=1.pt] (0,0) -- (-0.3,0.75);
\draw[lightgray, line width=1.pt] (0,0) -- (0.3,0.75);
\draw[lightgray, line width=1.pt] (0,0) -- (-0.1,0.75);
\draw[lightgray, line width=1.pt] (0,0) -- (0.1,0.75);

\draw[lightgray, line width=1.pt] (1,0) -- (0.7,0.75);
\draw[lightgray, line width=1.pt] (1,0) -- (1.3,0.75);
\draw[lightgray, line width=1.pt] (1,0) -- (0.9,0.75);
\draw[lightgray, line width=1.pt] (1,0) -- (1.1,0.75);

\draw[lightgray, line width=2.5pt] (-0.5,0.75) -- (1.5,0.75);
\draw[fill=lightgray] (0,0.16) circle (.15cm);
\draw[fill=black] (0,0) circle (.03cm);
\draw[fill=lightgray] (1,0.16) circle (.15cm);
\draw[fill=black] (1,0) circle (.03cm);
\node[scale=1] at (-.01,-.155) {$K_I\,,{\cal V}_I$};
\node[scale=1] at (1.0,-.155) {$K_{\bar{I}}\,,{\cal V}_{\bar{I}}$};
\node[scale=1] at (0.5,.12) {$s_{I}$};
\end{tikzpicture}
}\,.
\end{equation*}
The grey blobs are any completion of the graph and may contain an arbitrary number of exchanges. For exchange $I$, the energy flowing into the left vertex from the boundary is $K_I$ and the energy flowing into the right vertex is $K_{\bar{I}}$.\footnote{For example, for a four-point function we could refer to the $s$ channel as the $I = (12)$ channel, with $\bar{I} = (34)$.  The momentum flowing through the internal line is $\vec{s}_I = \vec{p}_1 + \vec{p}_2$.} By $I$ we therefore mean the set of external lines corresponding to the ``left" subgraph and by $\bar I$ we mean the set of external lines associated to the ``right" subgraph.
In addition, ${\cal V}_I$ and ${\cal V}_{\bar{I}}$ abstractly parametrize the vertex factors appearing on the left (resp. right) side of the exchange.  We can translate this into an expression as
\be
\psi_{\rm exc} = \frac{1}{2}\int \frac{\rd \omega}{2\pi i} \int\rd t_1\rd t_2\,  i{\cal V}_I(\vec k_I, \omega)e^{iK_It_1}\frac{\Big(e^{i\omega t_1} - e^{-i\omega t_1}\Big)\Big(e^{i\omega t_2} - e^{-i\omega t_2}\Big)}{\omega^2 - s^2_{I} + i\epsilon}e^{iK_{\bar{I}}t_2}\,i{\cal V}_{\bar{I}}(\vec k_{\bar{I}}, \omega)\,,
\label{eq:cuttimeint}
\ee
At the level of this expression, it is somewhat obscure from where the branch cuts in the final wavefunction originate. However, this may be made manifest by performing the time integrals:
\be
\psi_{\rm exc} = \frac{1}{2}\int\frac{\rd \omega}{2\pi i} i{\cal V}_I(\vec k_I, \omega)\frac{2\omega}{(K_{I}-i\epsilon)^2-\omega^2}  \frac{1}{\omega^2-s_I^2+i\epsilon}\frac{2\omega}{(K_{\bar{I}}-i\epsilon)^2-\omega^2}\,i{\cal V}_{\bar{I}}(\vec k_{\bar{I}}, \omega)\,.
\ee
Notice that at the level of this integrand, only $s_I^2$ appears, so $\psi(z)$ does not have a branch cut inside the integral. Instead, the cut arises when we actually perform the frequency integral. In particular, the $\omega$-integral may be computed by residues, and upon doing so, it is only the residue at $\omega = s_I$ which has a branch cut.
Therefore, it is only this residue which contributes to the integral along the branch cut in \eqref{correlatorrecursestart}.  
Looking back at the time integral~\eqref{eq:cuttimeint}, we see that evaluating the integrand on
the $\omega = s$ pole physically corresponds to putting the internal propagator on-shell, and serves to cut the internal line as in~\eqref{eq:cuteq}.  In particular, the time integrand is exactly the cut propagator in \eqref{eq:cutprop}, and carrying out the time integrals computes a product of shifted wavefunctions, so that we have\footnote{Notice the factor of $1/2$ difference between the integrand on the RHS and the cut diagram in e.g., \eqref{eq:cuteq}. This stems from the prefactor in \eqref{B2Bfrequency}, indicating that on the $\omega = s$ pole, the bulk-to-bulk propagator is half the cut propagator. }
\be
	\frac{1}{2\pi i}\oint_{\text{cut } I}\frac{\rd z}{z}\,\psi(z)  = -\frac{1}{2\pi i}\oint_{\text{cut $I$}}\frac{\rd z}{z}\,s_I(z)\,\tl\psi_{I}(z)\times \tl\psi_{\bar{I}}(z)\,.
\ee
We have written the integral along the branch cut of the full wavefunction as an integral along the branch cut of a product of lower point shifted wavefunction coefficients.  In this way, the wavefunction ``factorizes" into lower point objects around its branch cuts, much in the way that a scattering amplitude factors into lower point amplitudes on its poles.  Putting everything together, we can write the following recursion relation for the wavefunction:
\begin{tcolorbox}[colframe=white,arc=0pt,colback=greyish2]
\be
    \psi(0)  = \sum_I\frac{1}{2\pi i}\oint_{\text{cut $I$}}\frac{\rd z}{z}\,s_I(z)\,\tl\psi_{I}(z)\times \tl\psi_{\bar{I}}(z) - \frac{1}{2\pi i}\oint_{E(z)=0}\frac{\rd z}{z}\frac{A(z)}{E(z)}  - B_{\infty}\,.
    \label{eq:correlatorrecurse}
\ee
\end{tcolorbox}
\noindent Given this general formula, we will now apply it to some specific examples. Typically we will want to choose our shifts so that the boundary term $B_\infty$ is absent, allowing us to recursively construct higher point functions from simpler building blocks.

\subsection{Simple examples}
In order to demonstrate the use of~\eqref{eq:correlatorrecurse}, we first consider two simple examples---the wavefunction coefficients in a theory with $\phi^3$ and $\phi^4$ interactions, and the four-point wavefunction coefficient in scalar QED.

\paragraph{Scalar theory with $\phi^3$ and $\phi^4$ interactions:} Possibly the simplest example of a nontrivial wavefunction coefficient is the 4-point wavefunction in a scalar theory with both contact and exchange contributions. Concretely we consider a theory with interactions ${\cal L}_{\rm int} \sim \frac{\lambda_3}{3!}\phi^3+\frac{\lambda_4}{4!}\phi^4$. We shift the momenta as
\begin{equation}
    \hat{\vec{p}}_a(z) = \vec{p}_a + z\vec{q}_a, \hspace{.5 cm}{\rm with}\hspace{.5cm}\vec{p}_a\cdot \vec{q}_a = p_aq_a,\hspace{.5 cm}{\rm and}\hspace{.5cm}\sum_{a}^{4}\vec{q}_a = 0\,.
    \label{democraticshifts}
\end{equation}
 Under these shifts, the energies are deformed as
\begin{equation}
    \hat p_a(z) = p_a + zq_a\,.
\end{equation}
Note that we are not requiring that the total energy is conserved by the shifts (in contrast to what is normally done for scattering amplitudes).\footnote{We can write these shifts in four-vector notation more simply as
\begin{equation}
    P_a(z) = P_a + zQ_a, \hspace{1 cm}P_{a}\cdot Q_{a} =0, \hspace{1 cm}Q_{i}^2 = 0, \hspace{1 cm}\sum_{a}^{4}\vec{q}_{a} = 0\,,
\end{equation}
where $P_a^\mu \equiv (p_a,\vec p_a)$ and $Q_a^\mu \equiv (q_a,\vec q_a)$.
}
This choice of shifts causes $z^{-1}\psi_4(z)$ to scale as $z^{-2}$ when $z\to \infty$, so that the boundary term is absent.

\vskip4pt
In order to use the formula~\eqref{eq:correlatorrecurse}, we will require as inputs the three-point wavefunction coefficient:
\be
\psi_3 = \frac{\lambda_3}{p_1+p_2+p_3}\,,
\ee
along with the four-point scattering amplitude
\be
A_4 = -\lambda_3^2\left(\frac{1}{S}+\frac{1}{T}+\frac{1}{U}\right)+\lambda_4\,,
\label{phi3amp}
\ee
where $S = p_{12}^2-s^2 = p_{34}^2-s^2$ on the total energy singularity (and similarly for $T, U$). For the $s$-channel, the shifted 3-point wavefunction coefficients are given by
\be
	\tilde{\psi}_{(12)} = \frac{\lambda_3}{p_{12}^2 - s^2} \hspace{1 cm} \tilde{\psi}_{(34)} = \frac{\lambda_3}{p_{34}^2 - s^2}
\ee
and similarly for the other channels. With this information, we can write~\eqref{eq:correlatorrecurse} as
\be
\begin{aligned}
    \psi_4 = &~\frac{1}{2\pi i}\oint_{\text{$s$ cut }}\frac{\rd z}{z} \frac{\lambda_3^2 \,\hat s}{(\hat p_{12}^2-\hat s^2)(\hat p_{34}^2-\hat s^2)}
    +\frac{1}{2\pi i}\oint_{\text{$t$ cut }}\frac{\rd z}{z} \frac{\lambda_3^2\,\hat t}{(\hat p_{14}^2-\hat t^2)(\hat p_{23}^2-\hat t^2)}\\
   &+\frac{1}{2\pi i} \oint_{\text{$u$ cut }}\frac{\rd z}{z}\frac{\lambda_3^2\,\hat u}{(\hat p_{13}^2-\hat u^2)(\hat p_{24}^2-\hat u^2)}-\frac{1}{2\pi i} \oint_{\hat E = 0}\frac{\rd z}{z}\frac{A_4(z)}{\hat E}\,,
    \label{phi3recursion}
\end{aligned}
\ee
where the hatted variables indicate that the energies are deformed by the complex parameter, $z$. One lingering point of concern might be that the answer depends on the choice of representation of the scattering amplitude. 
 In particular, there are many different representations which are all related by 4-momentum conservation, but which are inequivalent when energy is not conserved. Thus a natural question is which one should be used? Fortunately according to~\eqref{phi3recursion}, one only needs information about $A_4(z)$ on the $E(z) = 0$ singularity, that is, when energy is conserved. On this singularity, all amplitudes related by 4-momentum conservation are equivalent, so it does not matter which representation we take.

\vskip4pt
Our goal now is to evaluate the integrals in~\eqref{phi3recursion}.  This is straightforward to do. For now, focus on the $s$-channel integral.  The most efficient strategy is to deform the contour so that it surrounds poles corresponding to different kinematic relations, and then to use those relations to simplify the integrand.  This may be done until the only pole remaining is the pole at $z = 0$, which is easy to evaluate.  For instance, we may first deform the contour away from the $s$ cut and onto the poles at $z = 0, \hat p_{12} = \hat s$, and $\hat p_{34} = \hat s$.  Note that the last two are \textit{folded singularities}.  On each locus, we apply the corresponding kinematic identity to simplify the integrand, which yields
\be
\begin{aligned}
\psi_{s\,{\rm cut}} &= \frac{1}{2\pi i}\oint_{\text{$s$ cut }}\frac{\rd z}{z} \frac{\lambda_3^2 \,\hat s}{(\hat p_{12}^2-\hat s^2)(\hat p_{34}^2-\hat s^2)}\,, \\
&= -\frac{\lambda_3^2 \, s}{( p_{12}^2-s^2)( p_{34}^2-s^2)} -\frac{1}{2\pi i}\oint_{\hat p_{12} = \hat s}\frac{\rd z}{z} \frac{\lambda_3^2  \,\hat p_{12}}{\hat E(\hat p_{12}^2-\hat{s}^2)(\hat p_{34}-\hat p_{12})} \\
&\,~~~- \frac{1}{2\pi i}\oint_{\hat p_{34} = \hat s}\frac{\rd z}{z} \frac{\lambda_3^2 \,\hat k_{34}}{\hat E(\hat p_{12}-\hat p_{34})(\hat p_{34}^2-\hat s^2)}\,.
\end{aligned}
\ee
Now we simply iterate the process by deforming the contour again to obtain:
\be
\begin{aligned}
	\psi_{s\,{\rm cut}} = &-\frac{\lambda_3^2 \, s}{( p_{12}^2-s^2)( p_{34}^2-s^2)} +\frac{\lambda_3^2  \, p_{12}}{( p_{12}^2-s^2)E( p_{34}- p_{12})} +\frac{\lambda_3^2 \,p_{34}}{ E( p_{12}-p_{34})( p_{34}^2- s^2)}\\
& +\frac{1}{2\pi i}\oint_{\hat p_{12} = \hat p_{34}}\frac{\rd z}{z} \frac{\lambda_3^2  \,\hat p_{12}}{\hat E(\hat p_{12}^2-\hat{s}^2)(\hat p_{34}-\hat p_{12})} + \frac{1}{2\pi i}\oint_{\hat p_{34} = \hat p_{12}}\frac{\rd z}{z} \frac{\lambda_3^2 \,\hat p_{34}}{\hat E(\hat p_{12}-\hat p_{34})(\hat p_{34}^2-\hat s^2)}\\
& +\frac{1}{2\pi i}\oint_{\hat E = 0}\frac{\rd z}{z} \frac{\lambda_3^2  \,\hat p_{12}}{\hat E(\hat p_{12}^2-\hat{s}^2)(\hat p_{34}-\hat p_{12})} + \frac{1}{2\pi i}\oint_{\hat E = 0}\frac{\rd z}{z} \frac{\lambda_3^2 \,\hat p_{34}}{\hat E(\hat p_{12}-\hat p_{34})(\hat p_{34}^2-\hat s^2)}\,.
\end{aligned}
\ee
In the second line, the $\hat p_{12} = \hat p_{34}$ residues cancel one another. The total energy residues may also be simplified, leaving us with 
\begin{align}
	\psi_{s\,{\rm cut}}  = &-\frac{\lambda_3^2 \, s}{( p_{12}^2-s^2)( p_{34}^2-s^2)} +\frac{\lambda_3^2  \, p_{12}}{( p_{12}^2-s^2)E( p_{34}- p_{12})} +\frac{\lambda_3^2 \,p_{34}}{ E( p_{12}-p_{34})( p_{34}^2- s^2)}\nonumber\\
& -\frac{1}{2\pi i}\oint_{\hat E = 0}\frac{\rd z}{z}\frac{\lambda_3}{\hat p_{12}^2 - \hat s^2}\,.
\end{align}
The second and third terms are (minus) the folded singularities of the first term, and are analytic in the $s$ branch cut region.\footnote{To be clear, this means that there are no odd powers of $s$, which contains a branch cut, or poles at $s = 0$, which are the branch points of the cut.}  
Simplifying this expression yields
\begin{equation}
	\psi_{s\,{\rm cut}} = \frac{\lambda_3^2}{E_{12}E_{34}E} -\frac{1}{2\pi i}\oint_{E = 0}\frac{\rd z}{z}\frac{\lambda_3}{\hat p_{12}^2 - \hat s^2},
\label{scut}
\end{equation}  
with a similar equation holding for the integrals along the $t$ and $u$ cuts.\footnote{An alternative procedure that is faster (but less systematic) is to simplify the integrands~\eqref{phi3recursion} before evaluating the branch cut integrals.  For instance, we may use the identity 
\be
\frac{\hat s}{(\hat p_{12}^2-\hat s^2)(\hat p_{34}^2-\hat s^2)} = -\frac{1}{\hat E_{12}\hat E_{34}\hat E} + \frac{1}{\hat{E}}\frac{\hat p_{12}\hat p_{34}+\hat s^2}{(\hat p_{12}^2-\hat s^2)(\hat p_{34}^2-\hat s^2)}\,,
\label{eq:identityschan}
\ee
and similarly for the $t, u$ channels. Since the second term in~\eqref{eq:identityschan} depends only on $\hat s^2$, it has no branch cuts in the $z$ plane, so that inside the integral along the $s$ branch cut we can replace
\be
\oint_{\text{$s$ cut }}\frac{\rd z}{z} \frac{\lambda_3^2 \,\hat s}{(\hat p_{12}^2-\hat s^2)(\hat p_{34}^2-\hat s^2)} = -\oint_{\text{$s$ cut }}\frac{\rd z}{z}\frac{\lambda_3^2}{\hat E\hat E_{12}\hat E_{34}} =  \frac{\lambda_3^2}{E_{12}E_{34}E} -\frac{1}{2\pi i}\oint_{\hat E = 0}\frac{\rd z}{z}\frac{\lambda_3}{\hat p_{12}^2 - \hat s^2},
\ee
which is precisely the same as in \eqref{scut}.}

\vskip4pt
Combining together all of the channels, one has
\be
\begin{aligned}
\psi_4 = ~&\frac{\lambda_3^2}{EE_{12}E_{34}} +\frac{\lambda_3^2}{EE_{14}E_{23}} +\frac{\lambda_3^2}{EE_{13}E_{24}}\\
&-\frac{1}{2\pi i}\oint_{\hat E = 0}\frac{\rd z}{z}\left(\frac{\lambda^2_3}{\hat p_{12}^2 - \hat s^2} +\frac{\lambda^2_3}{\hat p_{14}^2 - \hat t^2} +\frac{\lambda^2_3}{\hat p_{13}^2 - \hat u^2}\right) - \frac{1}{2\pi i}\oint_{\hat E = 0}\frac{\rd z}{z}\frac{A_4(z)}{\hat E}\,.
\end{aligned}
\ee
From~\eqref{phi3amp}, it is clear that all of the exchange pieces of the scattering amplitude cancel with the other term in the second line.  This leaves only the contact contribution, whose contour integral is straightforward to compute. Ultimately we end up with
\be
\psi_4 = \frac{\lambda_3^2}{EE_{12}E_{34}} +\frac{\lambda_3^2}{EE_{14}E_{23}} +\frac{\lambda_3^2}{EE_{13}E_{24}} + \frac{\lambda_4}{E}\,.
\ee
This formula has all of the correct singularities, and it agrees with the result of a direct bulk calculation. It is worth noting that the analogous scattering amplitude is not recursively constructible, essentially because the presence or absence of the contact $\lambda_4$ contribution cannot be determined from three-point information. In the way that we have proceeded, we have used the residue of the total energy singularity---the scattering amplitude---as an input, which fixes the coefficient of this contact term, making the wavefunction recursively constructible. 

\vskip4pt
We may apply precisely the same procedure to construct higher point wavefunction coefficients.   For instance, to recurse the 5-point wavefunction one simply needs to compute shifted four-point and 3-point coefficients and stitch them together according to the general recursion formula \eqref{eq:correlatorrecurse}.  The conceptual role of the integral along the branch cut continues to hold: an exchange channel $I$ is equal to the product of lower point shifted wavefunction coefficients, plus correction pieces which are analytic in the $s_{I}$-cut region and serve to subtract off folded singularities.  

\paragraph{Scalars with spin-1 exchange:} As a more complicated example, we consider the four-point wavefunction that arises from scalars exchanging a spin-1 field. This is the wavefunction coefficient in a theory of scalar QED. This example is interesting not only because it involves the exchange of a spinning particle, but also because the three-point coupling involved is conformally invariant so the wavefunction coefficient for conformally coupled scalars is the same in flat space and in de Sitter.

\vskip4pt
As before, one of the building blocks that we need is the four-point scattering amplitude\footnote{Like in the pure scalar example, this amplitude makes a particular choice of scalar contact interaction, whose coupling we have normalized to $1$.}
\begin{equation}
A_4 = 1 + \frac{2U}{S} \,,
\end{equation}
where, for simplicity, we will consider only the $s$-channel contribution to the wavefunction. This truncation is consistent because the corresponding wavefunction is gauge invariant.  We emphasize that any form of the amplitude related to this by 4-momentum conservation would work equally well. We also require the three-point wavefunction coefficient
\be
\psi_{J\varphi\varphi} = \frac{\vec{\xi}_3\cdot(\vec{p}_1 - \vec{p}_2)}{p_1 + p_2 + p_3}\,.
\ee
From this, we can compute the product of shifted wavefunction coefficients
\be
\tl\psi_{\varphi\varphi J}\times \tl\psi_{J\varphi\varphi} = \frac{s^2\Pi_{1,1}^{(s)}}{(p_{12}^2-s^2)(p_{34}^2-s^2)}\,,
\ee
where we have defined the sum over internal helicities
\be
s^2\Pi_{1,1}^{(s)}\equiv (\vec p_1-\vec p_2)^i \pi_{ij}^{(s)}(\vec p_3-\vec p_4)^j=u^2 - t^2 + \frac{p_{12}p_{34}(p_1 -p_2)(p_3 - p_4)}{s^2}\,,
\ee
by contracting the external momenta with the projector
\be
 \pi_{ij}^{(s)} \equiv \delta_{ij} - \frac{s_i s_j}{s^2} = \sum_{\lambda}\xi^{\lambda}_{i}\xi^{-\lambda}_{j}\,.
\ee
With these components, we can write~\eqref{eq:correlatorrecurse} as
\be
    \psi_{4} = \frac{1}{2\pi i}\oint_{\text{$s$ cut }}\frac{\rd z}{z}\,\hat s \frac{\hat s^2 \hat\Pi_{1,1}^{(s)}}{(\hat p_{12}^2-\hat s^2)(\hat p_{34}^2-\hat s^2)} - \frac{1}{2\pi i}\oint_{\hat E = 0}\frac{\rd z}{z}\frac{1}{\hat E}\left(1 + \frac{2\hat U}{\hat S} \right)\,.
\label{schannelspin1}
\ee
As in the $\phi^3$ example, we will now proceed to evaluate the integral along the branch cut in $s$.  The manipulations are exactly the same as in the scalar case, leading to the expression 
\be
\begin{aligned}
	\psi_{s\,{\rm cut}}  = &-s\frac{s^2 \Pi_{1,1}^{(s)}}{(p_{12}^2 - s^2)(p_{34}^2 - s^2)} + \frac{p_{12}(u^2 - t^2) + p_{34}(p_1 - p_2)(p_3 - p_4)}{(p_{12}^2 - s^2)E(p_{34} - p_{12})} \\
&+ \frac{k_{34}(u^2 - t^2) + p_{12}(p_1 - p_2)(p_3 - p_4)}{E(p_{12} - p_{34})(p_{34}^2 - s^2)} + \frac{1}{2\pi i}\oint_{\hat E = 0}\frac{\rd z}{z}\,\frac{1}{\hat E}\frac{(\hat p_1 - \hat p_2)(\hat p_3 - \hat p_4) + \hat t^2 - \hat u^2}{\hat p_{12}^2 - \hat s^2}\,,
\end{aligned}
\ee
where $\psi_{s\,{\rm cut}} $ stands for the first term in~\eqref{schannelspin1}.
The structure of this formula is the same as in the scalar case. The first term is a product of shifted lower-point wavefunction coefficients, and the second and third terms subtract off the folded singularities, and are analytic in the $s$-cut region.  The final term is a total energy correction which is exactly canceled by the scattering amplitude piece in \eqref{schannelspin1}.  Performing some algebra to clean up the result, we are left with
\begin{align}
	\psi_4^{(1)} =  \frac{s^2 \Pi^{(s)}_{1, 1}}{ E  E_{12} E_{34}} - \frac{ \Pi_{1, 0}^{(s)}}{E}\,,
\end{align}
where we have defined the quantity
\begin{equation}
    \Pi_{1, 0}^{(s)}\equiv\frac{(p_1 - p_2)(p_3 - p_4)}{s^2}\,.
\end{equation}
This matches a direct calculation~\cite{Baumann:2020dch}.

\subsection{Recursion from the $S$-matrix}

In this subsection, we will apply the recursion procedure to compute six-point functions for the exceptional scalar field theories.  We are using  the same information as in the brute-force  bootstrap approach, but systematized using complex analysis. We will illustrate the recursion algorithm for $P(X)/{\rm DBI}$ theories and galileon theories.  As it turns out, NLSM is an exceptional case. The data we used in addition to scattering information was the ${\rm U}(1)$ decoupling identity, which unlike soft theorems, cannot be formulated in terms of analytic structures in an obvious way. For this reason, it is easier to bootstrap the wavefunction as in Section~\ref{nlsmbootstrapsection}.

\subsubsection{$P(X)$ and DBI}
\label{DBIrecursesection}

First we demonstrate how to recurse the six-point wavefunction coefficient for a general $P(X)$ theory.  For an arbitrary $P(X)$ the required input is the scattering amplitude, the product of four-point wavefunctions corresponding to cutting the internal propagator, and the 
 $\mathcal{O}(p^0)$ soft theorem, which is an Adler zero condition.  For DBI, one may substitute the enhanced soft theorem for the scattering amplitude.  The latter procedure (though conceptually interesting) is more cumbersome and requires the full DBI soft theorem.  We have included an example computation in Section~\ref{DBIrecursionfromsoft}.

\vskip4pt
To use scattering information, we employ the following shifts, which are suitable for probing soft behavior:
\be
    \vec{p}_a(z) = (1 - c_az)\vec{p}_a, \hspace{1 cm}p_a(z) = (1 - c_a z)p_a, \hspace{1 cm}\sum_{i = 1}^{6}\vec{p}_a = 0\,.
	\label{softshifts}
\ee
Note that the shifted momenta do not obey energy conservation, and therefore probe the total energy singularity.  

\vskip4pt
Similar to~\cite{Cheung:2016drk}, it is necessary to introduce a mollifying function to improve the large $z$ behavior of the wavefunction coefficient. We will take this to be
\be
    F(z) = \prod_{a= 1}^{n}(1 - c_az)\,,
\ee
so that the quantity $(zF(z))^{-1}\psi_6(z)$ scales as $z^{-2}$ when $z\rightarrow \infty$, causing the boundary term to vanish.\footnote{In the recursion of DBI scattering amplitudes, one actually needs $F(z)\sim \prod(1 - za_i)^2$.  However, the presence of the total energy singularity in the wavefunction coefficient allows us to use one fewer power of $z$.}
In general, the cost of introducing $F(z)$ is that one must account for its singularities in the derivation of \eqref{eq:correlatorrecurse}. For a general $n>4$ point function, the modified recursion formula is 
\begin{equation}
    \psi_n(0) =  \sum_{I}\frac{1}{2\pi i}\oint_{\text{cut } I}\frac{\rd z}{zF(z)}s_I\tl{\psi}_I(z)\times\tl{\psi}_{\bar{I}}(z) - \frac{1}{2\pi i}\oint_{E(z) = 0}\frac{\rd z}{zF(z)}\frac{A_n(z)}{E(z)} - \frac{1}{2\pi i}\oint_{F(z) = 0}\frac{\rd z}{z F(z)}\psi_{n}(z)\,.
    \label{PXrecursion}
\end{equation}
The contour integral in the last term is a sum over the residues of the $F(z) = 0$ poles. However, due to the $\mathcal{O}(p)$ Adler zero, the soft limit of the wavefunction vanishes like
\be
    \lim_{z\rightarrow 1/c_a}\psi_n(z)\sim(1 - c_az)\,.
\ee
Thus the residue of the $F(z)=0$ poles vanish, so that the last term in~\eqref{PXrecursion} is zero.

\vskip4pt
Now all we need to do is compute the building blocks for the recursion formula.  A convenient form of the four-point wavefunction coefficient is~\eqref{eq:simplepx4pt}.\footnote{This form is not manifestly Bose symmetric, but it is convenient to treat one of the lines differently because it is the one that will be shifted.}
From this, we can compute the product of shifted four-point wavefunction coefficients. For instance, for the $(123)$ factorization channel, we get
\be
\begin{aligned}
	\tl{\psi}_{(123)}\times\tl{\psi}_{(456)} &=\frac{4}{f^8}  \frac{\tl{N}_{123}}{p_{123}^2 - s_{123}^2}\frac{\tl{N}_{456}}{p_{456}^2 - s_{123}^2}\,,\\
	{\rm where}~~\tl N_{123} &=  P_1\cdot P_2 ~P_2\cdot P_3 + P_2\cdot P_3~ P_3\cdot P_1 + P_3\cdot P_1~ P_1\cdot P_2\,.
\label{DBIshifted}
\end{aligned}
\ee
Quantities in the other channels are defined analogously.  We will also need an expression for the scattering amplitude
\be
    A_6 = \sum_I \frac{4}{f^8}\frac{\tl{N}_{I}\tl{N}_{\bar{I}}}{P_I^2} +  \frac{3}{f^8} P_1\cdot P_2 P_3\cdot P_4 P_5\cdot P_6 + \text{perms.},
\ee
where we have defined $P^2_{I} \equiv -p_I^2+s_I^2$,
which is the square of the sum of four-momenta associated to the exchanged particle. As an example, for the $I = (123)$ channel $P_{123}^2 = (P_1 + P_2 + P_3)^2$.  Putting these pieces together in the recursion formula gives
\be
\begin{aligned}
    \psi_6^{({\rm dbi})} = &~\frac{1}{2\pi i}\sum_{I}\oint_{\text{cut }I}\frac{\rd z}{zF(z)}\frac{4}{f^8}s_{I}(z)\frac{{\tl{N}}_{I}(z){\tl{N}}_{\bar{I}}(z)}{(\hat{p}_I^2 - \hat{s}_I^2)(\hat{p}_{\bar{I}}^2 - \hat{s}_{\bar{I}}^2)} \nonumber\\
	&- \frac{1}{2\pi i}\oint_{E(z) = 0}\frac{\rd z}{z F(z)}\frac{1}{\hat{E}}\left(\sum_{I}\frac{4}{f^8}\frac{\tl{N}_I(z)\tl{N}_{\bar{I}}(z)}{\hat{P}_I^2} +  \frac{3}{f^8} \hat{P}_1\cdot \hat{P}_2 \hat{P}_3\cdot\hat{P}_4 \hat{P}_5\cdot \hat{P}_6 + \text{perms.}\right)\,.
\end{aligned}
\ee
We can employ the analogue of the identity~\eqref{eq:identityschan}  to reduce this expression to 
\be
\begin{aligned}
    \psi_6^{({\rm dbi})} =&\frac{1}{2\pi i}\sum_{I}\oint_{\text{cut }I}\frac{\rd z}{zF(z)}\frac{4}{f^8}\frac{\tl{N}_{I}(z)\tl{N}_{\bar{I}}(z)}{\hat{E}_I\hat{E}\hat{E}_{\bar{I}}} \\ 
    &- \frac{1}{2\pi i}\oint_{E(z) = 0}\frac{\rd z}{z F(z)}\frac{1}{\hat{E}}\left(\sum_{I}\frac{4}{f^8}\frac{\tl{N}_I(z)\tl{N}_{\bar{I}}(z)}{\hat{P}_I^2} + \frac{3}{f^8}\hat{P}_1\cdot \hat{P}_2\hat{P}_3\cdot \hat{P}_4\hat{P}_5\cdot \hat{P}_6 + \text{perms.}\right),
\end{aligned}
\ee
and again add a contour at infinity in the first integral and deform the contour off the branch cuts to sum over the residues of the simple poles in the complex plane.  These are the poles at $z = 0, F(z) = 0$, and $E(z) = 0$.  The $E(z) = 0$ pole of the first integral exactly cancels the exchange part in the second integral.  Moreover, because of our choice of representation for the shifted four-point wavefunction coefficients in the first integral, the factors of $(1 - c_az)$ in the numerator and those in $F(z)$ cancel.\footnote{The same cancellation happens for the contact part of the amplitude in the second integral. Note that this is the only form of the amplitude such that this the case.  If we had chosen a different form of the amplitude related to this one by energy conservation, the poles of $F(z)$ in the second integral would \textit{not} cancel, and their residues must be computed, though the final answer would be the same.  It is also interesting to point out that the natural form of the scattering amplitude which we have chosen is also the one such that the $\mathcal{O}(p)$ Adler zero is manifestly obeyed.}

\vskip4pt
Making these cancellations leaves us with
\be
    \psi_6^{({\rm dbi})} = \frac{1}{2\pi i}\sum_{I}\oint_{z = 0}\frac{\rd z}{z}\frac{4}{f^8}\frac{\tl{N}_{I}(z)\tl{N}_{\bar{I}}(z)}{\hat{E}_I \hat{E} \hat{E}_{\bar{I}}} - \frac{1}{2\pi i}\oint_{E(z) = 0}\frac{\rd z}{z}\frac{3}{f^8 \hat{E}}\hat P_1\cdot \hat P_2 \hat P_3\cdot \hat P_4 \hat P_5\cdot \hat P_6 + \text{perms.}
\ee
Finally, we add a contour at infinity in the second integral, and deform the contour so that it only picks up the pole at $z=0$. Evaluating the residue and combining it with the residue of the first integral, the final answer is
\begin{tcolorbox}[colframe=white,arc=0pt,colback=greyish2]
\vspace{-4 pt}
\begin{align}
    \psi_6^{({\rm dbi})} = \sum_I\frac{4}{f^8}\frac{\tl{N}_I\tl{N}_{\bar{I}}}{E_I E E_{\bar{I}}} + \frac{3}{f^8E}P_1\cdot P_2 P_3\cdot P_4 P_5\cdot P_6 + \text{perms.},
\end{align}
\end{tcolorbox}
\noindent 
which matches a direct computation.  The algorithm can be readily generalized to higher points. 

\subsubsection{(Special) galileon}

Now we will demonstrate how to recursively construct the six-point wavefunction coefficient for the general galileon. The procedure is very similar to the $P(X)$ discussion in the prior subsection.  For the general galileon, one may recurse the wavefunction coefficient using information about factorization along its unitarity cut, the scattering amplitude, and the $\mathcal{O}(p)$ soft theorem.  For $n$-point wavefunctions with $n>4$, this is an Adler zero condition.  Similar to how DBI stands out as a privileged $P(X)$ theory, the special galileon is a distinguished point in the space of general galileon theories.  In particular, one may replace scattering information with the enhanced, $\mathcal{O}(p^2)$ soft theorem.

\vskip4pt
In order to recurse a general galileon using scattering information, we will use the same shifts as in~\eqref{softshifts}.  To preclude the boundary term, one must introduce the mollifying function\footnote{For special galileon scattering amplitudes, one may derive additional recursion relations by defining $F(z)$ with an additional power of $(1 - c_az)$.}
\begin{equation}
    F(z) = \prod_{a = 1}^{n}(1 - c_a z)^2\,.
\end{equation}
The cost of introducing $F(z)$ is that we have to account for its singularities in the derivation of~\eqref{eq:correlatorrecurse}.  The modified recursion formula is the same as in~\eqref{PXrecursion}, and again the final contour integral is a sum over the $F(z) = 0$ poles.  However, for $n>4$, the soft limit vanishes like
\begin{equation}
    \lim_{z\rightarrow 1/c_a}\psi_{n}(z)\sim(1 - c_a z)^2\,.
\end{equation}
Thus the residue of the $F(z)$ poles vanishes, so the term again is zero.  

\vskip4pt
Now all we need to do is compute the building blocks for the recursion formula.  The form of the four-point wavefunction which makes the computation simple is~\eqref{eq:simplegal4pt}.
Happily, when it comes to computing the shifted wavefunction coefficient the long tail of terms in this expression which are analytic in the total energy does not contribute.  For the $(123)$ channel,
\be
\begin{aligned}
	\tl{\psi}^{(123)}_4\times\tl{\psi}^{(123)}_4 &= \frac{4}{f^{12}} \frac{\tl{N}_{123}\tl{N}_{456}}{(p_{123}^2 - s_{123}^2)(p^2_{456} - s_{456}^2)}\\
	\tl{N}_{123} &= P_1\cdot P_2~P_2\cdot P_3~P_3\cdot P_1\,.
\end{aligned}
\ee
Quantities in the other channels are defined analogously.  We will also need the amplitude\footnote{It is actually possible to recursively construct this wavefunction coefficient without using scattering amplitude information.  One may define shifts that are insensitive to the total energy singularity but still cause the deformed wavefunction to vanish sufficiently fast at infinity that there is no boundary term.} 
\begin{equation}
    A_6 = \sum_{I}\frac{4}{f^{12}}\frac{\tl{N}_I\tl{N}_{\bar{I}}}{P_{I}^2}\,,
    \label{galamplitude}
\end{equation}
Putting these pieces together in the recursion formula gives 
\be
    \psi_6^{({\rm gal})} = \frac{1}{2\pi i}\sum_{I}\oint_{\text{cut }I}\frac{\rd z}{zF(z)}\frac{4}{f^{12}}\hat{s}_{I}\frac{\tl{N}_{I}(z)\tl{N}_{\bar{I}}(z)}{(\hat{p}_I^2 - \hat{s}_I^2)(\hat{p}_{\bar{I}}^2 - \hat{s}_{\bar{I}}^2)} - \frac{1}{2\pi i}\oint_{E(z) = 0}\frac{\rd z}{z F(z)}\frac{1}{\hat{E}}\sum_{I}\frac{4}{f^{12}}\frac{\tl{N}_I(z)\tilde{N}_{\bar{I}}(z)}{\hat{P}_I^2}\,.
\ee
We employ the same identity as in \eqref{eq:identityschan} to reduce this expression to 
\be
    \psi_6^{({\rm gal})} = -\frac{1}{2\pi i}\sum_{I}\oint_{\text{cut }I}\frac{\rd z}{zF(z)}\frac{4}{f^{12}}\hat{s}_{I}\frac{\tl{N}_{I}(z)\tl{N}_{\bar{I}}(z)}{\hat{E}_I\hat{E}\hat{E}_{\bar{I}}} - \frac{1}{2\pi i}\oint_{E(z) = 0}\frac{\rd z}{z F(z)}\frac{1}{\hat{E}}\sum_{I}\frac{4}{f^{12}}\frac{\tl{N}_{I}(z)\tl{N}_{\bar{I}}(z)}{\hat{P}_I^2}\,,
\ee
and again add a contour at infinity in the first integral and deform the contour so that we can evaluate it as a sum over the resides of its poles at $z = 0, F(z) = 0$, and $E(z) = 0$. The $E(z) = 0$ pole of the first integral exactly cancels the second integral.  Because we have made a convenient choice for the form of our shifted four-point wavefunctions, all of the factors of $(1 - c_a z)$ contained in $F(z)$ cancel with factors in the numerator.  Thus, all that remains is the $z = 0$ pole, with residue  
\begin{tcolorbox}[colframe=white,arc=0pt,colback=greyish2]
\begin{equation}
    \psi_6^{({\rm gal})} = \sum_{I}\frac{4}{f^{12}}\frac{\tl{N}_I\tl{N}_{\bar{I}}}{E_I E E_{\bar{I}}}\,.
\end{equation}
\end{tcolorbox}
\noindent This computation clearly demonstrates the advantage of the recursion procedure over doing a direct perturbative computation with the action~\eqref{quarticgalaction}.  From the perspective of this action, it is completely non-obvious that the numerators will organize themselves into a simple Lorentz invariant form, with no analytic pieces left over.

\subsection{Recursion from soft theorems}
In the previous sections, we implemented recursion relations for the wavefunction using the corresponding scattering amplitudes as input. The advantage of this approach is that the full information about the soft limits of the wavefunction is not needed. However, from a conceptual viewpoint, we might want to construct the wavefunction without inputting  scattering information explicitly. This is indeed possible, but we have to use the full information about the soft limit. This is technically more involved, but may be useful in some situations, particularly in the cosmological context. In this section, we demonstrate how this works for the simple examples of the NLSM and DBI.

\subsubsection{Nonlinear sigma model}
\label{nlsmrecursefromsoft}

We first consider the nonlinear sigma model. In order to input the full soft behavior of the wavefunction, we will use the following shifts:
\be
    \vec{p}_a = (1 - c_a z)\vec{p}_a, \hspace{1 cm}p_a(z) = (1 - c_az)p_a, \hspace{1 cm}\sum_{a}c_a\vec{p}_a = \sum_{a}c_ap_a = 0\,.
\label{softshiftnoenergy}
\ee
The final condition ensures that our shifts are not sensitive to the total energy pole, so that we will not need to make use of scattering information. The cost of this modification is that $\psi_n(z)$ has more divergent large-$z$ behavior than if we allowed the shifts to probe the total energy singularity. In fact, it has the same large-$z$ behavior as the corresponding scattering amplitude. To improve this behavior we need to introduce the mollifying function
\be
    F(z) = \prod_{a = 1}^n(1 - c_a z),
\ee
which is the same as is required to recurse the amplitude.  The corresponding modified recursion formula reads
\be
    \psi_n(0) =  \sum_{I}\frac{1}{2\pi i}\oint_{\text{cut } I}\frac{\rd z}{zF(z)}s_I(z)\tl{\psi}_I(z)\times\tl{\psi}_{\bar{I}}(z) - \frac{1}{2\pi i}\oint_{F(z) = 0}\frac{\rd z}{z F(z)}\psi_{n}(z)\,.
    \label{nlsmsoftrecurse}
\ee
In order to evaluate this formula, we need the following building blocks.
First, there is the four-point wavefunction coefficient.  Its most convenient form is~\eqref{eq:NLSM4pt}, so that the product of shifted wavefunction coefficients is
\be
\begin{aligned}
	\tl{\psi}_{(123)}\times&\tl{\psi}_{(456)}=\frac{1}{9f^4} \frac{\tl{N}_{123}\tl{N}_{456}}{(p_{123}^2 - s_{123}^2)(p_{456}^2 - s_{456}^2)}\,,\\
	&\tl{N}_{123} = P_1\cdot P_2 + P_2\cdot P_3 - 2P_1\cdot P_3\,.
\end{aligned}
\ee
To compute the residues at $F(z) = 0$, we will need the six-point soft theorem, given by \eqref{nslm6pointsofttheorem}.  
Then we may write the residues of at $F(z) = 0$ in \eqref{nlsmsoftrecurse} as
\be
    \oint_{F(z) = 0}\frac{\rd z}{zF(z)}\psi_n(z) = \oint_{F(z) = 0}\frac{\rd z}{zF(z)} \lim_{\vec{p}_a\rightarrow 0}\psi_{n}(z) \,,
\ee
because $F(z) = 0$ probes the soft limit of the wavefunction.
Overall, this reduces \eqref{nlsmsoftrecurse} to
\be
    \psi_6 = \frac{1}{2\pi i}\sum_{I}\frac{1}{9f^4}\oint_{\text{cut }I}\frac{\rd z}{zF(z)}s_I(z)\frac{\tl N_{I}(z)\tl N_{\bar{I}}(z)}{(\hat{p}^2_{I} - \hat{s}^2_{I})(\hat{p}^2_{\bar{I}} - \hat{s}^2_{\bar{I}})} - \oint_{F(z) = 0}\frac{\rd z}{zF(z)}\lim_{\vec{p}_a\rightarrow 0}\psi_{6}(z)\,.
\ee
To simplify the contour integral over the cut region, we use the analogue of \eqref{eq:identityschan}:
\be
    \psi_6 = -\frac{1}{2\pi i}\sum_{I}\frac{1}{9f^4}\oint_{\text{cut }I}\frac{\rd z}{zF(z)}\frac{\tl{N}_{I}(z)\tl{N}_{\bar{I}}(z)}{E\hat{E}_{I}\hat{E}_{\bar{I}}}- \oint_{F(z) = 0}\frac{\rd z}{zF(z)}\lim_{\vec{p}_a\rightarrow 0}\psi_{6}(z)\,.
\ee
To compute the first integral, we add a contour at infinity and deform the contour to pick up the residues of the simple poles of the integrand. These are at $z = 0$ and $F(z) = 0$. This gives
\be
    \psi_6 =\frac{1}{9f^4}\sum_{I}\frac{\tl N_{I}\tl N_{\bar{I}}}{EE_{I}E_{\bar{I}}} +\frac{1}{9f^4}\sum_{I}\sum_{a}\Res_{z=1/c_a}\left[\frac{1}{zF(z)}\frac{\tl{N}_{I}(z)\tl{N}_{\bar{I}}(z)}{E\hat{E}_{I}\hat{E}_{\bar{I}}}\right]- \sum_{a}\Res_{z=1/c_a}\left[\frac{1}{zF(z)}\lim_{\vec{p}_a\rightarrow 0}\psi_{n}(z)\right]\,.
    \label{nlsm6pointnoamplitude}
\ee
The first sum captures the exchange pieces, in the sense that it is exactly what one finds from computing exchange diagrams from a bulk computation.  Thus, the last two sums must combine to reconstruct the bulk contact piece. We will now verify that this is indeed correct. 
We can compute the second two terms in~\eqref{nlsm6pointnoamplitude} at for example the $z = 1/c_1$ soft pole:
\be
\begin{aligned}
    &\sum_I\Res_{z=1/c_1}\left[\frac{1}{zF(z)}\frac{\tl N_{I}(z)\tl N_{\bar I}(z)}{E\hat{E}_{I}\hat{E}_{\bar I}}\right] - \Res_{z=1/c_1}\left[\frac{1}{zF(z)}\lim_{\vec p_a\to 0}\psi_{6}(z)\right] \\
&~~~~~= \Res_{z=1/c_1}\left[\frac{1}{zF(z)}\right]\frac{1}{f^4}\left(\frac{1}{18}\frac{1}{E}\Big(\tl N_{456}(z) + \tl N_{234}(z) - 2\tl N_{345}(z)\Big) - \frac{1}{45}\Big(\hat{p}_{26} - 4\hat{p}_{35} + 6\hat{p}_4  \Big)\right)\bigg\rvert_{z = \frac{1}{c_1}}\,.
\end{aligned}
\ee
This expression can be simplified using the following identity
\be
\begin{aligned}
	&\Big(\hat{C}_1 - 4\hat{C}_2 + 6\hat{C}_3\Big)\Big|_{z = \frac{1}{c_1}} = -\left[5\left(\tl N_{456}(z) + \tl N_{234}(z) - 2\tl N_{345}(z)\right) - 2\Big(\hat{p}_{26} - 4\hat{p}_{35} + 6\hat{p}_4 \Big)\right]\bigg\rvert_{z = \frac{1}{c_1}}\,.
\label{nlsmcontactID}
\end{aligned}
\ee
where we have used the cyclic building blocks defined in \eqref{cyclic6}. A similar identity holds for the other residues.  Now summing over all of the residues, we finally arrive at
\be
\begin{aligned}
    \sum_{I}\sum_{a}\Res_{z=1/c_a}\Big[\frac{1}{zF(z)}\frac{\tl N_{I}(z)\tl N_{\bar{I}}(z)}{E\hat{E}_{I}\hat{E}_{\bar{I}}}\Big] &- \sum_{a}\Res_{z=1/c_a}\Big[\frac{1}{zF(z)}\lim_{\vec p_a\to 0}\psi_{6}(z)\Big] \\
&= -\sum_{a}\Res_{z=1/c_a}\Big[\frac{1}{zF(z)}\frac{1}{90f^4}\frac{1}{E}\Big(\hat{C}_1 - 4\hat{C}_2 + 6\hat{C}_3\Big)\Big]\,.
\end{aligned}
\ee
This set of residues can be written as a contour integral encircling the locations where $F(z) = 0$. If we add an arc at infinity, we can deform the contour to only pick up the $z=0$ pole, whose residue is easy to evaluate.
Performing this computation and substituting the  result into \eqref{nlsm6pointnoamplitude}, we obtain
\begin{tcolorbox}[colframe=white,arc=0pt,colback=greyish2]
\be
	\psi_6^{({\rm nlsm})} = \frac{1}{9f^4}\sum_{I}\frac{\tl N_I\tl N_{\bar{I}}}{E E_I E_{\bar{I}}} + \frac{1}{90f^4}\frac{1}{E}\Big(C_1 - 4C_2 + 6C_3\Big)\,,
\ee
\end{tcolorbox}
\noindent which matches the bulk perturbative computation, along with the bootstrap procedure in section~\ref{nlsmbootstrap} which utilized the scattering amplitude.
As is clear from this example, the computation is a bit cumbersome without using scattering information.  Apart from knowing the soft theorem, the limiting factor in this procedure's utility is knowing how to compute the sum of resides in~\eqref{nlsm6pointnoamplitude}.  
This approach in general requires nontrivial kinematic identities as in~\eqref{nlsmcontactID}.
Nevertheless, this construction is conceptually useful, as it shows that higher point functions can be reconstructed from soft information alone, without explicitly using scattering amplitudes. 

\subsubsection{DBI}
\label{DBIrecursionfromsoft}
As another example, we can construct DBI wavefunction coefficients using soft limits. 
The set up is largely the same as for the NLSM. We will use the same shifts as in~\eqref{softshiftnoenergy}. To improve the large $z$ behavior, we divide by
\be
	F(z) = \prod_{a = 1}^n(1 - c_a z)^2\,,
\ee
which is the same as is required to recurse the scattering amplitude.  Formally, the recursion relation is also identical to \eqref{nlsmsoftrecurse}.  Moreover, the shifted wavefunction coefficients are already computed in \eqref{DBIshifted}.  To input soft information, consider expanding $\psi(z)$ around one of the $z = 1/c_a$ poles:
\be
\psi_n(z\sim1/c_a) = \Big(\vec{p}_a(z)\cdot\vec{\partial}_{p_a} + p_a(z)\partial_{p_a}\Big) \psi_n(\vec{p}_a(z), p_a(z))\Big|_{\vec{p}_a(z) = 0} + \mathcal{O}(p_a^2(z))\,,
\ee
Then in the recursion formula, we can write the residues of $F(z) = 0$ as
\be
	\oint_{F(z) = 0}\frac{\rd z}{zF(z)}\,\psi_n(z) = \oint_{F(z) = 0}\frac{\rd z}{zF(z)}\Big(\vec{p}_a(z)\cdot\vec{\partial}_{p_a} + p_a(z)\partial_{p_a}\Big) \psi_6(\vec{p}_a(z), p_a(z))\Big|_{\vec{p}_a(z) = 0} \,,
\ee
where the right hand side is controlled by the spatial and temporal DBI soft theorems. Performing the same contour manipulations as in the NLSM case, we arrive at an analogous residue formula:
\be
\begin{aligned}
    \psi_6^{({\rm dbi})} =\frac{4}{f^8}\sum_{I}\frac{\tl N_{I}\tl N_{\bar{I}}}{EE_{I}E_{\bar{I}}} &+\frac{4}{f^8}\sum_{I}\sum_{a}\Res_{z=1/c_a}\left[\frac{1}{zF(z)}\frac{\tl N_{I}(z)\tl N_{\bar{I}}(z)}{E\hat{E}_{I}\hat{E}_{\bar{I}}}\right] \\
    &- \sum_{a}\Res_{z=1/c_a}\left[\frac{1}{zF(z)}\Big(\vec{p}_a(z)\cdot\vec{\partial}_{p_a} + p_a(z)\partial_{p_a}\Big) \psi_6(\vec{p}_a(z), p_a(z))\right]\,.
    \label{DBI6pointnoamplitude}
\end{aligned}
\ee
where we have defined the quantity
\be
	\tl{N}_{123} = P_1\cdot P_2 P_2\cdot P_3 + P_2\cdot P_3 P_3\cdot P_1 + P_3\cdot P_1 P_1\cdot P_2
\ee
As in the NLSM case, the first sum in \eqref{DBI6pointnoamplitude} correctly produces the exchange contribution, and the challenge is to show that the last two terms conspire to give the correct contact contribution. Combining everything, we obtain\footnote{Note that this step is rather nontrivial and requires the use of the kinematic identity:
\be
\begin{aligned}
	&\bigg(2(\hat{P}^{\mu}_{2} + \hat{P}^{\mu}_{3})\tl N_{456}(z) + (\hat{P}^{\mu}_{2} + \hat{P}^{\mu}_{3})\Big(\hat p_4 \hat{P}_5\cdot \hat{P}_6 + \hat p_5 \hat{P}_4\cdot \hat{P}_6 + \hat p_6 \hat{P}_4\cdot \hat{P}_5\Big)E \\
&\qquad\qquad+ \delta^{\mu}_{0}\Big(\hat{\vec{p}}_2\cdot\hat{\vec{p}}_3\hat{\vec{p}}_4\cdot\hat{\vec{p}}_5 - \hat{\vec{p}}_2\cdot\hat{\vec{p}}_3 \hat{p}_4 \hat{p}_5 + 3 \hat{p}_2 \hat{p}_3 \hat{p}_4 \hat{p}_5\Big)E+ \text{perms.} \bigg)\bigg\rvert_{z = \frac{1}{c_1}}
= -3\hat{P}^{\mu}_2 \,\hat{P}_3\cdot \hat{P}_4\, \hat{P}_5\cdot \hat{P}_6\bigg\rvert_{z = \frac{1}{c_1}} + \text{perms.}\,,
\end{aligned}
\ee
along with its analogues for the other $z = 1/c_a$ residues, and where perms. indicates we should sum over permutations of the $2,3,4,5,6$ lines.
}
\be
\begin{aligned}
\frac{4}{f^8}\sum_{I}\sum_{a}\Res_{z=\frac{1}{c_a}}\left[\frac{1}{zF(z)}\frac{\tl N_{I}(z)\tl N_{\bar{I}}(z)}{E\hat{E}_{I}\hat{E}_{\bar{I}}}\right] 
    &- \sum_{a}\Res_{z=\frac{1}{c_a}}\left[\frac{1}{zF(z)}\Big(\vec{p}_a(z)\cdot\vec{\partial}_{p_a} + p_a(z)\partial_{p_a}\Big) \psi_6(\vec{p}_a(z), p_a(z))\right]\\
&= -3\sum_a\Res_{z=\frac{1}{c_a}}\Big[\frac{1}{zF(z)}\hat{P}_1\cdot \hat{P}_2 \hat{P}_3\cdot \hat{P}_4 \hat{P}_5\cdot \hat{P}_6\Big]\,.
\end{aligned}
\ee
As in the NLSM case, we may easily compute this final set of residues by writing it as a contour integral, and deforming the contour onto the $z = 0$ pole.  Performing this procedure and combining with the exchange contribution, the final result for the six-point wavefunction coefficient is
\begin{tcolorbox}[colframe=white,arc=0pt,colback=greyish2]
\be
    \psi^{({\rm dbi})}_6 = \sum_I\frac{4}{f^8}\frac{\tl N_I \tl N_{\bar{I}}}{EE_I  E_{\bar{I}}} + \frac{3}{f^8E}P_1\cdot P_2 P_3\cdot P_4 P_5\cdot P_6 + \text{perms.}\,,
\ee
\end{tcolorbox}
\noindent which matches the bulk perturbative computation, along with the recursion procedure in \ref{DBIrecursesection} which utilized the scattering amplitude.

\newpage
\section{Conclusions}
\label{sec:conclusions}

We have studied the soft structure of the wavefunction of exceptional scalar field theories in flat space. We find that, while most of the structure survives, there are interesting and important differences with the scattering amplitude case. Foremost, wavefunction coefficients of exceptional scalar theories like the NLSM, DBI, and the special galileon satisfy nontrivial soft theorems in the limit where one of their external momenta is taken to zero, in contrast to their amplitude counterparts which display Adler zeroes. 

\vskip4pt
The soft theorems obeyed by these shift-symmetric scalar theories can be used to reconstruct wavefunction coefficients in two ways. The most straightforward approach is to input information about the total energy singularity of the wavefunction, which has the corresponding scattering amplitude as a residue. In order to completely specify the wavefunction we additionally need to use part of the soft limit of the wavefunction: theories that have a shift symmetry that scales like ${\cal O}(x^n)$ in coordinates have a soft theorem that fixes the ${\cal O}(q^{n})$ part of wavefunction, where $\vec q$ is the soft momentum. If we input scattering information, we only require the ${\cal O}(q^{n-1})$ soft theorem. Alternatively, we can use the highest-order soft theorem to reconstruct the wavefunction, which does not require the scattering amplitude as an input. We have also derived recursion relations that systematize each of these constructions. These recursion relations directly deform the momentum variables that the wavefunction depends on, and we have described how to handle the resulting subtleties involving the analytic structure.

\vskip4pt
Our investigation suggests a number of interesting directions for future study:
\begin{itemize}

\item We have focused on the properties of flat space wavefunction coefficients. From a formal standpoint, this situation is already of interest, but of course our ultimate aim is to understand the properties of the wavefunction in cosmological spacetimes. The most obvious and straightforward extension is to study the analogue of our results in a fixed de Sitter background for the exceptional scalar theories constructed in~\cite{Bonifacio:2018zex,Bonifacio:2021mrf}. The fields appearing in these theories have particular masses in de Sitter space, so that the relevant time integrals are related to those in flat space in a simple fashion. Consequently, It should be possible to uplift parts of our analysis directly to de Sitter space along the lines of~\cite{Benincasa:2019vqr,Baumann:2020dch,Hillman:2021bnk}.

\item In Section~\ref{sec:softrecurs}, we studied recursion relations for wavefunction coefficients that rely on complex deformations of spatial momenta. The analytic structure of the deformed wavefunction is slightly different in this case than when the energies are deformed~\cite{Arkani-Hamed:2017fdk,Jazayeri:2021fvk,Baumann:2021fxj}. It would be interesting to investigate these recursion relations in the de Sitter context (and to compare them with those in~\cite{Raju:2012zr}). Previous constructions have mostly avoided dealing directly with the branch cuts that appear when complexifying the $s_I$ variables, and the techniques developed here may be useful to recursively construct de Sitter wavefunctions, especially in cases where external particles have spin.

\item An important reason to study the structure of these theories is that they have interesting inter-relations and relations to Yang--Mills and gravity. At the level of scattering amplitudes, aspects of these relations have been systematized using the double copy~\cite{Cachazo:2014xea} and by operations that transform amplitudes of the various theories into each other~\cite{Cheung:2017ems}. It would be very interesting to understand both of these things from the perspective of the wavefunction. It is natural to suspect that the transmutation operations of~\cite{Cheung:2017ems} have some relation to the weight-shifting operators studied in~\cite{Karateev:2017jgd,Baumann:2019oyu}, and perhaps analogues can be found also for the flat space wavefunction. 

Some aspects of the double copy are understood for correlators and the wavefunction~\cite{Farrow:2018yni,Armstrong:2020woi,Albayrak:2020fyp,Alday:2021odx,Zhou:2021gnu,Sivaramakrishnan:2021srm,Herderschee:2022ntr,Cheung:2022pdk}, but much remains to be learned. The study of the flat space wavefunction could serve as a useful testing ground to better understand the intricacies of the double copy when energy is no longer conserved. In Appendix~\ref{app:Dcopy}, we briefly discuss the simplest double copy relation between the NLSM and special galileon for the flat space wavefunction. We expect that these scalar theories well help elucidate the
underlying structure and extend it both to de Sitter space and to other theories.

\item In cosmology, a central challenge is to understand the emergence of time from the late-time boundary where we define the wavefunction, or view observables. The static nature of the future boundary provides new conceptual challenges compared to AdS holography. One avenue toward progress could be to ask a different question to which the wavefunction is the answer. In~\cite{Arkani-Hamed:2017fdk}, such a problem was outlined, where the wavefunction of a scalar field with polynomial interactions in flat space is interpreted as a volume of a suitably defined polytope. This analysis proceeds diagram-by-diagram in perturbation theory, and it is important to understand how these contributions fit together into a more invariant structure. In this regard, the scalar theories that we have studied may be helpful.
The soft theorems require a conspiracy between various diagrammatic contributions, so it would be very interesting to find a definition of the wavefunction in these theories from a geometric perspective. 

\end{itemize}

Many rich and interesting structures have already been uncovered in the study of correlation functions in cosmological spacetimes, and there are many future discoveries to be made. In this journey, the flat space wavefunction, and in particular wavefunction coefficients of exceptional scalar theories, will serve as illuminating guideposts. We have already seen that they possess interesting structures akin to those in scattering amplitudes, and we expect that they will provide further structural insights into the nature of the wavefunction and of quantum field theory in cosmology.

\paragraph{Acknowledgements:} Thanks to Daniel Baumann, Rajendra Beekie, James Bonifacio, Carlos Duaso Pueyo, Tanguy Grall, Kurt Hinterbichler, Lam Hui, Sadra Jazayeri, Hayden Lee, Enrico Pajer, Guilherme Pimentel,  Diederik Roest, Rachel Rosen, Luca Santoni, David Stefanyszyn, and Sam Wong for helpful discussions. We are especially grateful to James Bonifacio, Tanguy Grall, and David Stefanyszyn for detailed comments on a draft. NB is supported by Simons Foundation Award Number 555117.

\appendix

\newpage
\section{Exceptional scalar theories}
\label{app:scalartheories}
In this Appendix, we give a brief overview of the different exceptional scalar field theories discussed in the main text.  These are the $\rm{SU}(N)$ nonlinear sigma model, Dirac--Born--Infeld theory, and the special galileon. 
These theories are exceptional in the broader class of scalar EFTs, because they have nonlinearly realized symmetries that control the structure of the theory and lead to enhanced Adler zeroes, meaning their amplitudes vanish faster in the soft limit than one would expect from derivative counting. These exceptional scalar theories have a large amount of structure and interesting inter-relations and relations to Yang--Mills and gravity.

\subsection{The nonlinear sigma model}
\label{app:NLSM}

The nonlinear sigma model is the low energy effective field theory corresponding to the symmetry breaking pattern $G_L\times G_R\rightarrow G_V$, where $G_{L, R}$ are two copies of some Lie group (in our case ${\rm SU}(N)$ or ${\rm U}(N)$) and $G_V$ is a diagonal subgroup.  We will follow the conventions of \cite{Kampf:2013vha}.\footnote{In particular, we take $[t^a, t^b] = i \sqrt{2} f^{abc}t^c$, where $f^{abc}$ are the totally anti-symmetric structure constants. Moreover the generators are normalized such that $\Tr[t^a t^b] = \delta^{ab}$.}  On a group element $U\in G$, the symmetry acts like
\begin{equation}
    U\longrightarrow V_R U V_{L}^{-1}, \hspace{1 cm}V_{L,R}\in G_{L, R}\,.
\end{equation}
At lowest order in derivatives, the Lagrangian invariant under this symmetry is simply 
\begin{equation}
    \mathcal{L} = -\frac{f^2}{4}\Tr[U^{-1}\partial_{\mu}U U^{-1}\partial^{\mu}U], \hspace{1 cm}U = \exp\Big(\sqrt{2}\frac{i}{f}\phi^a t^a\Big),
\end{equation}
where $t^a$ are the generators of $G$ in the fundamental representation.  For $G ={\rm SU}(N)$, we will use $a = 1,..., N^2 - 1$.  For $G ={\rm U}(N)$, we append an extra generator $t^0\propto\mathds{1}$, which spans the additional ${\rm U}(1)$ direction and commutes with all of the ${\rm SU}(N)$ generators.  In addition, note that we have used the \textit{exponential parametrization} to represent the Goldstone fields, see~\cite{Cronin:1967jq}.

\vskip 4 pt
Expanding in the Goldstone fields $\phi^a$, we may write the Lagrangian as
\be
\begin{aligned}
    \mathcal{L} = -\frac{1}{2}\partial_\mu\phi^{a}\partial^\mu\phi^{a} &+ \frac{1}{6f^2}\partial_\mu\phi^{a_1}\partial^\mu\phi^{a_2}\phi^{a_3}\phi^{a_4}f^{a_1a_3b_1}f^{a_2a_4b_1}\\
& + \frac{1}{45f^{4}}\partial_\mu\phi^{a_1}\partial^\mu\phi^{a_2}\phi^{a_3}\phi^{a_4}\phi^{a_5}\phi^{a_6}f^{a_1a_3b_1}f^{a_4b_1b_2}f^{a_5b_2b_3}f^{a_2a_6b_3} + \mathcal{O}(\phi^8)\,.
    \label{NLSMlagrangian}
\end{aligned}
\ee
The benefit of this choice of field variables is that if we embed the ${\rm SU}(N)$ model in a ${\rm U}(N)$ model, the decoupling of the ${\rm U}(1)$ degree of freedom occurs manifestly at the level of the Lagrangian.\footnote{To see this, note that any structure constant with a zero index vanishes $f^{0bc} = 0$ because the ${\rm U}(1)$ generator is proportional to the identity. Because all of the fields in the Lagrangian are contracted with structure constants, all interaction vertices involving the ${\rm U}(1)$ mode $\phi^0$ must be zero. Thus the mode decouples.}

The diagonal subgroup $G_V$ acts linearly on the fields $\phi^a$, which transform in the adjoint representation.  However, the axial subgroup of $G_L\times G_R$ acts on $U$ by
\begin{equation}
    U\longrightarrow V U V, \hspace{1 cm}V\in G,
\end{equation}
and is realized on the Goldstone modes nonlinearly as
\be
	\delta\phi^{c} = B^{c} - \frac{1}{3f^{2}}B^{a_1}\phi^{b_1}\phi^{b_2}f^{a_1 b_1 b_3}f^{c\, b_2 b_3} - \frac{1}{45f^{4}}B^{a_1}\phi^{b_1}\phi^{b_2}\phi^{b_3}\phi^{b_4}f^{a_1 b_1 \,c}f^{a_2 b_2 a_3}f^{b_3 c a_4}f^{b_4 a_3 a_4} + \mathcal{O}(\phi^6)\,.
    \label{nlsmsymmetry}
\ee
The Ward identity corresponding to this symmetry is responsible for controlling the soft behavior of the NLSM scattering amplitudes and wavefunctions. NLSM scattering amplitudes exhibit an $\mathcal{O}(p)$ Adler zero, despite having fewer than one derivative per field at the level of the action.

\vskip4pt
There is an alternative representation of the Goldstone fields which is also commonly used, known as the \textit{Cayley parametrization}. In this case the group element $U$ is given by
\be
	U = \frac{1 + \frac{i}{\sqrt{2}f}\phi}{1 - \frac{i}{\sqrt{2}f}\phi}\,,
\ee
and expanding out the Lagrangian gives
\be
	\mathcal{L} = \Tr\Big[-\frac{1}{2}\partial_\mu\phi\partial^\mu\phi + \frac{1}{2f^2}\phi^2\partial_{\mu}\phi\partial^{\mu}\phi + \mathcal{O}(\phi^6)\Big]\,.
\ee
The axial symmetry is realized very simply in this representation:
\be
	\delta\phi = B + \frac{1}{2f^2}\phi B\phi\,.
\ee
However, the disadvantage is that the ${\rm U}(1)$ mode does not decouple at the level of the Lagrangian, though it does at the level of scattering amplitudes.\footnote{This of course must be the case, because the exponential and Cayley parametrization are related by a field redefinition.} This non-decoupling does however appear in the wavefunction. For this reason, we do not use the Cayley parametrization, but it would be interesting to study the wavefunction of the U$(N)$ NLSM further.

\subsubsection{A brief tour of flavortown}
\label{app:flavor}

Computing scattering amplitudes and wavefunction coefficients for the nonlinear sigma model is made somewhat cumbersome by the presence of group theoretic flavor structures.
For the ${\rm SU}(N)$ nonlinear sigma model, one may circumvent this difficulty by working instead with so-called \textit{flavor ordered} objects.  By prescribing a particular ordering of the legs of a Feynman diagram, one may effectively remove the group-theoretic structures and reduce the number of channels necessary to describe a process.  While the full ``dressed" object will be a sum over channels composed of every combination of external momenta, the flavor ordered version only sums channels composed of consecutive momenta, drastically reducing the number of channels that need to be accounted for.  In addition, flavor ordered objects possess the same analytic structure and Ward identities (or a flavor ordered version thereof) as their dressed counterparts.  This is precisely the same as how color ordered amplitudes for Yang--Mills maintain the correct singularity structure of the full amplitude and also obey the same Ward identities enforced by gauge invariance.  We will begin by outlining the general procedure for computing a flavor ordered wavefunction from a dressed wavefunction. Then we will derive some useful facts about flavor ordered wavefunction which are used in the text. 

\vskip4pt
We begin with a flavor dressed object (e.g., a scattering amplitude or wavefunction coefficient) $\mathcal{M}^{a_1\cdots a_n}$.  In the exponential parametrization, the group theoretic factors will always be products of structure constants.\footnote{This remains true for amplitudes in any choice of field variables, but not for the wavefunction.  Regardless, the group factors for each interaction vertex may always be reduced to a single trace for any choice of field variables.}  For tree level diagrams, it turns out to always be possible to reduce the products of structure constants to a single trace over group generators $t^a$ by successively applying only the following two identities
\be
[t^a, t^b] = \frac{i}{\sqrt{2}}f^{abc}t^c, \hspace{1 cm} f^{abc} = -\frac{i}{\sqrt{2}}\Tr\big[[t^a, t^b]\,t^c\big]\,.
\label{iterate}
\ee
In this way, we may write a flavor dressed object in terms of the following expansion (for either ${\rm U}(N)$ or ${\rm SU}(N)$):
\be
	\mathcal{M}^{a_1a_2\cdots a_n}(p_1, p_2,\cdots, p_n) = \sum_{\sigma\in S_n/Z_n}\Tr[t^{\sigma(a_1)}t^{\sigma(a_2)}\cdots t^{\sigma(a_n)}]\mathcal{M}_{\sigma}(p_1,p_2,\cdots,p_n)
\label{flavorexpansion}
\ee
where $\sigma$ is a non-cyclic permutation.  Due to the Bose symmetry of the dressed object, all of the $\mathcal{M}_{\sigma}$ are related to one another via permutations of the momenta. More explicitly,
\be
	\mathcal{M}_{\sigma}(p_1,p_2,\cdots,p_n) = \mathcal{M}\big(\sigma(p_1), \sigma(p_2),\cdots,\sigma(p_n)\big)\,,
\ee
where $\mathcal{M}(p_1,p_2,\cdots,p_n)$ is defined to be the coefficient of $\Tr[t^{a_1}t^{a_2}\cdots t^{a_n}]$. Each of the quantities $\mathcal{M}\big(\sigma(p_1), \sigma(p_2),\cdots ,\sigma(p_n)\big)$ are called flavor ordered objects, and once just one is known, we can apply this formula to generate the full flavor dressed object.  Therefore, we need only keep track of and work with one of the orderings as opposed to the fully dressed object.  In effect, we may view the set of possible single trace structures as a basis of flavor structures.  For ${\rm SU}(N)$, this basis is complete in the sense that\footnote{This is actually just a rewriting of the ${\rm U}(N)$ completeness relation
\begin{align}
	\frac{1}{N}\Tr[X]\Tr[Y] + \sum_{a = 1}^{N^2 - 1}\Tr[X, t^a]\Tr[t^a, Y] = \Tr[XY],
\end{align}
which may be derived from the fact that any complex matrix may be written as a complex linear combination of hermitian matrices.
}
\begin{align}
	\sum_{a = 1}^{N^2 - 1}\Tr[X, t^a]\Tr[t^a, Y] = \Tr[XY] - \frac{1}{N}\Tr[X]\Tr[Y],
\end{align}
where $X$ and $Y$ are arbitrary matrices.  Using this relation, one may also prove that the trace structures are also approximately orthogonal:
\begin{align}
	\sum_{a_1,...a_n = 1}^{N^2 - 1}\Tr[t^{\sigma(a_1)}t^{\sigma(a_2)}\cdots t^{\sigma(a_n)}]\Tr[t^{\rho(a_1)}t^{\rho(a_2)}\cdots t^{\rho(a_n)}] = N^{n-2}(N^2 - 1)\Big(\delta_{\sigma\rho} + \mathcal{O}\Big(\frac{1}{N^2}\Big)\Big)
\end{align}
So long as the flavor ordered objects do not depend on $N$ (which they cannot, because they were generated by applying \eqref{iterate}, which does not have any $N$ dependence), this is sufficient to extract the flavor ordered object from the dressed object.  

\vskip4pt
The flavor ordered objects enjoy a number of properties (see \cite{Kampf:2013vha,Mangano:1990by,Elvang:2015rqa} for more details) and interrelations.  In the main text, we use two in an essential way:\footnote{Note that ${\rm U}(1)$ decoupling \textit{does not} hold for the wavefunction in the Cayley parametrization, for which the flavor-ordered wavefunction reads
\be
	\psi^{\text{Cayley}}_4(\vec{p}_1, \vec{p}_2, \vec{p}_3, \vec{p}_4) = \frac{1}{2f^2 E}\Big(P_1\cdot P_2 + P_1\cdot P_4 + P_2\cdot P_3 + P_3\cdot P_4\Big)\,.
\ee
This is to be expected, because the ${\rm U}(1)$ mode does not decouple at the level of the Cayley Lagrangian. 
}
\begin{itemize}
	\item {\bf Cyclicity:} The flavor ordered objects obey $\mathcal{M}\big(p_1, p_2,\cdots,p_n\big) = \mathcal{M}\big(\sigma(p_1), \sigma(p_2),\cdots,\sigma(p_n)\big)$, where $\sigma$ is a cyclic permutation.
\item ${\rm U}(1)$ {\bf Decoupling:} The flavor ordered objects obey the relations
\be
\begin{aligned}
\mathcal{M}(p_1, p_2, p_3,\cdots,p_n) + \mathcal{M}(p_2, p_1, p_3,\cdots,p_n) &+ \mathcal{M}(p_2, p_3, p_1,\cdots,p_n) \\
&+ \cdots + \mathcal{M}(p_2,p_3,\cdots,p_1,p_n) = 0\,.
\end{aligned}
\ee
\end{itemize} 
The first identity follows from cyclicity of the trace. To prove the second identity, carry out the expansion in \eqref{flavorexpansion} for  ${\rm U}(N)$. Then set $a_1 = 0$, and all of the other indices to ${\rm SU}(N)$ values.  On the left hand side, $a_1$ always appears as the index of a structure constant.  Thus the left hand side vanishes.  On the right hand side, this sets $t^{a_1}\propto \mathds{1}$, which simplifies the traces, allowing flavor ordered objects to be grouped together. This is best demonstrated through a simple example.  At three points, we have
\be
	\mathcal{M}^{a_1 a_2 a_3} = \Tr[t^{a_1}t^{a_2}t^{a_3}]\mathcal{M}\big(p_1, p_2, p_3\big) + \Tr[t^{a_1}t^{a_3}t^{a_2}]\mathcal{M}\big(p_1, p_3, p_2\big)\,.
\ee
After setting $a_1 = 0$, this may be simplified to
\be
	\Tr[t^{a_2}t^{a_3}]\Big(\mathcal{M}(p_1, p_2, p_3) + \mathcal{M}(p_1, p_3, p_2)\Big)=0\,,
\ee
which forces the term in parentheses to vanish. After applying cyclicity, this may be brought to the form of the ${\rm U}(1)$ decoupling relation.

\vskip4pt
For convenience, below we have included single trace representations for the products of 2 and 4 structure constants:
\begin{align}
    f^{a_1 a_2 e}f^{a_3 a_4 e} &= \frac{1}{2}\Big(-\Tr[1234] + \Tr[1243] + \Tr[2134] - \Tr[2143]\Big)\,,\\
    f^{a_1 a_2 e}f^{a_3 e g}f^{a_4 g h}f^{a_5 a_6 h} &= \frac{1}{4}\bigg( \Tr[123456] - \Tr[123465] - \Tr[123564] + \Tr[123654]\nonumber \\
    &\hspace{1cm}-\Tr[124563] + \Tr[124653] + \Tr[125643] - \Tr[126543]\nonumber \\
    &\hspace{1cm}-\Tr[213456] + \Tr[213465] + \Tr[213564] - \Tr[213654]\nonumber \\
    &\hspace{1cm}+\Tr[214563] - \Tr[214653] - \Tr[215643] + \Tr[216543]~\bigg)\,.
\end{align}

\subsection{Dirac--Born--Infeld}
Dirac--Born--Infeld  is the theory of a $D$-dimensional brane probing a flat $(D+1)$-dimensional bulk. At leading order in derivatives, the action is given by
\be
	S = -f^4\int \rd^Dx\,\left(\sqrt{1 + \frac{1}{f^4}(\partial\phi)^2} -1 \right)
\ee
This action nonlinearly realizes the ${\rm ISO}(D+1,1)$ symmetry of the bulk. Translations in the transverse bulk direction correspond to  the simple shift symmetry $\phi\mapsto \phi + c$, whereas the extra boost symmetry generates the transformation
\be
	\delta\phi = v_{\mu}x^{\mu} + \frac{1}{f^4}v^{\mu}\phi\partial_{\mu}\phi\,.
\ee
The Ward identity for the enhanced shift symmetry implies an $\mathcal{O}(p^2)$ Adler zero for scattering amplitudes.  For our application, we will need the boundary term generated by the enhanced shift symmetry in order to compute the corresponding temporal charge. From the probe brane formalism~\cite{deRham:2010eu}, this can be easily seen to be
\begin{equation}
	\delta\mathcal{L} = v^{\mu}\partial_{\mu}\big(\phi\mathcal{L}\big)\,.
\end{equation}
For the spatial symmetry, which has $v^{t} = 0$, the boundary term is purely spatial and may be dropped. On the other hand, for the temporal symmetry the boundary term has a time component and thus contributes to the charge.  This is described in more detail in Appendix~\ref{app:generators}.

\subsection{(Special) galileon}
\label{galreview}
Galileon theories are those which are invariant under the enhanced shift symmetry~\cite{Nicolis:2008in}
\begin{equation}
    \delta\varphi = c + b_{\mu}x^{\mu},
\end{equation}
and which have second-order equations of motion.
The Ward identity resulting from this symmetry controls the soft behavior of the galileon scattering amplitudes, which exhibit a $\mathcal{O}(p^2)$ Adler zero.  In $D$ spacetime dimensions, a generic galileon action may be written as
\begin{align}
	\mathcal{L} = \sum_{n = 1}^{D+1}c_n\mathcal{L}_n, \hspace{1 cm} \mathcal{L}_n = (\partial\phi)^2\mathcal{L}_{n-2}^{\text{TD}},
\end{align}
where $c_n$ are arbitrary coefficients and $\mathcal{L}^{\text{TD}}_n$ is the unique total derivative that may be formed with $n$ copies of $\phi$ and two derivatives per field.  Abstractly, these can be written
\begin{equation}
    \mathcal{L}^{\text{TD}}_{n} = \sum_{\sigma}(-1)^{\sigma}\eta^{\mu_1\sigma(\nu_1)}\cdots\eta^{\mu_1\sigma(\nu_n)}\partial_{\mu_1}\partial_{\nu_1}\phi\cdots\partial_{\mu_n}\partial_{\nu_n}\phi\,,
\end{equation}
where the sum is over all permutations $\sigma$.  The first few of these take the explicit form
\begin{align}
    &\mathcal{L}_{1}^{\text{TD}} = \square\phi\\
    &\mathcal{L}_{2}^{\text{TD}} = (\square\phi)^2 - \partial_{\mu}\partial_{\nu}\phi\partial^{\mu}\partial^{\nu}\phi\\
    &\mathcal{L}_{3}^{\text{TD}} = (\square\phi)^3 - 3\square\phi\partial_{\mu}\partial_{\nu}\varphi\partial^{\mu}\partial^{\nu}\phi + 2\partial_{\mu}\partial_{\nu}\phi\partial^{\nu}\partial^{\alpha}\phi\partial_{\alpha}\partial^{\mu}\phi\,.
\end{align}
The standard forms of the galileon action where Lorentz invariance is manifest all have more than one time derivative per field.  For Dirichlet boundary conditions, the action as written therefore does not have a well-posed variational principle, and evaluating the on-shell action computes wavefunction coefficients with operator insertions present (see Appendix \ref{app:boundaryterms}).  To cure this, we must add a boundary term to the action, which breaks manifest  Lorentz invariance.  We can do this explicitly for the quartic galileon, which is relevant to the discussion in the main text.  The bulk and boundary contributions to the action are
\begin{align}
    S^{(4)}_{\text{bulk}} &= \int \rd^{4}x\,(\partial\phi)^2\Big[(\Box\phi)^2 - (\partial_{\mu}\partial_{\nu}\phi)^2\Big]\,,\\
    S^{(4)}_{\text{boundary}} &= -2\int \rd^3x\,\nabla^2\phi\Big[\frac{1}{3}\dot\phi^3 - (\nabla\phi)^2\dot\phi\Big]\,.
    \label{quarticbulkboundary}
\end{align}
The boundary contribution may be absorbed into the bulk contribution by writing it as a total time derivative. After doing this and performing spatial integrations by parts, we arrive at the following simple form:
\be
	S^{(4)} = -\frac{1}{4f^6}\int \rd^{4}x\left(\dot\phi^2\left[ (\nabla^2\phi)^2 - (\nabla_i\nabla_j\varphi)^2\right] - \frac{1}{3}\partial_i\phi\partial^i\phi\left[ (\nabla^2\phi)^2 - (\nabla_i\nabla_j\phi)^2\right]\right)\,.
\label{quarticgalaction}
\ee
This way of writing the action behaves simply under the galileon shift symmetry.  First, the action $S^{(4)}$ does not pick up a temporal boundary term under the spatial galileon symmetry.  This is not the case for the bulk action in~\eqref{quarticbulkboundary}, which does develop a temporal boundary term.  In addition, under the temporal shift symmetry $\delta\phi = b_{0}t$, the action $S^{(4)}$ does pick up a temporal boundary term, which is given by
\be
\delta S = -2b_{0}\int \rd^{3}x\,\phi\Big[(\nabla\phi)^2 - (\partial_{i}\partial_{j}\phi)^2\Big].
\ee
More generally, after fixing the boundary term the action for the $n$th galileon term obeys
\be
\begin{aligned}
\delta\phi &= \vec{b}\cdot\vec{x}, &  \hspace{1.5cm}\delta S^{(n)} &= 0\,,\\
\delta\phi &= b_{0}t,  &  \delta S^{(n)} &= -2\int \rd^{3}x\,\phi\mathcal{L}^{\text{TD}}_{n-2}[\partial_{i}\phi]\,,
\end{aligned}
\ee
where we have discarded spatial boundary terms and $\mathcal{L}_{n}^{\text{TD}}[\partial_i\phi]$ is a galileon term comprised of spatial derivatives only. That is, the variation of galileon terms does not generate a temporal boundary term under the spatial symmetry $\delta\phi = \vec{b}\cdot\vec{x}$, but does generate a temporal boundary term for the  $\delta\phi = b_{0}t$ symmetry.

\vskip4pt
In the space of galileon theories, there is a subset whose actions are invariant under an enhanced field dependent shift symmetry.  The simplest of these is the quartic galileon~\cite{Hinterbichler:2015pqa}
\begin{equation}
    S_{\text{sgal}} = \int \rd^{4}x\,-\frac{1}{2}(\partial\phi)^2 + \frac{1}{12 f^{6}}\Big[(\Box\phi)^2 - (\partial_{\mu}\partial_{\nu})\phi)^2\Big]\,,
\end{equation}
which is often called the {\it special galileon}.
It is invariant under the symmetry
\begin{equation}
    \delta\phi = s_{\mu\nu}x^{\mu}x^{\nu} + \frac{1}{f^{6}}s^{\mu\nu}\partial_{\mu}\phi\partial_{\nu}\phi\,,
\end{equation}
for $s_{\mu\nu}$ a traceless symmetric tensor.  As a result of this symmetry, the amplitudes of the special galileon have a $\mathcal{O}(p^3)$ Adler zero.  Being a subset of the galileon, we have already described which boundary terms one must add to the action in order for the variational principle to be well-posed.  For tensors $s^{\mu\nu}$ that have no temporal component, i.e., $s^{0\mu} = 0$, the action (with the appropriate boundary terms) does not develop a temporal boundary term under the symmetry, in accordance with the logic laid out in  \ref{app:generators}.  On the other hand if these components are not zero, the action does pick up a temporal boundary term. This boundary term is cumbersome to specify and not essential for our main argument, so we have omitted it.\footnote{There is one way of writing of the special galileon action where the boundary term is simple to deduce:
\begin{align}
  S_{\text{sgal}}  = -\frac{f^6}{8}\int \rd^4x\left(x^2 + \frac{1}{f^6}\partial\phi\cdot\partial\phi\right)\left(1 - \frac{1}{2f^6}\mathcal{L}_2^{\text{TD}} + \frac{1}{24 f^{12}}\mathcal{L}_4^{\text{TD}}\right)\,.
\end{align}
Under special galileon symmetry, the Lagrangian changes by the total derivative
\begin{equation}
	\delta\mathcal{L} = \frac{2}{f^6}\partial_{\alpha}\Big(s^{\alpha\beta}\partial_{\beta}\phi\mathcal{L}\Big)\,,
\end{equation}
which is made explicit by the probe brane formalism of~\cite{Novotny:2016jkh}.  However, this way of writing the action does not have a well-posed variational principle, and it turns out that the resulting charge will have time derivatives of canonical momentum, so that it cannot be quantized in a straightforward fashion.  It would be interesting to determine whether or not there is a way to apply the probe brane formalism to determine the temporal symmetry boundary term from the fixed action~\eqref{quarticgalaction} in a simple manner.}

\newpage
\section{A boundary view on the wavefunction}
\label{app:technical}

In the main text, we take a boundary point of view on the computation of the wavefunction, in the sense that we try to define and compute this object directly on some particular time slice. However, we know from the bulk perspective that there are various ambiguities and subtleties in the definition of the wavefunction associated to, for example, boundary terms. These subtleties should have an avatar from the boundary perspective. In this Appendix we explore these technical subtleties and their resolution.

\subsection{Boundary terms and the variational principle}
\label{app:boundaryterms}

An important subtlety that we have to address is the presence of boundary terms in the action.  When it comes to computing scattering amplitudes, such boundary terms are typically harmless. However, they may affect the form of wavefunction coefficients, and thus it is worth asking whether there is a natural choice at the level of the action.  There are two types of boundary terms which we will consider. The first type contain derivatives which are normal to the future time slice on which we are evaluating the wavefunction. That is, they contain time derivatives.  We will refer to such boundary terms as {\it nonlocal}. The second type of boundary term does not contain normal derivatives, and we will refer to such terms as {\it contact terms}.\footnote{This nomenclature is motivated by the way that these boundary terms contribute to the wavefunction.}  We will argue that there is a natural way to fix nonlocal boundary terms, and that contact terms may be precluded by derivative counting.

\vskip4pt
Consider first nonlocal boundary terms, which contain a time derivative. Boundary terms of this type may be generated through integration by parts in time, or they may present on their own. The natural choice of nonlocal boundary terms are those which make the variational principle of the action well-posed.  Furthermore, this is actually the only choice of boundary terms for which the on-shell action is computing a vacuum  wavefunctional (a transition amplitude from the vacuum to a Heisenberg picture eigenstate) as opposed to a transition amplitude with an operator insertion (or equivalently a transition amplitude between an excited state and a Heisenberg picture eigenstate).

\vskip4pt
As a simple instructional example, consider a model which describes a free particle.  Traditionally, we take the action to be
\begin{equation}
    S_1 = \frac{1}{2}\int_{t_1}^{t_2}\rd t\,\dot q^2,
\end{equation}
which has one time derivative per field.  This is also the action that naturally appears when computing a transition amplitude for a model described by the Hamiltonian $H = p^2/2$:
\be
 \braket{x_2, t_2|x_1, t_1} =\int\limits_{\substack{\hspace{-0.3cm}q(t_2) \,=\,x_2\\ \hspace{-0.3cm} q(t_1)\,=\,x_1}} 
\hspace{-0.5cm} \raisebox{-.05cm}{ ${\cal D} q\, e^{iS_1[q]}$ }\,.
\label{correcttransition}
\ee
If we imagine that terms that differ by integrations by parts are indistinguishable, there is no reason to prefer this writing of the action  over
\begin{equation}
    S_2 = -\frac{1}{2}\int_{t_1}^{t_2}\rd t\,q\ddot q = \frac{1}{2}\int_{t_1}^{t_2}\rd t\,\dot q^2 - \frac{1}{2}q\dot q\Big\rvert^{t_2}_{t_1} \,.
\end{equation}
However, actions with more than one time derivative per field typically do not have a well-posed variational principle.  Following the logic of \cite{Dyer:2008hb}, we can check this with our simple point-particle example.  The variation of $S_2$ is given by
\begin{align}
    \delta S_2 = -\frac{1}{2}\int_{t_1}^{t_2}\rd t\,\delta q \ddot q - \frac{1}{2}\int_{t_1}^{t_2}\rd t\, q \delta \ddot q = -\int_{t_1}^{t_2}\rd t\,\delta q \ddot q  - q\delta\dot q\Big|_{t_1}^{t_2} + \dot q \delta q\Big|_{t_2}^{t_1}\,.
    \label{pointboundaryexample}
\end{align}
The variation $\delta q$ is defined to vanish on the boundary, but there is no such constraint on $\delta \dot q(t_{1,2})$, which may take on any value.  Thus in order for the variational problem to be well-posed, one must add a boundary term to the action whose variation cancels the $\delta\dot q(t_{1,2})$ term in \eqref{pointboundaryexample}.  The proper boundary term is 
\begin{equation}
    S_{\text{boundary}} = \frac{1}{2}\Big(q(t_2)\dot q(t_2) - q(t_1)\dot q(t_1)\Big),
\end{equation}
and altogether the action with a well-posed variational principle is
\begin{equation}
    S_2 + S_{\text{boundary}} \equiv S_1,
\end{equation}
leading us back to the writing of the action with one time derivative per field.

\vskip4pt
We can also study what happens when we compute transition amplitudes.  Traditionally, we expect transition amplitudes for a model with Hamiltonian $H = p^2/2$ to be computed by \eqref{correcttransition}.  On the other hand, if we take $S_2$ and plug it into the path integral, then we have
\be
\int\limits_{\substack{\hspace{-0.3cm}q(t_2) \,=\,x_2\\ \hspace{-0.3cm} q(t_1)\,=\,x_1}} 
\hspace{-0.5cm} \raisebox{-.05cm}{ ${\cal D} q\, e^{iS_2[q]}$ }=\int\limits_{\substack{\hspace{-0.3cm}q(t_2) \,=\,x_2\\ \hspace{-0.3cm} q(t_1)\,=\,x_1}} 
\hspace{-0.5cm} \raisebox{-.05cm}{ ${\cal D} q\, e^{iS_1[q]}e^{-\frac{i}{2}(q(t_2)\dot q(t_2) - q(t_1)\dot q(t_1))}$ } = \bra{x_2, t_2}e^{\frac{i}{2}(x_2p_2 - x_1 p_1)}\ket{x_1, t_1}\,.
\ee
This is still a transition amplitude in the theory with the Hamiltonian $H = p^2/2$, but now with an extra operator insertion present.  Alternatively, we can view the boundary term as modifying the states that we are computing a transition amplitude between.  When it comes to computing wavefunction coefficients, we do not want there to be any spurious operator insertions as we time evolve the ground state.  Therefore, we \textit{only} want to study actions which have at most one time derivative per field.  In the theories that we are interested in studying, this is enough to completely resolve the ambiguities of nonlocal  boundary terms.

\vskip4pt
In practice, we can essentially resolve the problem of nonlocal boundary terms by taking an action and performing integrations by parts in time until there is at most one time derivative per field. Then the variational principle is manifestly well-posed.  For the theories that we study in the main text, it is usually the case that the natural and most familiar form of the action is already written so that there is no more than one time derivative per field.  However, this is not the case for the (special) galileon.  Most of the standard forms of the galileon are manifestly Lorentz invariant, and as it turns out there is no writing of the action which is both manifestly Lorentz invariant and has at most one time derivative per field.

\vskip4pt
Now consider the case of contact boundary terms, which do not contain derivatives normal to the boundary. For instance, for DBI we might add a term like
\be
	S\supset \int \rd^3p_1 \rd^3p_2\rd^3p_3\rd^3p_4\,\delta^{(3)}(\vec{p}_1 + \vec{p}_2 + \vec{p}_3 + \vec{p}_4)\,\vec{p}_1\cdot\vec{p}_2\vec{p}_3\cdot\vec{p}_4\,\vp_{\vec{p}_1}(t_f)\vp_{\vec{p}_2}(t_f)\vp_{\vec{p}_3}(t_f)\vp_{\vec{p}_4}(t_f)\,.
\ee
Such a term is called a contact term because its contribution to the wavefunction is proportional to a $\delta$-function in position space.  For the theories we are studying, such terms are forbidden by power counting.  For instance, the contribution to the DBI four-point function from the bulk action contains 4 derivatives, and so boundary terms should only contain 3 derivatives at this order in fields.  However, the only way to form a rotationally invariant boundary term with 3 derivatives is to use the magnitude of one of the momenta.  This in effect would require the boundary term to have a normal derivative, and thus it is nonlocal, and covered by the discussion above.  In this way, we may fix the contact terms in the action for all of the theories that we study in the main text.     

\vskip4pt
At the level of the action, we have addressed how to fix the boundary terms in order to compute proper vacuum wavefunction coefficients.  However in the bootstrap philosophy the opposite question is more natural: If someone hands us a putative wavefunction coefficient that has all of the expected singularities (or, if we bootstrap it ourselves), how do we know it came from an action with a well-posed variational principle, or equivalently, that it was computed with no spurious operator insertions in the path integral?
We may test for this by deriving the Ward identities obeyed by vacuum wavefunction coefficients, and check that whatever we have bootstrapped obeys these Ward identities.  If it does not, then there must be an operator insertion.  However the converse is not true; just because the object we bootstrapped \textit{does} obey the Ward identity does not mean there is no operator insertion.  The proper strategy is then to impose so many Ward identities that there is only one object satisfying them all. Then we may be sure that the object we have bootstrapped is a wavefunction coefficient.

\vskip4pt
As a simple example, consider the following two objects, computed with a DBI Hamiltonian at four points:
\begin{align}
	\psi^{(a)}_4 &= \frac{1}{f^4E}\Big(P_1\cdot P_2 P_3\cdot P_4 + P_1\cdot P_3 P_2\cdot P_4 + P_1\cdot P_4 P_2\cdot P_3\Big)\,,\\
	\psi^{(b)}_4 &= -\frac{1}{2f^4E}\Big(P_1\cdot P_2 P_2\cdot P_3 + \text{perms.}\Big)\,.
\end{align}
Both of these have the correct singularity structure and appear to be perfectly reasonable wavefunction coefficients computed by evaluating the on-shell DBI action.  However, one of these was computed from an action with a well-posed variational principle, while the other is not. To detect which is the true wavefunction coefficient, we test each for the $\mathcal{O}(p)$ Adler zero.  The first object has the Adler zero, and the second does not, telling us that $\psi^{(a)}_4$ is the true vacuum wavefunction coefficient, whereas $\psi^{(b)}_4$ was computed in the presence of operator insertions.  Indeed, each object comes from the respective action
\begin{align}
	S_{(a)}&\supset\int \rd^4x \,\partial\phi^4,\\
	S_{(b)}&\supset \int \rd^4x\,\phi\partial_{\mu}\left(\partial^{\mu}\phi\partial\phi^2\right)\,,
\end{align}
which only differ by an integration by parts.  However, only the first has a well-posed variational principle.  In this way, we can use Ward identities to detect well-posedness of the variational principle.

\subsection{Symmetry generators}
\label{app:generators}

In the main text, we are interested in deriving Ward identities for charges that generate nonlinearly realized symmetries.  If we know the action, it is a simple matter to compute the associated current and derive the charge which generates a particular symmetry.  However, from the on-shell point of view, in many cases we want to be able to derive  Ward identities without starting from the action. This means that we need a way to write down the symmetry charges without utilizing the action.  As it turns out, we may often intuit the structure of the charges and their realization on fields just by knowing the symmetry transformation $\delta\phi$ alone.

\vskip4pt
Our starting point will be the action
\begin{equation}
    S[\phi] = \int \rd^4x\,\mathcal{L}[\phi, \dot\phi]\,.
\end{equation}
Note that we are specifically writing the action so that there is at most one time derivative per field. There may be arbitrarily many spatial derivatives acting on a single field, and we have suppressed their dependence in the Lagrangian.  If $\delta\phi$ is a symmetry of the action, then the Lagrangian changes by at most a total derivative:
\begin{equation}
    \delta S[\phi] = \int \rd^4x\,\partial_{\mu}K^{\mu},
\end{equation}
and the conserved current and its corresponding charge are given by
\begin{align}
    J^{\mu} &= \delta\phi\frac{\partial\mathcal{L}}{\partial\partial_{\mu}\phi} - K^{\mu},\\
    Q &= \int \rd^3x\,\delta\phi\frac{\partial\mathcal{L}}{\partial\dot\phi} - K^{0} = \int \rd^3x\,\delta\phi\Pi^{(\phi)} - K^{0}\,.
    \label{noethercharge}
\end{align}
We have used that by definition, $\Pi^{(\phi)} = \partial\mathcal{L}/\partial\dot\phi$.  Note that the $K^{0}$ piece comes from the temporal boundary term generated by the symmetry.  The symmetry may also generate spatial boundary terms, but they do not contribute to the charge.

\vskip4pt
Our strategy will be to use commutation relations to determine the form of $K^0$.  The symmetry transformation $\delta\phi$ may contain powers of $\dot\phi$ (or equivalently $\Pi_{\phi}$) and have dependence on the spacetime coordinate $x$.  First, suppose that the symmetry $\delta\phi$ does not involve any time derivatives or the time coordinate itself.  $Q$ must act on $\delta\phi$ via $[Q, \phi] = i\delta \phi$.  From~\eqref{noethercharge}, this requires that 
\begin{equation}
    \int \rd^3x\,[K^{0}(\vec x, t), \phi(\vec{y}, t)] = -i\int \rd^{3}x\,\frac{\delta K^{0}(\vec x, t)}{\delta\Pi^{(\phi)}(\vec{y}, t)} = 0 \implies K^{0} = K^{0}[\phi]\,.
\end{equation}
Therefore $K^0$ does not have any powers of canonical momenta.  Next, we will make the following assertion, without proof: If $Q$ commutes with the Hamiltonian for a Lorentz invariant system, then the symmetry $\delta\phi$ \textit{does not} generate a \textit{temporal} boundary term in the action, and in turn $K^0 = 0$.  Said another way, if $Q$ commutes with the Hamiltonian for a Lorentz invariant system, the Lagrangian (not the Lagrangian density) is invariant.\footnote{We do not have a general proof of this assertion. However, we will take it as an input; we are only interested in studying theories where the charges are implemented on fields such that this condition is true.  It is worth noting that this is usually the case, so long as one properly fixes the boundary term so that there is no more than one time derivative per field as prescribed in Appendix \ref{app:boundaryterms}. One can indeed check this for spacetime symmetries, as well as more exotic symmetries such as the shift symmetry, the spatial DBI symmetry, and the spatial (special) galileon symmetry, where ``spatial" indicates that the symmetry generated commutes with the Hamiltonian.  In fact, it would be interesting if there were a theory with a symmetry charge such that this is \textit{not} the case, but we are not aware of any examples.} 

\vskip4pt
Now consider the case when the symmetry $\delta\phi$ does contain the time coordinate, but not time derivatives.  Such charges typically do not commute with the Hamiltonian.  However, requiring again that $[Q, \phi] = i\delta\phi$, we can again conclude that $K^{0} = K^{0}[\phi]$.  However, if $Q$ and the Hamiltonian do not commute, we will \textit{not} assume that $K^{0} = 0$.  Despite this, it is sometimes the case that $K^0$ may be determined solely by the commutation relations of $Q$ with the other charges in the algebra, and does not need to be computed from the action at all (see \eqref{eq:temporalchargegal} for an example involving the general galileon).

\vskip4pt
Finally, consider the case when $\delta\phi$ contains time derivatives.  In such cases, it is difficult to write down the charges without starting from the action itself.  Because $\delta\phi$ contains $\Pi^{(\phi)}$, even implementing $[Q, \phi] = i\delta\phi$ becomes nontrivial. From~\eqref{noethercharge} we have
\begin{equation}
    [Q, \phi] = \int \rd^3x\,[\delta\phi(\vec{x}, t), \phi(\vec{y}, t)] + i\delta^{(3)}(\vec{x} - \vec{y})\delta\phi(\vec{x}, t) - [K^{0}(x), \phi(\vec{y}, t)]\,.
\end{equation}
In order for $Q$ to generate the symmetry properly, the first and third terms must cancel, which means that $K^{0}$ must contain factors of $\Pi^{(\phi)}$, and there is not a straightforward way to deduce the form of $K^{0}$ in such cases.  This is equivalent to the fact that we cannot anticipate the boundary term which the action develops under a temporal symmetry without simply computing it.  This is the fact that limits the utility of using the temporal DBI soft theorem and the temporal special galileon soft theorem in a bootstrap procedure.

\newpage
\section{Classical field profile and generating functionals}
\label{app:fieldasgenfunct}

In this Appendix, we highlight some features of the classical field profile which are used in the text to prove tree level soft theorems.  These observations essentially relate the classical field profile to combinations of wavefunction coefficients.  First, we will derive relations for the late time classical canonical momentum, which are similar to those pointed out in~\cite{Klebanov:1999tb} in the AdS context.  These are the analogue of the statement that the classical field profile with a Feynman pole prescription in the presence of a source is a generating functional for tree level in/out correlators~\cite{Boulware:1968zz}.  In the wavefunction context, we can actually derive additional relations, using the Bunch--Davies condition to relate early time classical field profiles to late time canonical momentum profiles in a simple and universal way.

\vskip4pt
To begin, we demonstrate the analogous statement for flat space in/out correlators: the classical field profile with a Feynman pole prescription is a generating functional for tree level Feynman diagrams.  We start by defining $W[J]$, the generating functional for connected in/out correlators:
\begin{equation}
    e^{iW[J]} \equiv \int \mathcal{D}\phi\,e^{iS[\phi] + i\int \rd^4x\,\phi J}\,.
\end{equation}
Taking a functional derivative of this form of the generating functional, we have
\begin{equation}
    \frac{\delta W[J]}{\delta J} = e^{-iW[J]}\int \mathcal{D}\phi\,e^{iS[\phi] + i\int \rd^4x\,\phi J}\phi\equiv\langle\phi\rangle_{J}\,.
\end{equation}
Thus, $\langle\phi\rangle_{J}$ is a generating functional for in/out correlators.  If we take the tree level approximation, then the result follows:
\begin{equation}
    \phi_{\text{cl}}[J] = \frac{\delta W[J]}{\delta J}\,,
\end{equation}
where $ \phi_{\text{cl}}[J] $ is the classical field profile in the presence of the source, $J$.

\vskip4pt
We want to carry out at a similar computation for the classical field profile with Dirichlet boundary conditions. We begin with the generating functional for connected Witten diagrams, which we will also call $W[\varphi]$. Of course, this is the logarithm of the wavefunction itself:
\begin{equation}
    e^{W[\varphi]}\equiv\Psi[\varphi, t_f] = \int^{\phi(t_f) = \varphi}\mathcal{D}\phi(t)\,e^{iS[\phi]}\,.
\end{equation}
Similar to the in/out computation, we can take a functional derivative of the generating functional:
\begin{equation}
    -i\frac{\delta W[\varphi]}{\delta\varphi(\vec{x})} = \frac{1}{\braket{\varphi, t_f| \Psi}}\bra{\varphi, t_f}\Pi^{(\phi)}(\vec{x},t_f)\ket{\Psi}\,.
\end{equation}
Therefore, the connected part of the canonical momentum, computed with Dirichlet boundary conditions, is a generating functional for connected Witten diagrams.  At tree level, we have
\begin{tcolorbox}[colframe=white,arc=0pt,colback=greyish2]
\be
\begin{aligned}
    &\Pi^{(\phi)}_{\text{cl}}(\vec{p}, t_f) = -i(2\pi)^3\frac{\delta W[\varphi]}{\delta\varphi_{-\vec{p}}} \\
&~~~= \sum_{n = 2}\frac{-i}{(n-1)!}\int \frac{\rd^3p_1\cdots \rd^{3}p_{n-1}}{(2\pi)^{3(n-2)}}\varphi_{\vec{p}_1}\cdots\varphi_{\vec{p}_{n-1}}\delta^{(3)}(\vec{p}_1 + \cdots + \vec{p}_{n-1} - \vec{p})\psi_{n}(\vec{p}_1,\cdots,\vec{p}_{n-1}, -\vec{p})\,.
\label{canonicalgenerator}
\end{aligned}
\ee
\end{tcolorbox}

\vskip4pt
To relate this expression to the field profile itself, first  write the canonical momentum as
\begin{equation}
    \Pi^{(\phi)}(\vec{p}, t) \equiv \frac{\partial \mathcal{L}}{\partial\dot\phi} = \dot\phi + (2\pi)^3\frac{\partial\mathcal{L}_{\text{int}}}{\partial\dot\phi}\,.
    \label{eq:pidefapp}
\end{equation}
We will also need the following recursively defined expression for the classical field profile:
\begin{equation}
    \phi_{\vec{p}}(t) = \mathcal{K}(p, t)\varphi_{\vec{p}} +i (2\pi)^3\int_{-\infty}^{0} \rd t'\,\mathcal{G}(t, t',\vec{p})\frac{\delta S_{\text{int}}}{\delta\phi_{-\vec{p}}(t')}\,.
    \label{eq:classicalfieldprofileapp}
\end{equation}
Using the following property of the bulk-to-bulk propagator
\begin{equation}
    \partial_t\mathcal{G}(p;t, t')\Big|_{t = 0} = -i\mathcal{K}(p,t')\,,
\end{equation}
leads to an expression for the time derivative of the classical field on the future boundary:
\begin{equation}
    \dot\phi_{\vec{p}}(t_f) = ip\,\vp_{\vec{p}} + (2\pi)^3\int_{-\infty}^{0}\rd t'\,\mathcal{K}(p, t')\frac{\delta S_{\text{int}}}{\delta\phi_{-\vec{p}}(t')}\,.
    \label{dotvarphi}
\end{equation}
If the theory does not have interactions involving time derivatives, then $\Pi^{(\phi)} = \dot\phi$ exactly, and recursively computing the above expression provides an alternative way to compute wavefunction coefficients. If the theory \textit{does} have time derivative interactions, then this expression is not a generating functional for wavefunction coefficients, and one must include corrections to account for the difference between $\Pi^{(\phi)}$ and $\dot\phi$.
Using~\eqref{dotvarphi} and~\eqref{eq:pidefapp}, we can write the following expression for $\Pi^{(\phi)}$ evaluated on the future boundary:
\be
    \Pi^{(\phi)}(\vec{p}, t_f) = ip\varphi_{\vec{p}} + (2\pi)^3\frac{\partial\mathcal{L}_{\text{int}}}{\partial\dot\phi_{-\vec{p}}(t)}\bigg\rvert_{t = t_f} +  (2\pi)^3\int_{-\infty}^{0}\rd t'\,\mathcal{K}(p, t')\frac{\delta S_{\text{int}}}{\delta\phi_{-\vec{p}}(t')}\,.
    \label{eq:pi1}
\ee
Writing out explicitly the Euler--Lagrange derivative inside the integral we get 
\be
    \Pi^{(\phi)}(\vec{p}, t_f) = ip\varphi_{\vec{p}} + (2\pi)^3\frac{\partial\mathcal{L}_{\text{int}}}{\partial\dot\phi_{-\vec{p}}(t)}\bigg\rvert_{t = t_f} +  (2\pi)^3\int_{-\infty}^{0}\rd t'\,\mathcal{K}(p, t')\left[\frac{\partial\mathcal{L}_{\text{int}}}{\partial\phi_{-\vec{p}}(t')} - \frac{\rd}{\rd t}\frac{\partial\mathcal{L}_{\text{int}}}{\partial\dot\phi_{-\vec{p}}(t')}\right]\,,
\ee
and after integrating the time derivative by parts, we find
\be
    \Pi^{(\phi)}(\vec{p}, t_f)  = ip\varphi_{\vec{p}} +  (2\pi)^3\int_{-\infty}^{0}\rd t'\,\mathcal{K}(p, t')\left[\frac{\partial\mathcal{L}_{\text{int}}}{\partial\phi_{-\vec{p}}(t')} + ip\frac{\partial\mathcal{L}_{\text{int}}}{\partial\dot\phi_{-\vec{p}}(t')}\right]\,.
    \label{pigenerator}
\ee
This expression may be used to compute wavefunction coefficients for theories with arbitrary scalar interactions.

\vskip4pt
As it turns out, we may use the above expression to derive statements about the classical field profile at \textit{early} times as well.  Using the form of the bulk-to-bulk propagator in \eqref{eq:bbulkprop} we have that at early times
\begin{equation}
    \lim_{t\to -\infty}\mathcal{G}(p;t, t') = \frac{1}{2p}\Big(\mathcal{K}(-p, t') - \mathcal{K}(p, t')\Big)\mathcal{K}(p, t)\,.
\end{equation}
Thus, at early times, the classical field profile~\eqref{eq:classicalfieldprofileapp} reduces to
\newpage
\be
\begin{aligned}
\lim_{t\to -\infty}\phi_{\vec{p}}(t) &= {\cal K}(p, t)\Bigg(\varphi_{\vec{p}} + \frac{i}{2p}(2\pi)^3\int_{-\infty}^{0}\rd t'\bigg[\Big(\mathcal{K}(-p, t') - \mathcal{K}(p, t')\Big)\frac{\partial\mathcal{L}_{\text{int}}}{\partial\phi_{-\vec{p}}(t')}\\
    &\hspace{6 cm}~~- \Big(\mathcal{K}(-p, t') +\mathcal{K}(p, t')\Big)ip\frac{\partial\mathcal{L}_{\text{int}}}{\partial\dot\phi_{-\vec{p}}(t')}\bigg]\Bigg)\,,
\end{aligned}
\label{eq:phiearlytime}
\ee
where we have made the same simplifications of the $\delta S/\delta\phi$ term that we performed to get from~\eqref{eq:pi1} to~\eqref{pigenerator}.
Note there is no future boundary term because of the relative minus sign between the bulk-to-boundary propagators.  

\vskip4pt
Interestingly~\eqref{eq:phiearlytime} is essentially the difference of two copies of~\eqref{pigenerator} with the sign of the external energy flipped:
\begin{equation}
   \lim_{t\to -\infty} \phi_{\vec{p}}(t) =   \lim_{t\to -\infty}  \frac{i}{2p}\Big(\Pi^{(\phi)}(\vec{p}, -p, t_f) - \Pi^{(\phi)}(\vec{p}, p, t_f) \Big)\mathcal{K}(p, t)\,.
\end{equation}
Therefore, the classical field profile at early times is a generating functional for shifted wavefunction coefficients:
\begin{tcolorbox}[colframe=white,arc=0pt,colback=greyish2]
\be
\begin{aligned}
    &  \lim_{t\to -\infty} \phi_{\vec{p}}(t) =   \lim_{t\to -\infty} \mathcal{K}(p, t)\sum_{n=2}\frac{1}{(n-1)!}\int \frac{\rd^{3}p_1\cdots \rd^3p_{n-1}}{(2\pi)^{3(n-2)}}\,\varphi_{\vec{p}_1}\cdots\varphi_{\vec{p}_{n-1}}\\
&\hspace{7 cm}\times\delta^{(3)}(\vec{p}_1 + \cdots  + \vec{p}_{n-1} - \vec{p})\tl{\psi}_n(\vec{p}_1,\cdots, \vec{p}_{n-1},-\vec{p})\,,
    \label{shiftedgenerator}
\end{aligned}
\ee
\end{tcolorbox}
\noindent where we have defined the quantity
\be
\tl{\psi}_n(\vec{p}) \equiv\frac{1}{2p} \Big( \psi_{n}(\vec{p}_1,\cdots,\vec{p}_{n-1},\vec{p}_n;-p_n)-\psi_{n}(\vec{p}_1,\cdots,\vec{p}_{n-1},\vec{p}_n;p_n)\Big).
\label{eq:shiftedcoeff}
\ee
We can in fact go one step further by noticing that a shifted wavefunction coefficient is in the form of a finite difference in the energy associated with the classical field.  In the soft limit, we can make the replacement
\begin{equation}
    \lim_{\vec{p}\rightarrow 0}\tl{\psi}_n(\vec{p}) = -\partial_{p}\psi_n(\vec{p})\Big|_{\vec{p} = 0} \,.
\end{equation}
Thus, the soft classical field profile at early times is the generating functional for partial derivatives of wavefunction coefficients with respect to energy:
\begin{tcolorbox}[colframe=white,arc=0pt,colback=greyish2]
\begin{equation}
    \lim_{\substack{\scriptsize
    	\begin{aligned}
	t&\to -\infty\\[-6pt]
	 \vec p&\to 0
	\end{aligned}
	}
	} \phi_{\vec{p}}(t) 
    = -\sum_{n = 2}\frac{1}{(n-1)!}\int \frac{\rd^3p_1\cdots \rd^3p_{n-1}}{(2\pi)^{3(n-2)}}\,\varphi_{\vec{p}_1}\cdots \varphi_{\vec{p}_{n-1}}\delta^{(3)}(\vec{p}_1 + \cdots + \vec{p}_{n-1})\partial_{p}\psi_{n}\Big|_{\vec{p} = 0}\,.
    \label{earlytimegenerator}
\end{equation}
\end{tcolorbox}
\noindent This final formula indicates that shifted wavefunction coefficients contain information about the system at early times.  This fact is essential in understanding the role that the initial state plays in various soft theorems.

\newpage

\newpage
\section{Navigating branch cuts}
\label{app:cuts}

In this Appendix, we will show that it is always possible to choose a branch of the complexified partial energy functions which does not have any zeros.  

\vskip4pt
We want to study the zeros of the multivalued functions of the form
\begin{equation}
    f(z) = A + B z + \sqrt{(z - c_-)(z - c_+)}\,,
\end{equation}
where $A, B$ are real and $c_{\pm}$ are complex.  The candidate zeros of the function are
\begin{equation}
    z_{\pm} = \frac{-(2 AB + c_{-} + c_{+}) \pm \sqrt{(2 AB + c_{-} + c_{+})^2 - 4(B^2 - 1)(A^2 - c_{-}c_{+})}}{2(B^2 - 1)}\,.
\end{equation}
We will prove that there is a branch of $f(z)$ with no zeros, and a branch with two zeros.  Define the two branches as
\be
\begin{aligned}
    f_{1}(z) &= A + Bz + |(z - c_-)(z - c_+)|^{\frac{1}{2}}\exp\Big(\frac{i}{2}[\arg(z - c_-) + \arg(z - c_+)]\Big)\,,\\
    f_{2}(z) &= A + Bz + |(z - c_-)(z - c_+)|^{\frac{1}{2}}\exp\Big(\frac{i}{2}[\arg(z - c_-) + \arg(z - c_+) + 2\pi]\Big)\,.
\end{aligned}
\ee
We define both $\arg$ functions in both branches to take values in the range $[0, 2\pi)$.  This unambiguously defines two single-valued functions which are (at least) meromorphic on the space $\mathbb{C} - [c_{-}, c_{+}]$.  On the putative zeros, each branch becomes
\be
\begin{aligned}
    f_{1}(z_{\pm}) &= A + B z_{\pm} + |A + B z_{\pm}|\exp\Big(\frac{i}{2}[\arg(z_{\pm} - c_-) + \arg(z_{\pm} - c_+)]\Big)\,,\\
    f_{2}(z_{\pm}) &= A + B z_{\pm} + |A + B z_{\pm}|\exp\Big(\frac{i}{2}[\arg(z_{\pm} - c_-) + \arg(z_{\pm} - c_+) + 2\pi]\Big)\,.
\end{aligned}
\ee
We can simplify this further by manipulating the $\arg$ functions, which are both defined mod $2\pi$.  We have that
\be
\begin{aligned}
    \arg(z_{\pm} - c_-) + \arg(z_{\pm} - c_+) &= \arg\Big((z_{\pm} - c_-)(z_{\pm} - c_+)\Big)\\
    &= \arg\Big((A + Bz_{\pm})^2\Big) = 2\arg\Big(A + Bz_{\pm}\Big)\,.
\end{aligned}
\ee
Substituting this into the prior equation, we come to
\be
\begin{aligned}
    &f_{1}(z_{\pm}) = A + B z_{\pm} + |A + B z_{\pm}|\exp\Big(i\arg(A + B z_{\pm})\Big) = A + Bz_{\pm} + A + Bz_{\pm}\neq 0\,,\\
    &f_{2}(z_{\pm}) = A + B z_{\pm} + |A + B z_{\pm}|\exp\Big(i\arg(A + B z_{\pm}) + i\pi\Big) = A + B z_{\pm} - (A + B z_{\pm}) = 0\,.
\end{aligned}
\ee
Thus, the branch $f_1(z)$ has no zeros while the branch $f_2(z)$ has two zeros.

\newpage
\section{Double copy}
\label{app:Dcopy}
In this Appendix, we include some brief comments about double copy relations between the wavefunction coefficients we construct in the main text. In the context of scattering amplitudes, exceptional scalar theories have a rich structure involving the double copy that connects them to many other quantum field theories~\cite{Bern:2019prr}. The simplest double copy involves relating two copies of the nonlinear sigma model to the special galileon~\cite{Cachazo:2014xea}, and so we focus on the wavefunction version of this double copy. We only consider properties of the corresponding four-point wavefunctions, but it would be interesting to investigate the structure at higher points.

\vskip4pt
In the exponential representation, the four-point wavefunction~\eqref{eq:NLSM4pt} for the nonlinear sigma model, with flavor factors restored, is given by\footnote{Here we have defined some analogues of the flat space Mandelstam variables:
\begin{align}
{\cal S} &\equiv -(k_{12}+s)(k_{34}+s)\,,\\
{\cal T} &\equiv - (k_{14}+t)(k_{23}+t)\,,\\
{\cal U} &\equiv - (k_{13}+u)(k_{24}+u)\,,
\end{align}
which serve to simplify some expressions and allow us to cast them in a form similar to amplitudes. In the $E\to0$ limit, these variables reduce to ordinary Mandelstam variables. One benefit of this is that one can take $E\to 0$ in all of our expressions and the formulas will reduce to their amplitude versions, divided by $E$.}
\be
\begin{aligned}
\psi^{({\rm nlsm})}_4 &= -f^{a_1a_2b}f^{a_3a_4 b} \frac{1}{6E}\bigg({\cal T}-{\cal U}+E(t-u)
\bigg)\\
&~~~-f^{a_1a_4b}f^{a_2a_3 b} \frac{1}{6E}\bigg( {\cal U}-{\cal S}+E(u-s)
\bigg)\\
&~~~-f^{a_3a_1b}f^{a_2a_4 b} \frac{1}{6E}\bigg({\cal S}-{\cal T}+E(s-t)
\bigg)\,.
\end{aligned}
\ee
In order to double copy this object, we first put it in the form
\be
\psi^{({\rm nlsm})}_4 = \frac{1}{E}\bigg(\frac{c_s\,\bar n_s}{\cal S}+ \frac{c_t\,\bar n_t}{\cal T}+ \frac{c_u\,\bar n_u}{\cal U}\bigg)\,,
\label{eq:corrDC}
\ee
where we have defined the color numerators as
\begin{align}
c_s &= f^{a_1a_2b}f^{a_3a_4 b} & c_t &=f^{a_1a_4b}f^{a_2a_3 b} & c_u &=f^{a_3a_1b}f^{a_2a_4 b}\,.
\label{eq:colornumsBI}
\end{align}
As a consequence of the Jacobi identity, these satisfy
\be
c_s+c_t+c_u = 0\,.
\label{eq:colorfactorjacobi}
\ee
We similarly define the kinematic numerators in~\eqref{eq:corrDC} as
\begin{align}
\bar n_s &= -\frac{{\cal S}}{6}\Big({\cal T}-{\cal U}+E(t-u)
\Big)\,,\\  
\bar n_t &= -\frac{{\cal T}}{6}\Big( {\cal U}-{\cal S}+E(u-s)
\Big)\,, \\
 \bar n_u &= -\frac{{\cal U}}{6}\Big({\cal S}-{\cal T}+E(s-t)
\Big)\,.
\end{align}
In contrast to the amplitude story, the kinematic numerators so defined do {\it not} satisfy a kinematic Jacobi identity:
\be
\bar n_s +\bar n_t +\bar n_u = -\frac{E}{6}\bigg( {\cal S}(t-u)+{\cal T}(u-s)+{\cal U}(s-t)\bigg)\equiv -E\,\frac{{\cal E}^{\rm nlsm}}{6}\,,
\label{eq:kinnum}
\ee
which implicitly defines ${\cal E}^{\rm nlsm}$. As expected, in the $E=0$ limit the kinematic numerators do add up to zero. This failure of the kinematic Jacobi identity implies that, despite appearances, the presentation of the wavefunction~\eqref{eq:corrDC} is not actually color-kinematics symmetric. 
Fortunately, this is a solvable problem~\cite{Armstrong:2020woi,Albayrak:2020fyp}. 
Because the $c_a$ satisfy the Jacobi identity, the wavefunction coefficient~\eqref{eq:corrDC} is invariant under the following generalized gauge-like transformation
\begin{align}
\bar n_s &\mapsto \bar n_s + {\cal S}\,\chi\,, \\
\bar n_t &\mapsto \bar n_t + {\cal T}\,\chi\,, \\
\bar n_u &\mapsto \bar n_u + {\cal U}\,\chi\,,
\end{align}
where $\chi$ is an arbitrary function. We can use the freedom to make the kinematic numerators sum to zero. Shifting each of the numerators, we see that the equation we want to solve is
\be
 \bar n_s+ \bar n_t+ \bar n_u+ ({\cal S}+{\cal T}+{\cal U})\chi = -\frac{ E}{6}\bigg( {\cal S}(t-u)+{\cal T}(u-s)+{\cal U}(s-t)\bigg)+({\cal S}+{\cal T}+{\cal U})\chi  = 0\,,
\ee
which implies that
\be
\chi = -\frac{ E\,{\cal E}^{{\rm nlsm}}}{6({\cal S}+{\cal T}+{\cal U})}\,.
\ee
Thus, if we define the kinematic numerators
\begin{align}
n_s &= -\frac{ {\cal S}}{6}\Big({\cal T}-{\cal U}+E(t-u)\Big) +\frac{{\cal S}}{6} \frac{ E\,{\cal E}^{{\rm nlsm}}}{{\cal S}+{\cal T}+{\cal U}}\,, \\
n_t &=-\frac{{\cal T}}{6}\Big( {\cal U}-{\cal S}+E(u-s)\Big)+\frac{{\cal T}}{6}\frac{ E\,{\cal E}^{{\rm nlsm}}}{{\cal S}+{\cal T}+{\cal U}}\,,\\
n_u &= -\frac{{\cal U}}{6}\Big({\cal S}-{\cal T}+E(s-t)
\Big)+\frac{{\cal U}}{6}\frac{ E\,{\cal E}^{{\rm nlsm}}}{{\cal S}+{\cal T}+{\cal U}}\,,
\end{align}
they will satisfy the kinematic Jacobi identity $n_s+n_t+n_u=0$ and the wavefunction
\be
\psi^{({\rm nlsm})}_4 = \frac{1}{E}\bigg(\frac{c_s\,n_s}{\cal S}+ \frac{c_t\,n_t}{\cal T}+ \frac{c_u\, n_u}{\cal U}\bigg)\,,
\label{eq:corrDC2}
\ee
is color-kinematics dual.

\paragraph{Double copy:} We can now try double copying~\eqref{eq:corrDC2} by replacing flavor factors with kinematic ones to obtain\footnote{As was pointed out by~\cite{Armstrong:2020woi,Albayrak:2020fyp}, after expressing flavor ordered wavefunctions as:
\begin{align}
\label{eq:smtFOnlsm}
E\psi^{({\rm nlsm})}_{s-t} &\equiv\frac{n^{({\rm NLSM})}_s}{{\cal S}}-\frac{n^{({\rm NLSM})}_t}{{\cal T}}= {\cal S}+{\cal T}-2{\cal U}+E(s+t-2u)\,,\\
E\psi^{({\rm nlsm})}_{u-t} &\equiv \frac{n^{({\rm NLSM})}_u}{{\cal U}}-\frac{n^{({\rm NLSM})}_t}{{\cal T}}=2{\cal S}-{\cal T}-{\cal U}+E(2s-t-u)\,,
\end{align}
where~\eqref{eq:smtFOnlsm} is the same as~\eqref{eq:NLSM4pt},
we can actually solve for the kinematic numerators as a function of these flavor ordered wavefunctions. (This is not possible for amplitudes because of BCJ relations.) 
}
\be
\psi^{({\rm dc})}_4 = \frac{1}{E}\bigg(\frac{n_s^2}{\cal S}+ \frac{n_t^2}{\cal T}+ \frac{n_u^2}{\cal U}\bigg)\,.
\ee
The natural question to ask is what is this object? In the $E\to 0$ limit, we know that this is the scattering amplitude of the special galileon, so we might be temped to say that this is the four-point wavefunction of the special galileon~\eqref{eq:simplegal4pt}. However, this is not the case. We can write the difference between the double-copied wavefunction and the special galileon four-point wavefunction:
\be
\begin{aligned}
\psi_4^{({\rm sgal})}-\psi^{({\rm dc})}_4 = &-\frac{1}{4}\Big( k_3 {\cal S}^2+k_1{\cal T}^2+k_2{\cal U}^2\Big)\\
&-\frac{1}{48}\bigg[ (3k_{13}+2u+s+t){\cal S}{\cal T}+(3k_{23}+2t+s+u){\cal S}{\cal U}+(3k_{12}+2s+u+t){\cal T}{\cal U}\bigg]\\
&-\frac{E}{144}\bigg[~9(k_1k_2+k_3^2+k_{13}t+k_{23}u)+81k_{12}k_3+72k_3s+7tu+t^2+u^2\\
&\hspace{1.3cm}+9(k_3k_2+k_1^2+k_{13}s+k_{12}u)+81k_{23}k_1+72k_1t+7su+s^2+u^2\\[3pt]
&\hspace{1.3cm}+9(k_1k_3+k_2^2+k_{12}t+k_{23}s)+81k_{13}k_2+72k_2u+7ts+t^2+s^2\\[3pt]
&\hspace{1.3cm}+{\cal T}{\cal U}+{\cal S}{\cal T}+{\cal S}{\cal U}-2k_1k_2k_3(k_1+k_2+k_3+s+t+u)\bigg]\\
&-\frac{E^2}{16}\bigg[ (k_{12}+s)(k_{13}+u)(k_{23}+t)+4k_1(k_{23}+t)^2+4k_2 (k_{13}+u)^2\\
&\hspace{1.3cm}+4k_3(k_{12}+s)^2-16k_1k_2k_3\bigg]-\frac{1}{36} \frac{ E\,({\cal E}^{{\rm nlsm}})^2}{{\cal S}+{\cal T}+{\cal U}}\,.
\end{aligned}
\label{eq:wrongDC}
\ee
It is worth noting that the difference between the two wavefunctions is regular as $E\to 0$. This is a needed requirement for consistency with the amplitudes double copy: the $E\to 0$ singularity's residue is the corresponding amplitude. Another notable feature is that the shift by ${\cal E}^{\rm nlsm}$ does not help the double copy, in fact it introduces a spurious singularity. Using the identity ${\cal S}+{\cal T}+{\cal U} = -E(E+s+t+u)$ we see that this last term in~\eqref{eq:wrongDC} has an unwanted singularity at $E+s+t+u = 0$.

\vskip4pt
Since the particular choice of $\chi$ that we have made to satisfy color-kinematics duality does not help the double copy, we can consider double copying the kinematic numerators with $\chi$ left arbitrary. The only change from~\eqref{eq:wrongDC} will be to replace the last term with $({\cal S}+{\cal T}+{\cal U})\chi^2/E$. Since $\chi$ is a totally arbitrary function, we can of course choose it so that~\eqref{eq:wrongDC} vanishes. (A similar point was made in~\cite{Sivaramakrishnan:2021srm}.) However the resulting kinematic numerators are extremely complicated, and do not satisfy color-kinematics duality. It would be interesting to understand if this choice of kinematic numerators has an interpretation or independent construction.

\vskip4pt
Essentially the takeaway is that we find that the most naive version of the double copy does not work for the flat space wavefunction (this is broadly the same conclusion as~\cite{Albayrak:2020fyp,Sivaramakrishnan:2021srm,Herderschee:2022ntr,Cheung:2022pdk}). Since the relation between the NLSM and special galileon is one of the simplest double copy relations, we expect that further investigation will help elucidate the fate and structure of the double copy for the wavefunction.

\newpage
\renewcommand{\em}{}
\bibliographystyle{utphys}
\addcontentsline{toc}{section}{References}
\bibliography{softcorrbib}

\providecommand{\href}[2]{#2}\begingroup\raggedright\begin{thebibliography}{10}

\bibitem{Weinberg:1965nx}
S.~Weinberg, ``{Infrared photons and gravitons},''
  \href{http://dx.doi.org/10.1103/PhysRev.140.B516}{{\em Phys. Rev.} {\bf 140}
  (1965)  B516--B524}.

\bibitem{Britto:2005fq}
R.~Britto, F.~Cachazo, B.~Feng, and E.~Witten, ``{Direct proof of tree-level
  recursion relation in Yang-Mills theory},''
  \href{http://dx.doi.org/10.1103/PhysRevLett.94.181602}{{\em Phys. Rev. Lett.}
  {\bf 94} (2005)  181602}, \href{http://arxiv.org/abs/hep-th/0501052}{{\tt
  arXiv:hep-th/0501052}}.

\bibitem{Bern:2019prr}
Z.~Bern, J.~J. Carrasco, M.~Chiodaroli, H.~Johansson, and R.~Roiban, ``{The
  Duality Between Color and Kinematics and its Applications},''
  \href{http://arxiv.org/abs/1909.01358}{{\tt arXiv:1909.01358 [hep-th]}}.

\bibitem{Bern:2010ue}
Z.~Bern, J.~J.~M. Carrasco, and H.~Johansson, ``{Perturbative Quantum Gravity
  as a Double Copy of Gauge Theory},''
  \href{http://dx.doi.org/10.1103/PhysRevLett.105.061602}{{\em Phys. Rev.
  Lett.} {\bf 105} (2010)  061602}, \href{http://arxiv.org/abs/1004.0476}{{\tt
  arXiv:1004.0476 [hep-th]}}.

\bibitem{Cachazo:2013iea}
F.~Cachazo, S.~He, and E.~Y. Yuan, ``{Scattering of Massless Particles:
  Scalars, Gluons and Gravitons},''
  \href{http://dx.doi.org/10.1007/JHEP07(2014)033}{{\em JHEP} {\bf 07} (2014)
  033}, \href{http://arxiv.org/abs/1309.0885}{{\tt arXiv:1309.0885 [hep-th]}}.

\bibitem{Cachazo:2014xea}
F.~Cachazo, S.~He, and E.~Y. Yuan, ``{Scattering Equations and Matrices: From
  Einstein To Yang-Mills, DBI and NLSM},''
  \href{http://dx.doi.org/10.1007/JHEP07(2015)149}{{\em JHEP} {\bf 07} (2015)
  149}, \href{http://arxiv.org/abs/1412.3479}{{\tt arXiv:1412.3479 [hep-th]}}.

\bibitem{Cheung:2017ems}
C.~Cheung, C.-H. Shen, and C.~Wen, ``{Unifying Relations for Scattering
  Amplitudes},'' \href{http://dx.doi.org/10.1007/JHEP02(2018)095}{{\em JHEP}
  {\bf 02} (2018)  095}, \href{http://arxiv.org/abs/1705.03025}{{\tt
  arXiv:1705.03025 [hep-th]}}.

\bibitem{Boulanger:2000rq}
N.~Boulanger, T.~Damour, L.~Gualtieri, and M.~Henneaux, ``{Inconsistency of
  interacting, multigraviton theories},''
  \href{http://dx.doi.org/10.1016/S0550-3213(00)00718-5}{{\em Nucl. Phys. B}
  {\bf 597} (2001)  127--171}, \href{http://arxiv.org/abs/hep-th/0007220}{{\tt
  arXiv:hep-th/0007220}}.

\bibitem{Benincasa:2007xk}
P.~Benincasa and F.~Cachazo, ``{Consistency Conditions on the S-Matrix of
  Massless Particles},'' \href{http://arxiv.org/abs/0705.4305}{{\tt
  arXiv:0705.4305 [hep-th]}}.

\bibitem{Schuster:2008nh}
P.~C. Schuster and N.~Toro, ``{Constructing the Tree-Level Yang-Mills S-Matrix
  Using Complex Factorization},''
  \href{http://dx.doi.org/10.1088/1126-6708/2009/06/079}{{\em JHEP} {\bf 06}
  (2009)  079}, \href{http://arxiv.org/abs/0811.3207}{{\tt arXiv:0811.3207
  [hep-th]}}.

\bibitem{Porrati:2008rm}
M.~Porrati, ``{Universal Limits on Massless High-Spin Particles},''
  \href{http://dx.doi.org/10.1103/PhysRevD.78.065016}{{\em Phys. Rev. D} {\bf
  78} (2008)  065016}, \href{http://arxiv.org/abs/0804.4672}{{\tt
  arXiv:0804.4672 [hep-th]}}.

\bibitem{McGady:2013sga}
D.~A. McGady and L.~Rodina, ``{Higher-spin massless $S$-matrices in
  four-dimensions},'' \href{http://dx.doi.org/10.1103/PhysRevD.90.084048}{{\em
  Phys. Rev. D} {\bf 90} (2014) no.~8, 084048},
  \href{http://arxiv.org/abs/1311.2938}{{\tt arXiv:1311.2938 [hep-th]}}.

\bibitem{Cheung:2014dqa}
C.~Cheung, K.~Kampf, J.~Novotny, and J.~Trnka, ``{Effective Field Theories from
  Soft Limits of Scattering Amplitudes},''
  \href{http://dx.doi.org/10.1103/PhysRevLett.114.221602}{{\em Phys. Rev.
  Lett.} {\bf 114} (2015) no.~22, 221602},
  \href{http://arxiv.org/abs/1412.4095}{{\tt arXiv:1412.4095 [hep-th]}}.

\bibitem{Cheung:2016drk}
C.~Cheung, K.~Kampf, J.~Novotny, C.-H. Shen, and J.~Trnka, ``{A Periodic Table
  of Effective Field Theories},''
  \href{http://dx.doi.org/10.1007/JHEP02(2017)020}{{\em JHEP} {\bf 02} (2017)
  020}, \href{http://arxiv.org/abs/1611.03137}{{\tt arXiv:1611.03137
  [hep-th]}}.

\bibitem{Cheung:2017yef}
C.~Cheung, G.~N. Remmen, C.-H. Shen, and C.~Wen, ``{Pions as Gluons in Higher
  Dimensions},'' \href{http://dx.doi.org/10.1007/JHEP04(2018)129}{{\em JHEP}
  {\bf 04} (2018)  129}, \href{http://arxiv.org/abs/1709.04932}{{\tt
  arXiv:1709.04932 [hep-th]}}.

\bibitem{Cheung:2015ota}
C.~Cheung, K.~Kampf, J.~Novotny, C.-H. Shen, and J.~Trnka, ``{On-Shell
  Recursion Relations for Effective Field Theories},''
  \href{http://dx.doi.org/10.1103/PhysRevLett.116.041601}{{\em Phys. Rev.
  Lett.} {\bf 116} (2016) no.~4, 041601},
  \href{http://arxiv.org/abs/1509.03309}{{\tt arXiv:1509.03309 [hep-th]}}.

\bibitem{Padilla:2016mno}
A.~Padilla, D.~Stefanyszyn, and T.~Wilson, ``{Probing Scalar Effective Field
  Theories with the Soft Limits of Scattering Amplitudes},''
  \href{http://dx.doi.org/10.1007/JHEP04(2017)015}{{\em JHEP} {\bf 04} (2017)
  015}, \href{http://arxiv.org/abs/1612.04283}{{\tt arXiv:1612.04283
  [hep-th]}}.

\bibitem{Elvang:2018dco}
H.~Elvang, M.~Hadjiantonis, C.~R.~T. Jones, and S.~Paranjape, ``{Soft Bootstrap
  and Supersymmetry},'' \href{http://dx.doi.org/10.1007/JHEP01(2019)195}{{\em
  JHEP} {\bf 01} (2019)  195}, \href{http://arxiv.org/abs/1806.06079}{{\tt
  arXiv:1806.06079 [hep-th]}}.

\bibitem{Bonifacio:2019rpv}
J.~Bonifacio, K.~Hinterbichler, L.~A. Johnson, A.~Joyce, and R.~A. Rosen,
  ``{Matter Couplings and Equivalence Principles for Soft Scalars},''
  \href{http://dx.doi.org/10.1007/JHEP07(2020)056}{{\em JHEP} {\bf 07} (2020)
  056}, \href{http://arxiv.org/abs/1911.04490}{{\tt arXiv:1911.04490
  [hep-th]}}.

\bibitem{Maldacena:2011nz}
J.~M. Maldacena and G.~L. Pimentel, ``{On graviton non-Gaussianities during
  inflation},'' \href{http://dx.doi.org/10.1007/JHEP09(2011)045}{{\em JHEP}
  {\bf 09} (2011)  045}, \href{http://arxiv.org/abs/1104.2846}{{\tt
  arXiv:1104.2846 [hep-th]}}.

\bibitem{Raju:2012zr}
S.~Raju, ``{New Recursion Relations and a Flat Space Limit for AdS/CFT
  Correlators},'' \href{http://dx.doi.org/10.1103/PhysRevD.85.126009}{{\em
  Phys. Rev. D} {\bf 85} (2012)  126009},
  \href{http://arxiv.org/abs/1201.6449}{{\tt arXiv:1201.6449 [hep-th]}}.

\bibitem{Arkani-Hamed:2017fdk}
N.~Arkani-Hamed, P.~Benincasa, and A.~Postnikov, ``{Cosmological Polytopes and
  the Wavefunction of the Universe},''
  \href{http://arxiv.org/abs/1709.02813}{{\tt arXiv:1709.02813 [hep-th]}}.

\bibitem{Arkani-Hamed:2018kmz}
N.~Arkani-Hamed, D.~Baumann, H.~Lee, and G.~L. Pimentel, ``{The Cosmological
  Bootstrap: Inflationary Correlators from Symmetries and Singularities},''
  \href{http://dx.doi.org/10.1007/JHEP04(2020)105}{{\em JHEP} {\bf 04} (2020)
  105}, \href{http://arxiv.org/abs/1811.00024}{{\tt arXiv:1811.00024
  [hep-th]}}.

\bibitem{Baumann:2020dch}
D.~Baumann, C.~Duaso~Pueyo, A.~Joyce, H.~Lee, and G.~L. Pimentel, ``{The
  Cosmological Bootstrap: Spinning Correlators from Symmetries and
  Factorization},'' \href{http://dx.doi.org/10.21468/SciPostPhys.11.3.071}{{\em
  SciPost Phys.} {\bf 11} (2021)  071},
  \href{http://arxiv.org/abs/2005.04234}{{\tt arXiv:2005.04234 [hep-th]}}.

\bibitem{Baumann:2021fxj}
D.~Baumann, W.-M. Chen, C.~Duaso~Pueyo, A.~Joyce, H.~Lee, and G.~L. Pimentel,
  ``{Linking the Singularities of Cosmological Correlators},''
  \href{http://arxiv.org/abs/2106.05294}{{\tt arXiv:2106.05294 [hep-th]}}.

\bibitem{Goodhew:2020hob}
H.~Goodhew, S.~Jazayeri, and E.~Pajer, ``{The Cosmological Optical Theorem},''
  \href{http://dx.doi.org/10.1088/1475-7516/2021/04/021}{{\em JCAP} {\bf 04}
  (2021)  021}, \href{http://arxiv.org/abs/2009.02898}{{\tt arXiv:2009.02898
  [hep-th]}}.

\bibitem{Cespedes:2020xqq}
S.~C\'espedes, A.-C. Davis, and S.~Melville, ``{On the time evolution of
  cosmological correlators},''
  \href{http://dx.doi.org/10.1007/JHEP02(2021)012}{{\em JHEP} {\bf 02} (2021)
  012}, \href{http://arxiv.org/abs/2009.07874}{{\tt arXiv:2009.07874
  [hep-th]}}.

\bibitem{Benincasa:2020aoj}
P.~Benincasa, A.~J. McLeod, and C.~Vergu, ``{Steinmann Relations and the
  Wavefunction of the Universe},''
  \href{http://dx.doi.org/10.1103/PhysRevD.102.125004}{{\em Phys. Rev. D} {\bf
  102} (2020)  125004}, \href{http://arxiv.org/abs/2009.03047}{{\tt
  arXiv:2009.03047 [hep-th]}}.

\bibitem{Meltzer:2020qbr}
D.~Meltzer and A.~Sivaramakrishnan, ``{CFT unitarity and the AdS Cutkosky
  rules},'' \href{http://dx.doi.org/10.1007/JHEP11(2020)073}{{\em JHEP} {\bf
  11} (2020)  073}, \href{http://arxiv.org/abs/2008.11730}{{\tt
  arXiv:2008.11730 [hep-th]}}.

\bibitem{Sleight:2020obc}
C.~Sleight and M.~Taronna, ``{From AdS to dS Exchanges: Spectral
  Representation, Mellin Amplitudes and Crossing},''
  \href{http://arxiv.org/abs/2007.09993}{{\tt arXiv:2007.09993 [hep-th]}}.

\bibitem{Hogervorst:2021uvp}
M.~Hogervorst, J.~a. Penedones, and K.~S. Vaziri, ``{Towards the
  non-perturbative cosmological bootstrap},''
  \href{http://arxiv.org/abs/2107.13871}{{\tt arXiv:2107.13871 [hep-th]}}.

\bibitem{DiPietro:2021sjt}
L.~Di~Pietro, V.~Gorbenko, and S.~Komatsu, ``{Analyticity and Unitarity for
  Cosmological Correlators},'' \href{http://arxiv.org/abs/2108.01695}{{\tt
  arXiv:2108.01695 [hep-th]}}.

\bibitem{Sleight:2021plv}
C.~Sleight and M.~Taronna, ``{From dS to AdS and back},''
  \href{http://dx.doi.org/10.1007/JHEP12(2021)074}{{\em JHEP} {\bf 12} (2021)
  074}, \href{http://arxiv.org/abs/2109.02725}{{\tt arXiv:2109.02725
  [hep-th]}}.

\bibitem{Arkani-Hamed:2015bza}
N.~Arkani-Hamed and J.~Maldacena, ``{Cosmological Collider Physics},''
  \href{http://arxiv.org/abs/1503.08043}{{\tt arXiv:1503.08043 [hep-th]}}.

\bibitem{Baumann:2019oyu}
D.~Baumann, C.~Duaso~Pueyo, A.~Joyce, H.~Lee, and G.~L. Pimentel, ``{The
  cosmological bootstrap: weight-shifting operators and scalar seeds},''
  \href{http://dx.doi.org/10.1007/JHEP12(2020)204}{{\em JHEP} {\bf 12} (2020)
  204}, \href{http://arxiv.org/abs/1910.14051}{{\tt arXiv:1910.14051
  [hep-th]}}.

\bibitem{Sleight:2019mgd}
C.~Sleight, ``{A Mellin Space Approach to Cosmological Correlators},''
  \href{http://dx.doi.org/10.1007/JHEP01(2020)090}{{\em JHEP} {\bf 01} (2020)
  090}, \href{http://arxiv.org/abs/1906.12302}{{\tt arXiv:1906.12302
  [hep-th]}}.

\bibitem{Sleight:2019hfp}
C.~Sleight and M.~Taronna, ``{Bootstrapping Inflationary Correlators in Mellin
  Space},'' \href{http://dx.doi.org/10.1007/JHEP02(2020)098}{{\em JHEP} {\bf
  02} (2020)  098}, \href{http://arxiv.org/abs/1907.01143}{{\tt
  arXiv:1907.01143 [hep-th]}}.

\bibitem{Sleight:2021iix}
C.~Sleight and M.~Taronna, ``{On the consistency of (partially-)massless matter
  couplings in de Sitter space},''
  \href{http://dx.doi.org/10.1007/JHEP10(2021)156}{{\em JHEP} {\bf 10} (2021)
  156}, \href{http://arxiv.org/abs/2106.00366}{{\tt arXiv:2106.00366
  [hep-th]}}.

\bibitem{Albayrak:2018tam}
S.~Albayrak and S.~Kharel, ``{Towards the higher point holographic momentum
  space amplitudes},'' \href{http://dx.doi.org/10.1007/JHEP02(2019)040}{{\em
  JHEP} {\bf 02} (2019)  040}, \href{http://arxiv.org/abs/1810.12459}{{\tt
  arXiv:1810.12459 [hep-th]}}.

\bibitem{Albayrak:2019yve}
S.~Albayrak and S.~Kharel, ``{Towards the higher point holographic momentum
  space amplitudes. Part II. Gravitons},''
  \href{http://dx.doi.org/10.1007/JHEP12(2019)135}{{\em JHEP} {\bf 12} (2019)
  135}, \href{http://arxiv.org/abs/1908.01835}{{\tt arXiv:1908.01835
  [hep-th]}}.

\bibitem{Benincasa:2019vqr}
P.~Benincasa, ``{Cosmological Polytopes and the Wavefuncton of the Universe for
  Light States},'' \href{http://arxiv.org/abs/1909.02517}{{\tt arXiv:1909.02517
  [hep-th]}}.

\bibitem{Pajer:2020wxk}
E.~Pajer, ``{Building a Boostless Bootstrap for the Bispectrum},''
  \href{http://dx.doi.org/10.1088/1475-7516/2021/01/023}{{\em JCAP} {\bf 01}
  (2021)  023}, \href{http://arxiv.org/abs/2010.12818}{{\tt arXiv:2010.12818
  [hep-th]}}.

\bibitem{Bonifacio:2021azc}
J.~Bonifacio, E.~Pajer, and D.-G. Wang, ``{From Amplitudes to Contact
  Cosmological Correlators},'' \href{http://arxiv.org/abs/2106.15468}{{\tt
  arXiv:2106.15468 [hep-th]}}.

\bibitem{Cabass:2021fnw}
G.~Cabass, E.~Pajer, D.~Stefanyszyn, and J.~Supe\l{}, ``{Bootstrapping Large
  Graviton non-Gaussianities},'' \href{http://arxiv.org/abs/2109.10189}{{\tt
  arXiv:2109.10189 [hep-th]}}.

\bibitem{Hillman:2021bnk}
A.~Hillman and E.~Pajer, ``{A Differential Representation of Cosmological
  Wavefunctions},'' \href{http://arxiv.org/abs/2112.01619}{{\tt
  arXiv:2112.01619 [hep-th]}}.

\bibitem{Chen:2009zp}
X.~Chen and Y.~Wang, ``{Quasi-Single Field Inflation and Non-Gaussianities},''
  \href{http://dx.doi.org/10.1088/1475-7516/2010/04/027}{{\em JCAP} {\bf 04}
  (2010)  027}, \href{http://arxiv.org/abs/0911.3380}{{\tt arXiv:0911.3380
  [hep-th]}}.

\bibitem{Noumi:2012vr}
T.~Noumi, M.~Yamaguchi, and D.~Yokoyama, ``{Effective field theory approach to
  quasi-single field inflation and effects of heavy fields},''
  \href{http://dx.doi.org/10.1007/JHEP06(2013)051}{{\em JHEP} {\bf 06} (2013)
  051}, \href{http://arxiv.org/abs/1211.1624}{{\tt arXiv:1211.1624 [hep-th]}}.

\bibitem{Assassi:2012zq}
V.~Assassi, D.~Baumann, and D.~Green, ``{On Soft Limits of Inflationary
  Correlation Functions},''
  \href{http://dx.doi.org/10.1088/1475-7516/2012/11/047}{{\em JCAP} {\bf 11}
  (2012)  047}, \href{http://arxiv.org/abs/1204.4207}{{\tt arXiv:1204.4207
  [hep-th]}}.

\bibitem{Lee:2016vti}
H.~Lee, D.~Baumann, and G.~L. Pimentel, ``{Non-Gaussianity as a Particle
  Detector},'' \href{http://dx.doi.org/10.1007/JHEP12(2016)040}{{\em JHEP} {\bf
  12} (2016)  040}, \href{http://arxiv.org/abs/1607.03735}{{\tt
  arXiv:1607.03735 [hep-th]}}.

\bibitem{An:2017hlx}
H.~An, M.~McAneny, A.~K. Ridgway, and M.~B. Wise, ``{Quasi Single Field
  Inflation in the non-perturbative regime},''
  \href{http://dx.doi.org/10.1007/JHEP06(2018)105}{{\em JHEP} {\bf 06} (2018)
  105}, \href{http://arxiv.org/abs/1706.09971}{{\tt arXiv:1706.09971
  [hep-ph]}}.

\bibitem{Kumar:2017ecc}
S.~Kumar and R.~Sundrum, ``{Heavy-Lifting of Gauge Theories By Cosmic
  Inflation},'' \href{http://dx.doi.org/10.1007/JHEP05(2018)011}{{\em JHEP}
  {\bf 05} (2018)  011}, \href{http://arxiv.org/abs/1711.03988}{{\tt
  arXiv:1711.03988 [hep-ph]}}.

\bibitem{Alexander:2019vtb}
S.~Alexander, S.~J. Gates, L.~Jenks, K.~Koutrolikos, and E.~McDonough,
  ``{Higher Spin Supersymmetry at the Cosmological Collider: Sculpting SUSY
  Rilles in the CMB},'' \href{http://dx.doi.org/10.1007/JHEP10(2019)156}{{\em
  JHEP} {\bf 10} (2019)  156}, \href{http://arxiv.org/abs/1907.05829}{{\tt
  arXiv:1907.05829 [hep-th]}}.

\bibitem{Wang:2019gbi}
L.-T. Wang and Z.-Z. Xianyu, ``{In Search of Large Signals at the Cosmological
  Collider},'' \href{http://dx.doi.org/10.1007/JHEP02(2020)044}{{\em JHEP} {\bf
  02} (2020)  044}, \href{http://arxiv.org/abs/1910.12876}{{\tt
  arXiv:1910.12876 [hep-ph]}}.

\bibitem{Wang:2020ioa}
L.-T. Wang and Z.-Z. Xianyu, ``{Gauge Boson Signals at the Cosmological
  Collider},'' \href{http://dx.doi.org/10.1007/JHEP11(2020)082}{{\em JHEP} {\bf
  11} (2020)  082}, \href{http://arxiv.org/abs/2004.02887}{{\tt
  arXiv:2004.02887 [hep-ph]}}.

\bibitem{Wang:2021qez}
L.-T. Wang, Z.-Z. Xianyu, and Y.-M. Zhong, ``{Precision Calculation of
  Inflation Correlators at One Loop},''
  \href{http://arxiv.org/abs/2109.14635}{{\tt arXiv:2109.14635 [hep-ph]}}.

\bibitem{Lu:2021wxu}
Q.~Lu, M.~Reece, and Z.-Z. Xianyu, ``{Missing scalars at the cosmological
  collider},'' \href{http://dx.doi.org/10.1007/JHEP12(2021)098}{{\em JHEP} {\bf
  12} (2021)  098}, \href{http://arxiv.org/abs/2108.11385}{{\tt
  arXiv:2108.11385 [hep-ph]}}.

\bibitem{Tong:2021wai}
X.~Tong, Y.~Wang, and Y.~Zhu, ``{Cutting Rule for Cosmological Collider
  Signals: A Bulk Evolution Perspective},''
  \href{http://arxiv.org/abs/2112.03448}{{\tt arXiv:2112.03448 [hep-th]}}.

\bibitem{Arkani-Hamed:2018bjr}
N.~Arkani-Hamed and P.~Benincasa, ``{On the Emergence of Lorentz Invariance and
  Unitarity from the Scattering Facet of Cosmological Polytopes},''
  \href{http://arxiv.org/abs/1811.01125}{{\tt arXiv:1811.01125 [hep-th]}}.

\bibitem{Benincasa:2018ssx}
P.~Benincasa, ``{From the flat-space S-matrix to the Wavefunction of the
  Universe},'' \href{http://arxiv.org/abs/1811.02515}{{\tt arXiv:1811.02515
  [hep-th]}}.

\bibitem{Hillman:2019wgh}
A.~Hillman, ``{Symbol Recursion for the dS Wave Function},''
  \href{http://arxiv.org/abs/1912.09450}{{\tt arXiv:1912.09450 [hep-th]}}.

\bibitem{Callan:1970yg}
C.~G. Callan, Jr., ``{Broken scale invariance in scalar field theory},''
  \href{http://dx.doi.org/10.1103/PhysRevD.2.1541}{{\em Phys. Rev. D} {\bf 2}
  (1970)  1541--1547}.

\bibitem{Grall:2020ibl}
T.~Grall, S.~Jazayeri, and D.~Stefanyszyn, ``{The cosmological phonon:
  symmetries and amplitudes on sub-horizon scales},''
  \href{http://dx.doi.org/10.1007/JHEP11(2020)097}{{\em JHEP} {\bf 11} (2020)
  097}, \href{http://arxiv.org/abs/2005.12937}{{\tt arXiv:2005.12937
  [hep-th]}}.

\bibitem{Kampf:2019mcd}
K.~Kampf, J.~Novotny, M.~Shifman, and J.~Trnka, ``{New Soft Theorems for
  Goldstone Boson Amplitudes},''
  \href{http://dx.doi.org/10.1103/PhysRevLett.124.111601}{{\em Phys. Rev.
  Lett.} {\bf 124} (2020) no.~11, 111601},
  \href{http://arxiv.org/abs/1910.04766}{{\tt arXiv:1910.04766 [hep-th]}}.

\bibitem{Green:2020ebl}
D.~Green and E.~Pajer, ``{On the Symmetries of Cosmological Perturbations},''
  \href{http://dx.doi.org/10.1088/1475-7516/2020/09/032}{{\em JCAP} {\bf 09}
  (2020)  032}, \href{http://arxiv.org/abs/2004.09587}{{\tt arXiv:2004.09587
  [hep-th]}}.

\bibitem{Jazayeri:2021fvk}
S.~Jazayeri, E.~Pajer, and D.~Stefanyszyn, ``{From locality and unitarity to
  cosmological correlators},''
  \href{http://dx.doi.org/10.1007/JHEP10(2021)065}{{\em JHEP} {\bf 10} (2021)
  065}, \href{http://arxiv.org/abs/2103.08649}{{\tt arXiv:2103.08649
  [hep-th]}}.

\bibitem{Anninos:2014lwa}
D.~Anninos, T.~Anous, D.~Z. Freedman, and G.~Konstantinidis, ``{Late-time
  Structure of the Bunch-Davies De Sitter Wavefunction},''
  \href{http://dx.doi.org/10.1088/1475-7516/2015/11/048}{{\em JCAP} {\bf 11}
  (2015)  048}, \href{http://arxiv.org/abs/1406.5490}{{\tt arXiv:1406.5490
  [hep-th]}}.

\bibitem{Goon:2018fyu}
G.~Goon, K.~Hinterbichler, A.~Joyce, and M.~Trodden, ``{Shapes of gravity:
  Tensor non-Gaussianity and massive spin-2 fields},''
  \href{http://dx.doi.org/10.1007/JHEP10(2019)182}{{\em JHEP} {\bf 10} (2019)
  182}, \href{http://arxiv.org/abs/1812.07571}{{\tt arXiv:1812.07571
  [hep-th]}}.

\bibitem{Melville:2021lst}
S.~Melville and E.~Pajer, ``{Cosmological Cutting Rules},''
  \href{http://dx.doi.org/10.1007/JHEP05(2021)249}{{\em JHEP} {\bf 05} (2021)
  249}, \href{http://arxiv.org/abs/2103.09832}{{\tt arXiv:2103.09832
  [hep-th]}}.

\bibitem{Goodhew:2021oqg}
H.~Goodhew, S.~Jazayeri, M.~H. Gordon~Lee, and E.~Pajer, ``{Cutting
  cosmological correlators},''
  \href{http://dx.doi.org/10.1088/1475-7516/2021/08/003}{{\em JCAP} {\bf 08}
  (2021)  003}, \href{http://arxiv.org/abs/2104.06587}{{\tt arXiv:2104.06587
  [hep-th]}}.

\bibitem{Hinterbichler:2015pqa}
K.~Hinterbichler and A.~Joyce, ``{Hidden symmetry of the Galileon},''
  \href{http://dx.doi.org/10.1103/PhysRevD.92.023503}{{\em Phys. Rev. D} {\bf
  92} (2015) no.~2, 023503}, \href{http://arxiv.org/abs/1501.07600}{{\tt
  arXiv:1501.07600 [hep-th]}}.

\bibitem{Bonifacio:2018zex}
J.~Bonifacio, K.~Hinterbichler, A.~Joyce, and R.~A. Rosen, ``{Shift Symmetries
  in (Anti) de Sitter Space},''
  \href{http://dx.doi.org/10.1007/JHEP02(2019)178}{{\em JHEP} {\bf 02} (2019)
  178}, \href{http://arxiv.org/abs/1812.08167}{{\tt arXiv:1812.08167
  [hep-th]}}.

\bibitem{Bonifacio:2021mrf}
J.~Bonifacio, K.~Hinterbichler, A.~Joyce, and D.~Roest, ``{Exceptional scalar
  theories in de Sitter space},'' \href{http://arxiv.org/abs/2112.12151}{{\tt
  arXiv:2112.12151 [hep-th]}}.

\bibitem{Karateev:2017jgd}
D.~Karateev, P.~Kravchuk, and D.~Simmons-Duffin, ``{Weight Shifting Operators
  and Conformal Blocks},''
  \href{http://dx.doi.org/10.1007/JHEP02(2018)081}{{\em JHEP} {\bf 02} (2018)
  081}, \href{http://arxiv.org/abs/1706.07813}{{\tt arXiv:1706.07813
  [hep-th]}}.

\bibitem{Farrow:2018yni}
J.~A. Farrow, A.~E. Lipstein, and P.~McFadden, ``{Double copy structure of CFT
  correlators},'' \href{http://dx.doi.org/10.1007/JHEP02(2019)130}{{\em JHEP}
  {\bf 02} (2019)  130}, \href{http://arxiv.org/abs/1812.11129}{{\tt
  arXiv:1812.11129 [hep-th]}}.

\bibitem{Armstrong:2020woi}
C.~Armstrong, A.~E. Lipstein, and J.~Mei, ``{Color/kinematics duality in
  AdS$_{4}$},'' \href{http://dx.doi.org/10.1007/JHEP02(2021)194}{{\em JHEP}
  {\bf 02} (2021)  194}, \href{http://arxiv.org/abs/2012.02059}{{\tt
  arXiv:2012.02059 [hep-th]}}.

\bibitem{Albayrak:2020fyp}
S.~Albayrak, S.~Kharel, and D.~Meltzer, ``{On duality of color and kinematics
  in (A)dS momentum space},''
  \href{http://dx.doi.org/10.1007/JHEP03(2021)249}{{\em JHEP} {\bf 03} (2021)
  249}, \href{http://arxiv.org/abs/2012.10460}{{\tt arXiv:2012.10460
  [hep-th]}}.

\bibitem{Alday:2021odx}
L.~F. Alday, C.~Behan, P.~Ferrero, and X.~Zhou, ``{Gluon Scattering in AdS from
  CFT},'' \href{http://dx.doi.org/10.1007/JHEP06(2021)020}{{\em JHEP} {\bf 06}
  (2021)  020}, \href{http://arxiv.org/abs/2103.15830}{{\tt arXiv:2103.15830
  [hep-th]}}.

\bibitem{Zhou:2021gnu}
X.~Zhou, ``{Double Copy Relation in AdS Space},''
  \href{http://dx.doi.org/10.1103/PhysRevLett.127.141601}{{\em Phys. Rev.
  Lett.} {\bf 127} (2021) no.~14, 141601},
  \href{http://arxiv.org/abs/2106.07651}{{\tt arXiv:2106.07651 [hep-th]}}.

\bibitem{Sivaramakrishnan:2021srm}
A.~Sivaramakrishnan, ``{Towards color-kinematics duality in generic
  spacetimes},'' \href{http://arxiv.org/abs/2110.15356}{{\tt arXiv:2110.15356
  [hep-th]}}.

\bibitem{Herderschee:2022ntr}
A.~Herderschee, R.~Roiban, and F.~Teng, ``{On the Differential Representation
  and Color-Kinematics Duality of AdS Boundary Correlators},''
  \href{http://arxiv.org/abs/2201.05067}{{\tt arXiv:2201.05067 [hep-th]}}.

\bibitem{Cheung:2022pdk}
C.~Cheung, J.~Parra-Martinez, and A.~Sivaramakrishnan, ``{On-shell Correlators
  and Color-Kinematics Duality in Curved Symmetric Spacetimes},''
  \href{http://arxiv.org/abs/2201.05147}{{\tt arXiv:2201.05147 [hep-th]}}.

\bibitem{Kampf:2013vha}
K.~Kampf, J.~Novotny, and J.~Trnka, ``{Tree-level Amplitudes in the Nonlinear
  Sigma Model},'' \href{http://dx.doi.org/10.1007/JHEP05(2013)032}{{\em JHEP}
  {\bf 05} (2013)  032}, \href{http://arxiv.org/abs/1304.3048}{{\tt
  arXiv:1304.3048 [hep-th]}}.

\bibitem{Cronin:1967jq}
J.~A. Cronin, ``{Phenomenological model of strong and weak interactions in
  chiral U(3) x U(3)},'' \href{http://dx.doi.org/10.1103/PhysRev.161.1483}{{\em
  Phys. Rev.} {\bf 161} (1967)  1483--1494}.

\bibitem{Mangano:1990by}
M.~L. Mangano and S.~J. Parke, ``{Multiparton amplitudes in gauge theories},''
  \href{http://dx.doi.org/10.1016/0370-1573(91)90091-Y}{{\em Phys. Rept.} {\bf
  200} (1991)  301--367}, \href{http://arxiv.org/abs/hep-th/0509223}{{\tt
  arXiv:hep-th/0509223}}.

\bibitem{Elvang:2015rqa}
H.~Elvang and Y.-t. Huang, {\em {Scattering Amplitudes in Gauge Theory and
  Gravity}}.
\newblock Cambridge University Press, 4, 2015.

\bibitem{deRham:2010eu}
C.~de~Rham and A.~J. Tolley, ``{DBI and the Galileon reunited},''
  \href{http://dx.doi.org/10.1088/1475-7516/2010/05/015}{{\em JCAP} {\bf 05}
  (2010)  015}, \href{http://arxiv.org/abs/1003.5917}{{\tt arXiv:1003.5917
  [hep-th]}}.

\bibitem{Nicolis:2008in}
A.~Nicolis, R.~Rattazzi, and E.~Trincherini, ``{The Galileon as a local
  modification of gravity},''
  \href{http://dx.doi.org/10.1103/PhysRevD.79.064036}{{\em Phys. Rev. D} {\bf
  79} (2009)  064036}, \href{http://arxiv.org/abs/0811.2197}{{\tt
  arXiv:0811.2197 [hep-th]}}.

\bibitem{Novotny:2016jkh}
J.~Novotny, ``{Geometry of special Galileons},''
  \href{http://dx.doi.org/10.1103/PhysRevD.95.065019}{{\em Phys. Rev. D} {\bf
  95} (2017) no.~6, 065019}, \href{http://arxiv.org/abs/1612.01738}{{\tt
  arXiv:1612.01738 [hep-th]}}.

\bibitem{Dyer:2008hb}
E.~Dyer and K.~Hinterbichler, ``{Boundary Terms, Variational Principles and
  Higher Derivative Modified Gravity},''
  \href{http://dx.doi.org/10.1103/PhysRevD.79.024028}{{\em Phys. Rev. D} {\bf
  79} (2009)  024028}, \href{http://arxiv.org/abs/0809.4033}{{\tt
  arXiv:0809.4033 [gr-qc]}}.

\bibitem{Klebanov:1999tb}
I.~R. Klebanov and E.~Witten, ``{AdS / CFT correspondence and symmetry
  breaking},'' \href{http://dx.doi.org/10.1016/S0550-3213(99)00387-9}{{\em
  Nucl. Phys. B} {\bf 556} (1999)  89--114},
  \href{http://arxiv.org/abs/hep-th/9905104}{{\tt arXiv:hep-th/9905104}}.

\bibitem{Boulware:1968zz}
D.~G. Boulware and L.~S. Brown, ``{Tree Graphs and Classical Fields},''
  \href{http://dx.doi.org/10.1103/PhysRev.172.1628}{{\em Phys. Rev.} {\bf 172}
  (1968)  1628--1631}.

\end{thebibliography}\endgroup

\end{document}